\documentclass[pra,twocolumn,english,nofootinbib,longbibliography,superscriptaddress,floatfix,showkeys]{revtex4-2}
\pdfoutput=1 %

\usepackage{stylesheet}
\usepackage{comment}
\usepackage{xcolor}
\usepackage{eczoo}

\begin{document}

\title{
Quantum theory of molecular orientation
}

\author{Victor~V.~Albert}
\thanks{Equal contribution.}
\affiliation{\QUICS}

\author{Eric~Kubischta}
\thanks{Equal contribution.}
\affiliation{\QUICS}

\author{Mikhail~Lemeshko}
\affiliation{\IST}

\author{Lee~R.~Liu}
\affiliation{\JILA}

\date{\today}

\begin{abstract}
We formulate a quantum phase space
for rotational and nuclear-spin states of rigid molecules. 
\nw{
For each nuclear spin isomer, we re-derive the isomer's admissible angular momentum states from molecular geometry and nuclear-spin data,
introduce its angular position states using quantization theory, and develop a generalized Fourier transform converting between the two.
We classify molecules into three types --- asymmetric, rotationally symmetric, and perrotationally symmetric --- with the last type having no macroscopic analogue due to nuclear-spin statistics constraints.
We discuss two general features in perrotationally symmetric state spaces that are Hamiltonian-independent and induced solely by symmetry and spin statistics.
First, we quantify when and how an isomer's state space is completely rotation-spin entangled, meaning that it does not admit any separable states.
Second, we identify isomers whose position states house an internal pseudo-spin or ``fiber'' degree of freedom, and
the fiber's Berry phase or matrix after adiabatic changes in position
yields naturally robust operations, akin to
braiding anyonic quasiparticles or realizing fault-tolerant quantum gates.
We outline how the fiber can be used as a quantum error-correcting code and discuss scenarios where these features can be experimentally probed.
}

\end{abstract}

\maketitle

%

\newcommand{\PeelOffMain}{0}

\tableofcontents

\newpage
\section{Introduction}
%

The quantum-mechanical duality, or conjugacy, between position and momentum \cite{heisenberg_ober_1927}
has far-reaching consequences in physics, chemistry, and materials science.
Conjugate state pairs include states of fixed position and momentum of a continuous-variable system, eigenstates of a qubit's Pauli-\(Z\) and \(X\) matrices, and the Wannier and Bloch functions of an electron.
Conjugacy also occurs in angular systems, with the caveat that
angular position is continuous but periodic, while angular momentum
is unbounded but discrete.

\nw{
Despite over a century of research, a notion of conjugacy remains underdeveloped for many angular systems.
An asymmetry between position and momentum is especially prominent for rotational states of molecules, particularly molecules with symmetries. Molecular orientations, \textit{a.k.a.}\ angular position states or position states
for brevity, have been studied for asymmetric and linear
molecules due to their intuitive geometrical picture \cite{harter_frame_1978,harter_principles_1993,schmidt_rotation_2015,papendell_quantum_2017,stickler_rotational_2017,damme_linking_2017,stickler_rotational_2018,stickler_quantum_2021},
and have been analyzed in local regions of the configuration space, where they are known as pendular states \cite{Friedrich_spatial_1991,rost_pendular_1992,Slenczka_pendular_1994,Friedrich_alignment_1995}. 
But position states for the \textit{entire} configuration space of \textit{arbitrary} molecules have, to our knowledge, yet to be formulated.
}

\begin{figure}[t!]
\includegraphics[width=0.8\columnwidth]{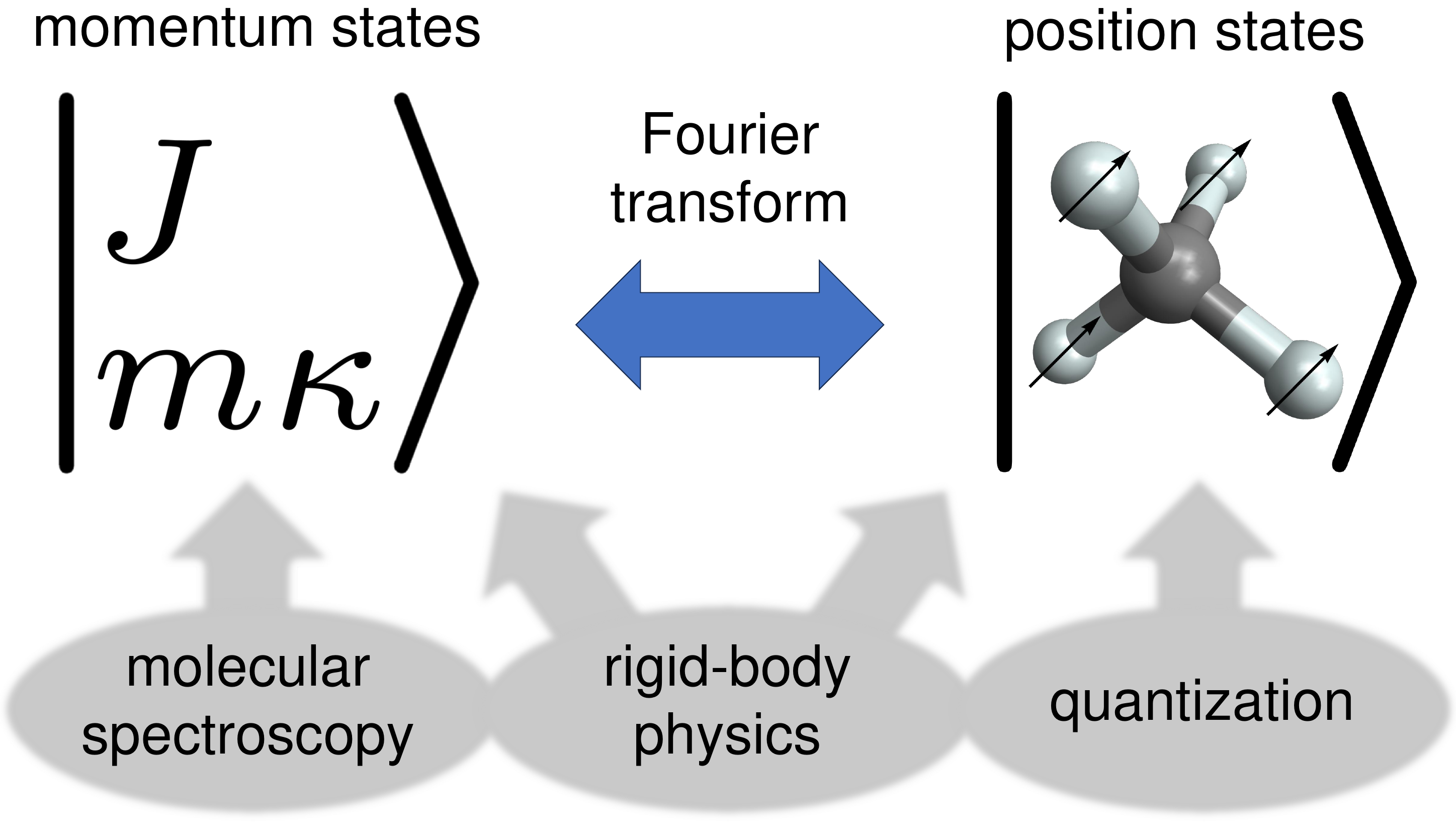}

\caption{\label{fig:outline}
\nw{
We use techniques from molecular spectroscopy \cite{Rasetti_incoherent_1929,Wilson_statistical_1935,Wilson_symmetry_1935,longuet-higgins_symmetry_1963,Van_Vleck_coupling_1951,hougen_interpretation_1971,louck_eckart_1976,harter_orbital_1977,Watson_aspects_1977,berger_classification_1977,ezra_symmetry_1982,Papousek_molecular_1982,bunker_molecular_1998,bunker_fundamentals_2004,brown_rotational_2003,biedenharn_angular_2010,Herzberg_spectra_2013,Herzberg_infrared_1987,Herzberg_electronic_1966,yurchenko2023computational} to develop a rigid-body-based
\cite{casimir1931rotation,Landau_mechanics_1976,Littlejohn_gauge_1997,chirikjian_engineering_2000,wormer_rigid_nodate,Lynch_modern_2017} classification of
molecular angular-momentum states. We show that each
molecular symmetry isomer corresponds to a particular quantization~\cite{isham_topological_1984,Landsman_geometry_1991,Tanimura_reduction_2000,levay_canonical_1996} of the configuration space, yielding \vva{angular position} states and a Fourier transform~\vva{\eqref{eq:fourier}}.
}
}
\end{figure}
\begin{figure*}[t!]
\includegraphics[width=0.95\textwidth]{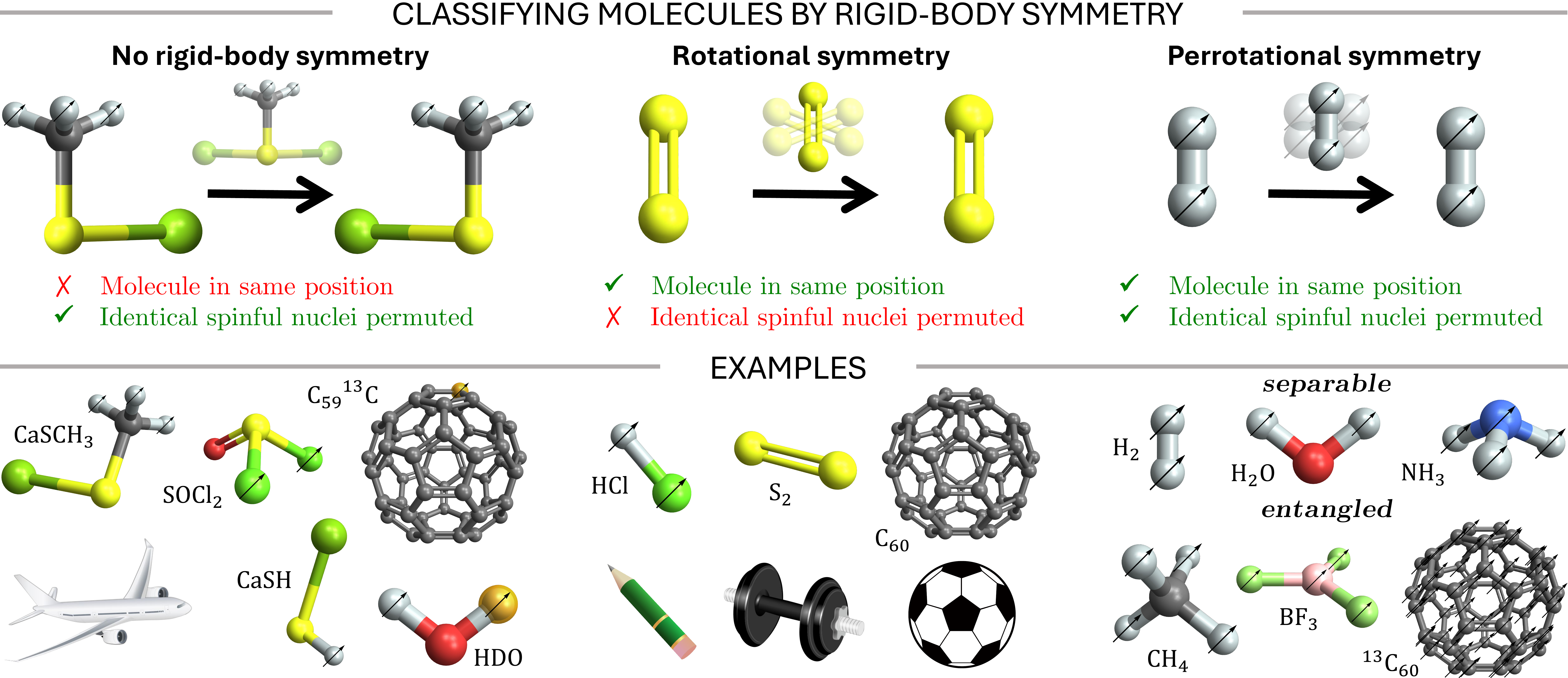}

\caption{\label{fig:header}Molecular rotational states
can be characterized by their behavior under orientation-preserving rotations.
Orientations, or angular positions, of \textit{asymmetric} molecules are the same as those of any rigid body (e.g., an airplane) whose center of mass is fixed.
While every rotation moves an asymmetric molecule to a different position, \textit{rotationally symmetric} molecules remain in the same position under some rotations.
Such molecules can have nonzero nuclear spin (marked by ``\(\nearrow\)''), but any rotations that permute identical spinful nuclei must also rotate the rest of the molecule into a different position.
Molecule-frame rotations that permute spinful nuclei and \textit{do} leave the rest of the molecule invariant
have to produce the molecule's nuclear-spin statistics.
We call any molecule admitting such permutation-rotations a \textit{perrotationally symmetric} molecule.
The rotational state space of such molecules has no classical analogue and is the main subject of this work.
Perrotationally symmetric molecules with non-commutative symmetries can admit \textit{separable} or \textit{entangled} nuclear spin isomers, with the latter exhibiting complete rotation-spin entanglement due to a combination of symmetry and nuclear-spin statistics.
}
\end{figure*}

\nw{
One reason for this is that momentum, not position, plays the dominant role in molecular spectroscopy.
Spectroscopic experiments with rotating and vibrating
molecular gases provide information about the natural
molecular Hamiltonian. This Hamiltonian is nearly diagonal in the momentum-state basis, making it maximally
incompatible with position states.

On the other hand, this century has brought forth the
ability to cool and trap \textit{individual} molecules \cite{Ospelkaus_efficient_2008,Ni_high_2008,Takekoshi_ultracold_2014,Barry_magneto-optical_2014,Park_ultracold_2015,Rvachov_long-lived_2017,anderegg_radio_2017,truppe_molecules_2017,collopy_3d_2018,seeselberg_modeling_2018,liu_building_2018,de_marco_degenerate_2019,liu_molecular_2019,urban_coherent_2019,Voges_ultracold_2020,Ding_sub-doppler_2020,Valtolina_dipolar_2020,Mitra_direct_2020,augenbraun_molecular_2020,Schindewolf_evaporation_2022,Stevenson_ultracold_2023,Lin_microwave_2023,Anderegg_quantum_2023,Glikin_systematic_2023,Vilas_magneto-optical_2022,Koller_electric-field-controlled_2022,Vilas_optical_2023,Lu_raman_2023,Bao_raman_2023,holland_-demand_2023,Bao_dipolar_2023,Singh_dynamics_2023} and
to orient molecules
using strong electric fields \cite{Loesch_brute_1990,Friedrich_spatial_1991,rost_pendular_1992,Slenczka_pendular_1994,Friedrich_alignment_1995,Seideman_rotational_1995,Kim_spectroscopy_1996,Fleischer_molecular_2011}
(see overviews \cite{Stapelfeldt_colloquium_2003,Malinovskaya_atomic_2015,fitch_laser-cooled_2021,committee_on_identifying_opportunities_at_the_interface_of_chemistry_and_quantum_information_science_advancing_2023,Augenbraun_direct_2023,Cornish_quantum_2024}).
Given this increased level of
control, it is both reasonable and necessary to move away
from the Hamiltonian-centric approach. For one, the natural Hamiltonian is not particularly useful, and it should
be possible to mitigate it or even engineer it away using,
e.g., external drives \cite{koch_quantum_2019,Pozzoli_classical_2022,Leibscher_full_2022}, creative coupling to ancillae \cite{lin_quantum_2020,todaro_state_2021}, or tools from other platforms \cite{frattini_3-wave_2017,elliott_designing_2018}. In
addition, since the natural Hamiltonian often has more symmetry than the molecule, the molecule’s state
space cannot be inferred from that Hamiltonian alone
(see Sec.~\ref{sec:prior_work}).

Moving toward a more symmetry-centric approach, we
determine the admissible angular phase space solely from
symmetry and nuclear spin-statistics constraints.
We expect this approach to work for all degrees of freedom,
but, for this work, assume that orbital, electron-spin,
and vibrational degrees of freedom are decoupled and
unaffected by symmetry rotations, leaving only the rotational and nuclear-spin degrees of freedom (assuming a
fixed center of mass). This rigid-rotor-like approximation
\cite[Sec. 10.5]{bunker_molecular_1998} captures the leading-order physics of small
molecules whose relevant intramolecular interactions are
either small on the energy scale of molecular rotations, or
can be avoided altogether by working in the vibrational
ground state.

We perform a rigid-body-based classification of molecular rotation-spin states (see Fig.~\ref{fig:header}), adapt spectroscopic
recipes for obtaining momentum states, introduce molecular position states via quantization theory, and express each set of states in terms of its conjugate via a
generalized Fourier transform (see Fig.~\ref{fig:outline}). Enforcing
only the minimal symmetry of the molecule, the resulting phase space can host any Hamiltonian of the same
or greater symmetry, failing only for “violent” Hamiltonians that, e.g., break molecules and/or model chemical
reactions.

Armed with a birds-eye view of molecular phase space, we proceed to study two of its exotic features.
We quantify the nature and strength of symmetry-induced rotation-spin entanglement and identify robust rotation-induced ``topological'' operations on position states.

We show that the state spaces of certain nuclear spin isomers form completely entangled subspaces~\cite{wallach2002unentangled,parthasarathy2004maximal,parthasarathy2005extremal,walgate2008generic} --- subspaces that do not admit any separable states with respect to the rotation-spin tensor-product decomposition. 
Rotation-spin entanglement is well known in the spectroscopy community, but it was never systematically quantified, to our knowledge (see Sec.~\ref{sec:prior_work}).
We relate this entanglement to the underlying symmetry-group data and show that it can be quite predominant, with entangled basis states spanning over \(98\%\) of the entire state space of certain molecules.
Its presence may not seem surprising from the Hamiltonian-centric point of view, as generic molecular eigenstates are highly entangled with respect to the various degrees of freedom. 
However, complete entanglement is a symmetry-induced property of the \textit{state space},
independent of and unaffected by any symmetric Hamiltonian.

``Topological'' operations in the form of Berry phases \cite{berry_quantal_1984} are present in isomers as simple as
\ifthenelse{\equal{\PeelOffMain}{0}}{ortho hydrogen (see Sec.~\ref{sec:toy-model}),}{ortho hydrogen,}
and position states of larger molecules admit non-Abelian operations in the form of Berry matrices \cite{simon_holonomy_1983,wilczek_appearance_1984,Zanardi_holonomic_1999}.
\vva{We prove that such holonomic operations persist, up to exponentially small corrections, when Gaussian wavepackets are substituted for bona-fide molecular position states\ifthenelse{\equal{\PeelOffMain}{0}}{ (see Sec.~\ref{subsec:Connection})}{}.}
\vva{These operations provide} another example of the ``holonomy$=$monodromy'' correspondence from condensed-matter physics~\cite{read_non-abelian_2009} and quantum information~\cite{gottesman_fibre_2017}.
This suggests a new way to robustly store and process quantum information, and we show that certain symmetric nuclear spin isomers provide a similar level of protection against noise as \eczoo[molecular codes]{molecular}~\cite{albert_robust_2020} do for asymmetric molecules (see Sec.~\ref{sec:fiber-codes}).

A full-fledged molecular phase space 
should serve as the backbone for molecular quantum science and technology, akin to the more conventional phase spaces associated with qubit and continuous-variable \cite{braunstein_quantum_2005} platforms.
This space is primed to be utilized for
quantum simulation \cite{Micheli_toolbox_2006,gorshkov_quantum_2011,wall_simulating_2013,Wall_realizing_2015,Karle_topological_2023},
metrology \cite{DeMille_probing_2017,Kondov_molecular_2019,Leung_terahertz_2023},
ultra-cold chemistry \cite{heazlewood_towards_2021,Liu_quantum_2023},
and collision physics \cite{Cheuk_observation_2020}.
In particular, fiber encodings and their topological operations are relevant to molecular quantum information processing \cite{demille_quantum_2002,tesch_quantum_2002,yelin_schemes_2006,wei_entanglement_2011,ni_dipolar_2018,yu_scalable_2019,Sawant_ultracold_2020,Zeppenfeld_robust_2023,Jain_AE_2023,Freedman_quantum_2019}.
We do not confirm the prevalence of other allegedly fragile quantum effects in large molecules \cite{fisher_quantum_2015} and chemical reactions \cite{quack1977detailed,fisher_quantum_2018,Muechler_topological_2020,Palma_topological_2021}, but our noticed features do complement this unique 
\ifthenelse{\equal{\PeelOffMain}{0}}{line of thought (see Sec.~\ref{sec:prior_work}).}{line of thought.}
}

\section{Summary of results}
\label{sec:summary}

\nw{
We develop a phase space for molecules whose orbital,
electron-spin, and vibrational factors (A) are decoupled
from the rotational and nuclear-spin factors and (B) are
unaffected by (read: transform trivially under) symmetry
rotations. A typical molecule to keep in mind is a rigid
one that has been cooled to its ground electronic and
vibrational state. Molecules in fixed excited electronic
and vibrational states are also allowed, as long as those
factors are invariant under symmetry rotations.
}

\subsection{Classification of molecular rotational states}

We group molecules into three classes, based on their behavior under orientation-preserving, or proper, rotations (see Fig.~\ref{fig:header}). 
Asymmetric and rotationally symmetric molecules are analogous to asymmetric and symmetric rigid bodies, respectively.
Molecules in the third ``perrotationally'' symmetric class have no macroscopic
analogue because their state space is determined by additional nuclear-spin
statistics constraints.
\nw{
\vva{The term ``perrotation'' is a contraction of ``permutation'' and ``rotation''~\cite{brester_Kristallsymmetrie_1923,wigner_uber_1930,gilles_internal_1972}.
It reflects that symmetry rotations of such molecules also permute identical spinful nuclei and thus have to respect their Bose or Fermi statistics.}
Previous treatments do not distinguish the rotational from the perrotational classes, interpreting spinless nuclei as spin-zero particles instead of as classical objects.
We build up to our result by first reviewing an established spectroscopic recipe for determining momentum states of all three classes.
}

\prg{No rigid-body symmetry}

Defining the rotational state space of an asymmetric molecule requires two frames: the observer's \textit{lab frame} and the \textit{molecule frame}, with the latter bolted to and moving with the molecule.
Each frame can be rotated, and the two types of lab-based and molecule-based rotations correspond to the two factors of the group $\so 3\times\so 3$
of proper rotations acting on
the molecule.
The basis of fixed-momentum or \textit{rotational} states $|_{m}^{\ell},\om\ket$
is labeled by
the total angular momentum $\ell$ and its projections, $m$ and $\om$, onto
the $\z$-axis of the lab and molecule frames, respectively.
\prg{Rotational symmetry}

For molecules with symmetries, positions that are equivalent under molecule-frame symmetry rotations have to be identified.
This identification
collapses the molecule frame, restricting the space \(\{|\om\ket\}\) of molecule-frame momenta to a particular subspace.
For a \textit{rotationally} symmetric molecule,
this admissible state space is the same as that of a rigid body with the same symmetry.

For example, the momentum states of a heteronuclear diatomic 
\ifthenelse{\equal{\PeelOffMain}{0}}{(equivalently, a pencil; see Example~\ref{ex:heteronuclear})}{(equivalently, a pencil)}
are $|_{m}^{\ell},\om=0\rangle$
because only these states are invariant under rotations
that form the molecule's symmetry group, $\C{\infty}\cong\so 2$. The state space
of $\D{\infty}\cong\OO 2$-symmetric spinless homonuclear diatomics such
as disulfur 
\ifthenelse{\equal{\PeelOffMain}{0}}{(equivalently, a dumbbell; see Example~\ref{ex:disulfur})}{(equivalently, a dumbbell)}
is further
restricted to even $\ell$ due to additional invariance under
rotations that permute the two nuclei. The state space of rotationally
$\C 3=\Z_{3}$-symmetric molecules like cobalt tetracarbonyl hydride (whose cobalt and carbon nuclei have no nuclear spin) admits
multiple $\om$'s for a given $\ell$; their admissible state set
consists of every \(\om \equiv 0\) modulo 3.
For other symmetry groups, such as the icosahedral symmetry group of the fullerene 
\ifthenelse{\equal{\PeelOffMain}{0}}{(equivalently, a soccer ball; see Example~\ref{ex:fullerene}),}{(equivalently, a soccer ball),}
admissible states are superpositions of
multiple $|\om\ket$ states.

In terms of representation theory, restricting to a subspace of symmetric
states is equivalent to picking states that transform according to
the trivial irreducible representation, or \textit{irrep}, of the
symmetry group \(\G\). We label the admissible momentum states of a
general rotationally symmetric molecule by $|_{m\k}^{\ell}\ket$,
where $\k$ indexes copies (\textit{a.k.a.}\ the \textit{multiplicity
space}) of the trivial $\G$-irrep present in the
space of $|\om\ket$ states for each $\ell$.

\prg{Perrotational symmetry}

Symmetry rotations acting on \textit{per}rotationally \cite{brester_Kristallsymmetrie_1923,wigner_uber_1930,gilles_internal_1972} symmetric molecules
\textit{per}mute identical spinful nuclei, correlating rotational states with the nuclear spin.
Molecule-frame symmetry rotations gain extra permutation operators to account for effects on the spins, and identification of symmetry-related positions yields a subspace of the \textit{joint} rotation-spin space.

Admissible tensor products of rotational and
nuclear-spin states are grouped into molecular \textit{isomers}. Each
isomer yields the same required Bose or Fermi spin statistics under molecule-based
symmetries, but differs from other isomers in how the \textit{individual}
rotational and nuclear-spin factors transform.

For example, consider \ce{H2} --- a perrotationally $\D{\infty}$-symmetric
molecule that \textit{cannot} be treated the same way as a dumbbell.
Molecule-frame rotations that permute its identical spin-half
hydrogen nuclei must result in a spin-statistics phase of $-1$. Admissible
states are split into two isomers --- para and ortho
--- with the required phase coming from either the nuclear-spin or the 
\ifthenelse{\equal{\PeelOffMain}{0}}{
rotational factor (see Sec.~\ref{sec:toy-model} and Example~\ref{ex:deuterium-a2}).
}{
rotational factor.
}

Basis states of para hydrogen consist of the same
even-$\ell$ rotational states as a dumbbell, tensored with the anti-symmetric (singlet) nuclear-spin state.
States of ortho
hydrogen consist of rotational states of \textit{odd}
angular momentum, tensored with symmetric (triplet) nuclear-spin states.
The existence of the odd-momentum ortho isomer shows that
perrotationally symmetric molecules (such as hydrogen and deuterium) can occupy momentum states inaccessible
to rotationally symmetric molecules with the same symmetry (such as disulfur).

In terms of representation theory, rotational states of para hydrogen transform according to the trivial irrep $\a_{1}$
of the symmetry group. On the other hand, rotational states of the
ortho isomer transform according \vva{to} the sign irrep $\a_{2}$. Together,
products of the rotational and nuclear-spin irreps of each isomer yield the
correct spin-statistics irrep, $\a_{1}\ot\a_{2}=\a_{2}$
and $\a_{2}\ot\a_{1}=\a_{2}$, respectively.

More generally, an isomer is uniquely determined by the triple of
\textit{rot}ational, \textit{nuc}lear-spin, and \textit{mol}ecular
spin-statistics irreps,
\begin{equation}
\l\ot\t\down\s\,,\label{eq:branching}
\end{equation}
where ``$\down$'' means that the tensor-product representation
has to be projected into the desired irrep $\s$. The projection
becomes an equality for scalar irreps ($\dim\l=\dim\t=1$), like those
of \ce{H2}, becoming relevant only for irreps of higher dimension.%
\ifthenelse{\equal{\PeelOffMain}{0}}{
We tabulate all possible irrep triples for all symmetry groups in Table \ref{tab:isomer-table}.
}{
We tabulate all possible irrep triples for all symmetry groups.
}

Despite complications due to spin statistics,
admissible rotational states of a perrotational isomer are still indexed by $\ell$, $m$, and $\k$, but with $\k$
now going over the multiplicity space of $\l$ for each momentum.
This long-known recipe, stemming from nearly a century of rigorous work in spectroscopy \cite{Rasetti_incoherent_1929,Wilson_statistical_1935,Wilson_symmetry_1935,longuet-higgins_symmetry_1963,Van_Vleck_coupling_1951,hougen_interpretation_1971,louck_eckart_1976,harter_orbital_1977,Watson_aspects_1977,berger_classification_1977,ezra_symmetry_1982,Papousek_molecular_1982,bunker_molecular_1998,bunker_fundamentals_2004,brown_rotational_2003,biedenharn_angular_2010,Herzberg_spectra_2013,Herzberg_infrared_1987,Herzberg_electronic_1966,yurchenko2023computational}, provides a momentum basis for any molecule.

\prg{Induced representations \& quantization}

\nw{
A main result of this work is identifying that momentum states form an \textit{induced representation} \cite{mackey1968induced,mackey_induced_1952,coleman_induced_1968,Inui_group_1990,carter_lectures_1995,chirikjian_engineering_2000,folland_course_2016} of the group of lab-based rotations, denoted by \(\text{Ind}_{\l}^{\G}\so{3}\) or $\l\up\so{3}$ when the symmetry group is evident.
Induced representations allow us to relate molecular spectroscopy to quantization theory \cite{isham_topological_1984,Landsman_geometry_1991,Tanimura_reduction_2000,levay_canonical_1996},
revealing that each nuclear spin isomer is in one-to-one correspondence with a particular quantization of the configuration space of molecular orientations.
This relation readily yields molecular position states and the Fourier transform, summarized in Sec.~\ref{sec:position-states-summary}.

\begin{theorem*}
The rotational state space of a $\G$-symmetric molecular
isomer
transforms as an induced representation $\l\up\so{3}$ under lab-based rotations,
where $\l$ is an irrep of the symmetry group.
\end{theorem*}


\vva{Concretely, \(\l\up\so{3}\) is the space of \((\dim\l)\)-component wavefunctions \(\psi(\gr)\) of the molecule's orientation \(\gr\in\so 3\) that transform according to \(\l\) under molecule-frame symmetry rotations \(\gr\to\gr\rg\), where \(\rg\in\G\). Lab-based rotations \(\gr\to\rh\gr\), where \(\rh\in\so 3\), map this space to itself.}
Nuclear spin isomers in all three classes are exhaustively classified using induced representations.
In the asymmetric case, $\G$ becomes the trivial group. 
In the rotationally
symmetric case, there is only one isomer, and $\l$ becomes the trivial
irrep.
In the perrotationally symmetric case, the combination of symmetry
and spin statistics allows for more general irreps $\l$, governed by condition (\ref{eq:branching}). For example, the odd-$\ell$
momentum states of ortho hydrogen make up the representation
$\text{Ind}_{\a_2}^{\D \infty}\so{3} = \a_{2}\up\so{\textnormal{3}}$.
The same classification can be made for molecules confined to move in only two dimensions (see Sec.~\ref{sec:toy-model}).

Our classification is similar to that of solid-state materials \cite{bradlyn_topological_2017} and nematic liquid crystals \cite{mermin_topological_1979,liu2016generalized}: all three are based on induced representations and distinguish isomers with ``topologically nontrivial'' features.
In the case of molecules, induced representations with nontrivial inducing irreps exhibit ``topological'' behavior under lab-based rotations, and those with vector irreps exhibit complete rotation-spin entanglement.
}

\subsection{Complete rotation-spin entanglement}

Since admissible states of symmetric molecules form a subspace of
the asymmetric molecular state space $\{|_{m}^{\ell},\om\ket\}$,
we can decompose the latter into the various induced representations
for a given symmetry group.

For abelian $\G$ (i.e., groups with commuting
elements), the asymmetric state-space identity operator decomposes as $\id_{\rot}=\bigoplus_{\ir}\,\ir\up\so 3$, where \(\Gamma\) goes over all \(\G\)-irreps.
Selecting an isomer projects this space into the $\ir=\l$ sector,
which is then tensored with the appropriate \vva{nuclear-spin} irrep $\t$
such that the spin-statistics condition
(\ref{eq:branching})
is satisfied. Each such isomer admits tensor-product states of the
form $|_{m\k}^{\ell}\ket_{\rot}|\chi\ket_{\nat}$, where $\chi\in\{1,2,\cdots,\mst\}$
indexes the multiplicity of the corresponding $\t$ irrep up to the statistical weight $\mst$ \cite{bunker_molecular_1998,bunker_fundamentals_2004}.

More interesting isomers occur for \textit{non-Abelian} symmetry
groups. Such groups admit both scalar ($\dim\ir=1$) and ``vector''
irreps ($\dim\ir>1$), with the latter hosting internal degrees of freedom.
The asymmetric state-space identity in such cases decomposes as
\cite{isham_topological_1984,Landsman_geometry_1991,Tanimura_reduction_2000,levay_canonical_1996}
\begin{equation}
\id_{\rot}=\mathord{\bigoplus}_{\ir}\,\left[\ir\up\so{3}\right] \ot\id_{\dim\ir}\,,
\end{equation}
where the induced representation is present with multiplicity
$\dim\ir$ for each irrep $\ir$.
Selecting an isomer once again projects the above space into the
$\ir=\l$ sector. 

An analogous selection of the $\t$ irrep occurs
on the nuclear-spin side, resulting in the bipartite irrep-multiplicity subspace $\id_{\dim\t}\ot\id_{\mst}$.
But since \(\l\) and \(\t\) may be vector irreps, their internal degrees of freedom
%
have to be \textit{projected}, via
Eq.~(\ref{eq:branching}), into a subspace transforming according
to the spin-statistics irrep $\s$. This subspace can be determined using \vva{Clebsch-Gordan}
tables \cite{altmann_point-group_1994,ceulemans_group_2013} and turns out to always be one-dimensional, defined by a state we denote by $|\s\ket$.

Basis states for a general isomer can be written as
\begin{equation}
|_{m\k}^{\ell}\ket_{\rot}|\s\ket|\chi\ket_{\nat}\,.\label{eq:basis}
\end{equation}
The three indices of the first factor belong to the induced representation, the
middle factor is supported on the composite rotation-spin space $\id_{\dim\l}\ot\id_{\dim\t}$,
and the fourth index \(\chi\) indexes the nuclear-spin multiplicity up
to the statistical weight.

Exhaustively determining all possible nuclear spin isomers, we observe
that the dimensions of participating irreps always coincide,
\begin{equation}
\dim\l=\dim\t\equiv\dm\quad\quad\text{(Schmidt rank)}\,,
\end{equation}
and that the state $|\s\ket$ is maximally entangled with respect
to the rotation-spin tensor product. Quantitatively, the state has
\textit{Schmidt rank} $\dm$ \cite{Ekert_entangled_1995}, i.e., it can only be written as
a superposition of at least $\dm$ of any basis states that span the
$\dm^{2}$-dimensional composite space $\id_{\dm}\ot\id_{\dm}$. As
such, we call nuclear spin isomers with $\dm>1$ \textit{entangled} and
denote $\dm=1$ isomers as \textit{separable}.
\vva{For example, both isomers of \ce{H2} are separable.
Even (odd) rotational states occur only with total nuclear spin zero (one), but each isomer is spanned by products of rotational and nuclear-spin states.}

\nw{
The state
$|\s\ket$ ``dresses'' each basis state in Eq.~\eqref{eq:basis} in a way that is impossible
to remove, even by superposition.
While basis-state superpositions or
changes of basis can certainly yield states of higher entanglement,
the Schmidt rank of \textit{any} state of the isomer cannot go below
$\dm$.
Any entangled isomer thus forms a \textit{completely
entangled subspace}  \cite{wallach2002unentangled,parthasarathy2004maximal,parthasarathy2005extremal,walgate2008generic}, i.e., a subspace that does not contain any separable states.
\vva{Since separable mixed states are mixtures of product states, every mixed state of such an isomer, including a thermal one, is entangled as well.}
This rotation-spin entanglement
is enforced by symmetry and spin statistics and cannot be removed
without transitioning to another isomer or breaking the assumptions of the model.
}

\ifthenelse{\equal{\PeelOffMain}{0}}{
Complete entanglement %
is quite widespread.
Entangled states make up over a third of all basis states from Eq.~(\ref{eq:basis}) for a series of 
dihedrally symmetric chains (see Table \ref{tab:fractions}).
Tetrahedrally symmetric molecules like methane admit three separable and one entangled isomer, but the
latter occupies over 56\% of the entire state space
(see Example~\ref{ex:methane-fraction}). 
An extreme case is isotopic \ce{$^{13}$C60} fullerene: all but one of its isomers and 98\% of its basis states are completely entangled (see Example~\ref{ex:fullerene-h}).
\vva{These fractions are also the abundances of entangled isomers at temperatures well above the rotational constants, where each isomer is populated in proportion to \(\dm\mst\).
For methane, the entangled fraction of \(9/16\) is the ortho share of the 5:9:2 meta:ortho:para ratio.
This ratio persists upon cooling because nuclear spin isomers interconvert slowly~\cite{chapovsky_nuclear_1999}.
Over half of the molecules in a typical methane sample are thus in a completely entangled isomer.}
}{
Complete entanglement %
is quite widespread.
Entangled states make up over a third of all basis states from Eq.~(\ref{eq:basis}) for a series of 
dihedrally symmetric chains.
Tetrahedrally symmetric molecules like methane admit three separable and one entangled isomer, but the
latter occupies over 56\% of the entire state space. 
An extreme case is isotopic \ce{$^{13}$C60} fullerene: all but one of its isomers and 98\% of its basis states are completely entangled.
}

\begin{figure}[t]
\centering{}
\includegraphics[width=1.0\columnwidth]{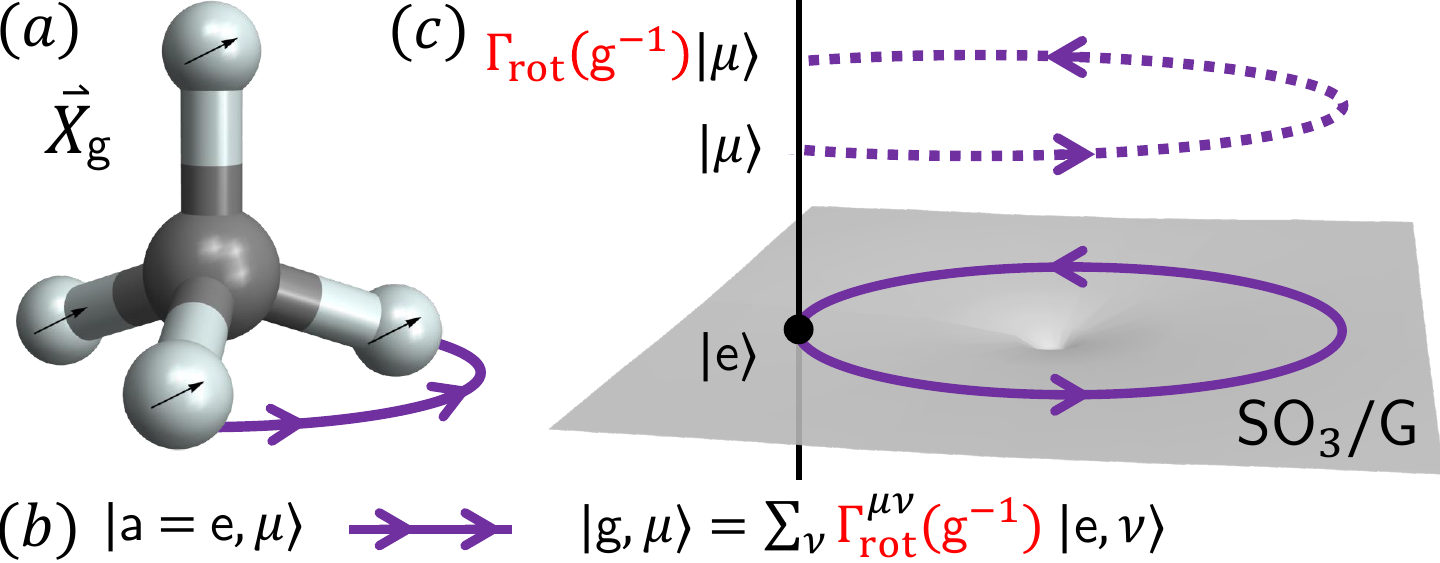}
\caption{\label{fA0:holonomy}
With the molecular frame collapsed due to identification of positions related by molecule-based symmetries, we are still free to rotate the lab frame.
(a) Lab-based rotations, \(\rr_{\rg}\) for \(\rg\) in a molecule's symmetry group, map the molecule to the same position.
For example, a rotation rotates the methane molecule by 120$^{\circ}$, with the path traversed by the forward-most nucleus depicted by two arrows.
(b) A position state \(|\cs=\re\ket\) (with \(\cs\) set to the identity element \(\re\) for simplicity) of a perrotationally symmetric molecule carries with it a \textit{fiber} space, spanned by the basis \(\{|\mu\ket\}\).
Symmetry rotations transform the internal state of fiber via the monodromy matrix \({\color{red}\l(\rg^{-1})}\).
(c) In a cartoon depiction, the fiber can be thought of as a vertical line at each point \(\cs\) in the position-state space \(\nicefrac{\so 3}{\G}\), with a non-contractible closed path resulting in a monodromy action on the fiber state.
}
\end{figure}

\subsection{Position states}\label{sec:position-states-summary}

Dual to the momentum states are molecular states that
label orientations, or angular positions, of a molecule.
Such idealized states are Dirac-\(\delta\) orthonormal, just
like position states of a harmonic oscillator or a two-dimensional rotor \cite{albert_general_2017}. 
As with those established systems, normalizable finite-energy approximations of position states have to be considered for applications, and
we \vva{show} that approximate states\vva{' overlaps and holonomy} approach
those of the ideal states exponentially
\ifthenelse{\equal{\PeelOffMain}{0}}{in their \vva{mean squared} angular momentum (see \vva{Sec.~\ref{subsec:Connection}}).}{in their \vva{mean squared} angular momentum.}
\vva{In practice, approximate position states are realized by pendular states in strong aligning fields or by transiently laser-aligned wavepackets~\cite{Friedrich_spatial_1991,Stapelfeldt_colloquium_2003,koch_quantum_2019}.
Nonlinear molecules require fixing all three axes.
Elliptically polarized laser fields achieve this for asymmetric tops~\cite{larsen_three_2000}, and crystal fields can do so for more symmetric molecules (see Sec.~\ref{subsec:expt}).
These approximations are accurate when their angular spread is small compared to the smallest angle of a symmetry rotation that permutes nuclei (e.g., \(120^\circ\) for ammonia).
Approximate states undergo the same monodromies as the ideal states, since the operator that suppresses their high angular momenta commutes with rotations (see Sec.~\ref{subsec:approximate-states}).}

Each position state of an asymmetric molecule is labeled by the proper rotation \(\gr\) required to orient the lab frame into the molecule frame at that position.
For asymmetric molecules, each position is distinct by definition, so all rotations are required to label all possible positions.
The states $|\gr\ket$ for \(\gr\in\so3\) span the
induced representation space of $\irt\up\so{3}$ (see theorem above), induced by the
trivial irrep \(\irt\) of the trivial symmetry group.

Rotationally symmetric molecules (and symmetric rigid bodies)
are invariant under a subgroup $\G$ of molecule-based rotations and do not
require the entire $\so 3$ space to be used as their configuration
space. Instead, their position states $|\cs\ket$ correspond to cosets
$\cs\G$ of the symmetry group in $\so 3$, with coset representatives \(\cs\) making up the \textit{coset space} \(\nicefrac{\so 3}{\G}\) \cite{postnikov_three-dimensional_1992,wulker_quantizing_2019}. The set \(\{|\cs\ket\}\) spans the representation
space of $\irt\up\so{3}$, induced by the trivial irrep of \(\G\).

Perrotationally $\G$-symmetric molecular position states span the
representation space of more general $\l\up\so{3}$.
In the context of quantization, induced representations for different \(\l\) correspond to different ways to ``quantize'' a particle on the coset space.
For separable isomers
($\dim\l=\dm=1$), positions are still labeled by coset representatives
$\cs\in\nicefrac{\so 3}{\G}$.

For entangled isomers (\(\dm >1\)), coset representatives are \textit{not} sufficient to describe the full state space.
Instead, each position carries with it an \textit{internal} degree of freedom
in the form of a $\dm$-dimensional vector space, $\{|\mu\ket,\,1\leq\m\leq\dm\}$.
From a geometrical perspective, the coset space acts as the base
space while the vector space serves as the \textit{fiber} of a fiber
bundle \cite{isham_topological_1984} [see Fig.~\ref{fA0:holonomy}(c)].

Position states $|\cs,\m\ket$ can be expressed as superpositions
of \textit{many} momentum states $|_{m\k}^{\ell}\ket$, and \vva{vice versa},
conforming to the Heisenberg uncertainty principle.
Both bases
are complete and orthogonal, and the Fourier transform
for a general entangled isomer is
\begin{subequations}
\label{eq:fourier}
\begin{align}
|\cs,\mu\ket&=\sum_{\ell\down\l}\sum_{|m|\leq\ell}\sum_{\k=1}^{\mult(\ell)}H_{m\k}^{\ell}(\cs,\m)|_{m\k}^{\ell}\ket\\|_{m\k}^{\ell}\ket&=\int_{\nicefrac{\so 3}{\G}}\dd\cs\sum_{\mu=1}^{\dm}H_{m\k}^{\ell\star}(\cs,\m)|\cs,\mu\ket~,
\end{align}%
%
\ifthenelse{\equal{\PeelOffMain}{0}}{
\vva{where the coset representative \(\cs\) is a rotation that can be parameterized by Euler angles.
Up to normalization, the induced harmonic \(H_{m\k}^{\ell}(\cs,\m)\) is the Wigner matrix element \(\bra m|D^{\ell}(\cs)|\m\k\ket\)~\cite{varshalovich_quantum_1988}, whose right index labels a \(\G\)-adapted basis (see Sec.~\ref{sec:position}).}
}{
where the induced harmonic $H$ is a cousin of the Wigner $D$-matrix.
}%
The shorthand
``$\ell\down\l$'' stands for all $\ell$ which house at least one
copy of $\l$, and $\mult(\ell)\geq1$ labels how many copies there are
for that momentum.
The above states are tensored with
the appropriate \(\mst\)-dimensional nuclear-spin factor \(\{|\chi\ket\}\) from Eq.~\eqref{eq:basis} to yield the correct spin statistics.
\end{subequations}

\subsection{``Topological'' holonomy}\label{subsec:holonomy-summary}
Lab-based rotations are ``passive'' in that they rotate the lab frame.
However, since molecular position is fully characterized by the molecule's orientation \textit{relative} to the lab frame, performing lab-based rotations
is equivalent to ``actively'' rotating the molecule about an axis defined in that frame.

Lab-based symmetry rotations rotate each nucleus into a different position \textit{without} permuting any nuclear-spin factors.
Acting purely on the rotational factor, lab-based symmetries induce non-trivial isomer-\textit{dependent} behavior, in contrast to the isomer-\textit{in}dependent spin statistics induced by molecule-based perrotations.

A lab-based rotation, \(\rr_{\rg}\) for some \(\rg\in\so3\), rotates a nuclear spin isomer from its initial position \(\cs\) to a position denoted by \(\rg\cs\), tracing out a path in the coset space \(\nicefrac{\so 3}{\G}\).
The fiber \(|\mu\ket\) is carried along this path, transforming in a way dictated by the isomer's inducing irrep \(\l\).
The nuclear factor \(|\chi\ket\) is also carried along, but transforms trivially and so is omitted from now on.

If \(\rg\) implements a symmetry of the molecule, then the final and initial positions are identified, \(\rg\cs\cong\cs\), but \textit{only} up to a residual evolution on the fiber (see Fig.~\ref{fA0:holonomy} for the \(\cs=\re\) case).
This evolution --- a \textit{monodromy} --- is the same for the class of paths related to the original path by smooth deformations, depending only on the global, or ``topological'', features of the path.

For general closed adiabatic paths in position space, monodromy can be thought of as a robust version of holonomy \cite{simon_holonomy_1983,wilczek_appearance_1984,Zanardi_holonomic_1999} (itself a generalization of the Berry phase \cite{berry_quantal_1984}).
\vva{We prove that position-state} holonomy does not depend on any local, or geometric, features of the path\ifthenelse{\equal{\PeelOffMain}{0}}{ \vva{(see Theorem~\ref{thm:flat}).}}{\vva{.}}
\vva{This independence is} an instance of the
``holonomy \(=\) monodromy'' correspondence \cite{read_non-abelian_2009,gottesman_fibre_2017}.
\ifthenelse{\equal{\PeelOffMain}{0}}{
With respect to past instantiations of ``topological'' holonomy (see Sec.~\ref{sec:prior_work}), the surprise here is its presence in ``classical'' rigid-body position states.
}{
With respect to past instantiations of ``topological'' holonomy, the surprise here is its presence in ``classical'' rigid-body position states.
}
For example, a lab-frame rotation exchanging the positions of the two hydrogens of \ce{H2}
induces a monodromy of $-1$ for the ortho isomer, in contrast to the trivial $+1$ monodromy of the para isomer.%
\ifthenelse{\equal{\PeelOffMain}{0}}{
This is easily derived by noting that the former (latter) isomer hosts only odd-momentum (even-momentum) rotational states; see Sec.~\ref{sec:toy-model} and Example~\ref{ex:deuterium-a2-monodromy}.
}{
This is easily derived by noting that the former (latter) isomer hosts only odd-momentum (even-momentum) rotational states.}%
Each isomer's monodromy is independent of the axis of rotation, as long as the rotation exchanges the two nuclear positions.
On the other hand, any rotation around the molecule's principal axis yields the trivial monodromy of \(+1\) for both isomers.

The monodromies of each isomer correspond to evaluations of symmetry-group elements in the isomer's inducing irreps.
The para-isomer irrep is trivial, while the ortho-isomer sign irrep yields the possible \(\pm 1\) monodromy.
Which monodromy is induced depends on the homotopy class of the path in position-state space, which in this case is the projective plane, $\nicefrac{\so 3}{\D{\infty}}\cong\rptwo$.

For an isomer with symmetry group \(\G\) and inducing irrep \(\l\),
the possible monodromies correspond to \(\l(\rg)\) for any \(\rg\in\G\) [see Fig.~\ref{fA0:holonomy}(c)].%
\ifthenelse{\equal{\PeelOffMain}{0}}{
This set generates the \textit{monodromy group}, and we determine such groups for all possible isomers in Table \ref{tab:monodromy}.
}{
This set generates the \textit{monodromy group}, and we determine such groups for all possible isomers.
}%

Separable isomers, like ortho \ce{H2}, admit root-of-unity monodromy.
The $^{2}\e\ot{}^{1}\e=\a$ separable isomer of perrotationally $\C 3$-symmetric
ammonia is the simplest to realize a monodromy that is not $\pm1$.%
\ifthenelse{\equal{\PeelOffMain}{0}}{
Cyclically permuting the isomer's three hydrogen nuclei yields a cube root of unity, as realized by the \(^2 \e\) irrep of \(\C 3\) (see Example \ref{ex:ammonia-monodromy}).
}{
Cyclically permuting the isomer's three hydrogen nuclei yields a cube root of unity, as realized by the \(^2 \e\) irrep of \(\C 3\).
}%
Such a monodromy cannot be directly attributed to Bose/Fermi spin statistics and is instead reminiscent
of anyonic statistics \cite{Leinaas_theory_1977,wilczek_quantum_1982,Wu_general_1984,arovas_fractional_1984,nayak_non-abelian_2008} (themselves stemming from an induced representation \cite{Goldin_particle_1980,isham_topological_1984}).

\begin{figure}[t]
\includegraphics[width=1.0\columnwidth]{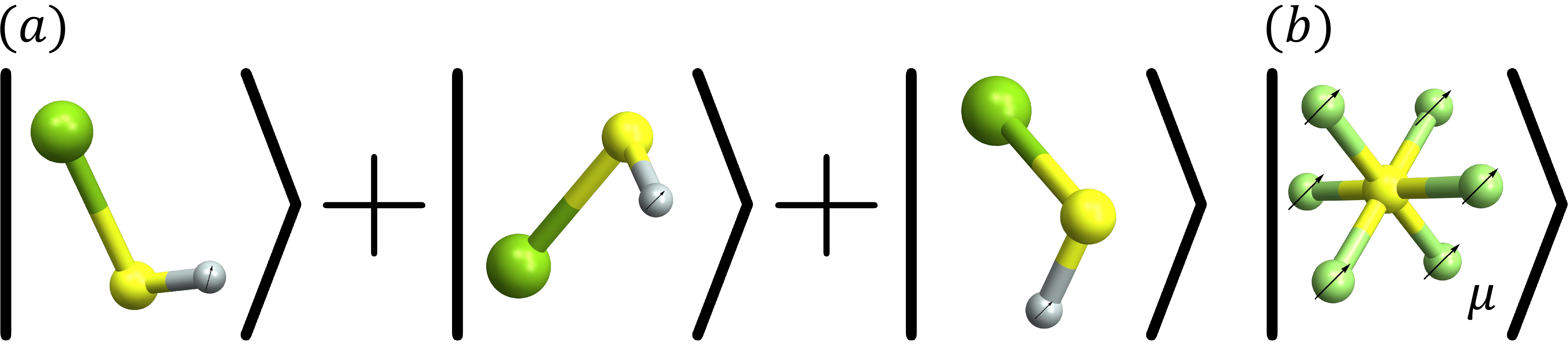}
\caption{\label{fig:qec}
\nw{
(a) Molecular code states \cite{albert_robust_2020} are superpositions of several position states of an asymmetric molecule, related by rotations in a subgroup \(\G\).
(b) Fiber codes \(\{|\re,\m\ket\}_{\m=1}^{\dm}\)~\eqref{eq:fourier} encode in a \textit{single} molecular position, \(\cs = \re\), of a perrotationally \(\G\)-symmetric entangled isomer (here, \(\O\)-symmetric \ce{SF6}).
The former induces the trivial induced representation, \(\a\up\so 3\), on the asymmetric molecular state space, while the latter forms a non-trivial induced representation, \(\l\up\so 3\), with a  fiber degree of freedom of dimension \(\dm = \dim \l\).
Both encodings are comparable in performance against shifts in the molecules' orientation and kicks in their momenta (see \vva{Sec.~\ref{sec:fiber-codes}}).
}
}
\end{figure}

For entangled isomers, monodromies are matrices acting on the \(\dm\)-dimensional fibers.
These generate \textit{non-Abelian} monodromy groups, akin to those obtained by
braiding non-Abelian anyons \cite{Wen_non-abelian_1991,Moore_nonabelions_1991,Levay_berry_1995,nayak_2n-quasihole_1996,read_non-abelian_2009} or realizing fault-tolerant quantum gates \cite{gottesman_fibre_2017}.
There exist several nuclear spin isomers realizing
representations of the dihedral,  tetrahedral, octahedral, and icosahedral groups.

\ifthenelse{\equal{\PeelOffMain}{0}}{
The $\e\ot\e\down\a_{2}$ entangled isomer of
dihedrally perrotation-symmetric boron trifluoride admits a two-dimensional fiber (see Example~\ref{ex:bf3-holonomy}).
Rotations permuting its three fluorine nuclei
realize the two-dimensional
irrep $\e$ of a dihedral group via monodromy.

The four entangled isomers of icosahedrally perrotation-symmetric \ce{$^{13}$C60} fullerene admit fibers of dimensions $\dm=3$, $3$, $4$, and $5$, respectively.
Monodromies of each isomer realize the 60-element icosahedral group
irreps $\ti_{1}$, $\ti_{2}$, $\g$, and $\h$, respectively (see Example~\ref{ex:fullerene-h-holonomy}).
}{
The $\e\ot\e\down\a_{2}$ entangled isomer of
dihedrally perrotation-symmetric boron trifluoride admits a two-dimensional fiber.
Rotations permuting its three fluorine nuclei
realize the two-dimensional
irrep $\e$ of a dihedral group via monodromy.

The four entangled isomers of icosahedrally perrotation-symmetric \ce{$^{13}$C60} fullerene admit fibers of dimensions $\dm=3$, $3$, $4$, and $5$, respectively.
Monodromies of each isomer realize the 60-element icosahedral group
irreps $\ti_{1}$, $\ti_{2}$, $\g$, and $\h$, respectively.
}%

\vva{Position-state monodromy differs from a conventional Berry phase~\cite{berry_quantal_1984}, which is acquired by a gapped eigenstate of a parameter-dependent Hamiltonian and depends on the geometry of the path.
Monodromy requires neither a Hamiltonian nor an encircled degeneracy (cf.\ conical intersections in Sec.~\ref{sec:prior_work}), and it depends only on the homotopy class of the path in \(\nicefrac{\so 3}{\G}\).
It is exact as long as assumptions (A) and (B) hold, the dynamics do not distinguish identical nuclear spins, and the molecule does not tunnel between symmetry-related orientations along the path.
In an isolated molecule, only weak hyperfine interactions distinguish identical nuclear spins~\cite{chapovsky_nuclear_1999}.
Approximate position states incur additional corrections that are exponentially small in their mean squared angular momentum (see Theorem~\ref{thm:flat}).
Rotating an aligning field does not change its spectrum, so no level crossings occur along such a path.
Adiabaticity matters only insofar as it keeps the orientation localized.}

While our predicted monodromy occurs after an adiabatic path in position-state space, it is not always necessary to initialize a molecule in a fixed position state to obtain the same effect as that arising from a position-state monodromy.%
\ifthenelse{\equal{\PeelOffMain}{0}}{
In the case of \ce{H2} confined to two dimensions, \textit{any} ortho or para state will do (see Sec.~\ref{sec:toy-model}).
}{
In the case of \ce{H2} confined to two dimensions, \textit{any} ortho or para state will do.
}
In three dimensions, one can set the molecule to rotate in the equatorial plane with some fixed angular momentum \(\ell\), corresponding to the \(\z\)-axis momentum projection \(m=0\).
\ifthenelse{\equal{\PeelOffMain}{0}}{
Any rotational state \(|^{\ell}_{m=0}\ket\) yields the same phase of \(-1\) under an equatorial \(\pi\)-rotation (see Example~\ref{ex:deuterium-a2-monodromy}).
}{
Any rotational state \(|^{\ell}_{m=0}\ket\) yields the same phase of \(-1\) under an equatorial \(\pi\)-rotation.
}

\subsection{Fibers as protected encodings}

\nw{
Molecules were postulated to be useful for quantum information processing over 20 years ago~\cite{demille_quantum_2002,tesch_quantum_2002,yelin_schemes_2006,wei_entanglement_2011,ni_dipolar_2018,yu_scalable_2019,Sawant_ultracold_2020,Zeppenfeld_robust_2023,Jain_AE_2023,Freedman_quantum_2019} due to their identical nature and large state space, the latter stemming from their many degrees of freedom.
Reference~\cite{albert_robust_2020} showed that superpositions of \textit{multiple} position states of asymmetric and rotationally \(\C \infty\)-symmetric molecules form an error-correcting \textit{molecular code}.

On the other hand, the fiber degrees of freedom \(\{|\m\ket\}\) of perrotationally \(\G\)-symmetric entangled isomers all lie in a \textit{single} molecular orientation, \(\cs=\re\) (see Fig.~\ref{fig:qec}).
We show \vva{in Sec.~\ref{sec:fiber-codes} that}, despite not consisting of a superposition, certain fiber encodings have a comparable degree of protection as molecular codes.
Moreover, fiber codes, in some cases, allow for a richer set of fault-tolerant gates via monodromy (see previous subsection).

Superposing position states of asymmetric molecules according to the molecular-code prescription~\cite{albert_robust_2020} induces the same coset structure as that of rotationally symmetric molecules, i.e., the representation \(\a \up \so 3\) induced by the trivial irrep of some \(\so 3\) subgroup.
As such, rotation errors acting on either molecular or fiber code states can be corrected as long \vva{as} those rotations do not implement a non-contractible path in the coset space.

Molecular codes also protect against sufficiently small kicks in an asymmetric molecule's momentum.
Because their code states are superpositions of different position states, momentum kick operators --- diagonal in the position-state basis --- modulate the code states in a way that can be detected and sometimes corrected.

Fiber codes protect against momentum kicks in a different way.
Since a fiber lies within a particular isomer, any kicks that transition the state out of the given isomer can be detected.
Moreover, shifting the momentum of the molecule \textit{without} transitioning to another isomer can only be done for certain momenta \(\ell\).
We identify the isomers that can only be affected by momentum kicks by \(\ell\geq\ell_{\text{min}}>1\), meaning that those isomers are immune to intra-isomer momentum kicks by \vva{up to} \(\ell_{\text{min}}-1\).
This restriction narrows down six types of isomer whose fibers form useful encodings:
the \(\e_{i>1}\) irreps of the dihedral,
the \(\e\) and \(\ti_2\) irreps of the octahedral,
and the \(\ti_2\), \(\g\), and \(\h\) irreps of the icosahedral groups (see Table~\ref{tab:monodromy}). 
\vva{Perturbations that depend only on the molecule's orientation, such as couplings of its multipoles or polarizability to stray fields, act trivially on the fiber.
In contrast, couplings that distinguish identical nuclear spins can act on the fiber (see Sec.~\ref{sec:fiber-codes}).}
}

\begin{figure}[t]
\includegraphics[width=1\columnwidth]{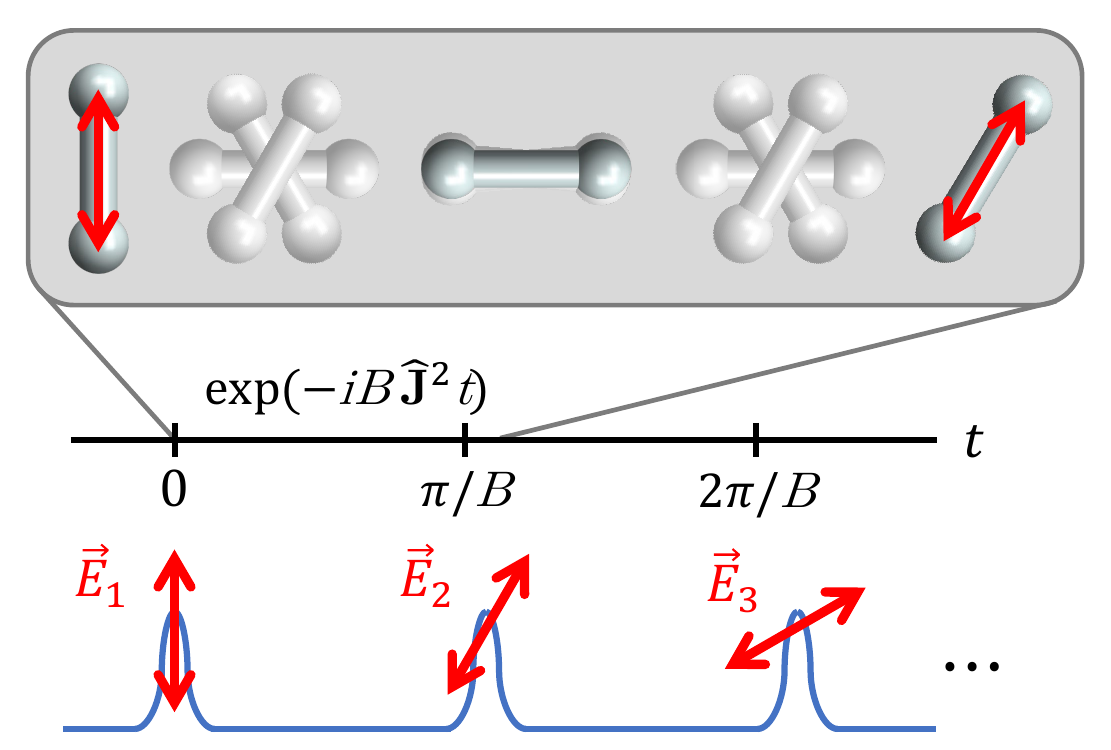}

\caption{\label{fig:stroboscopic}
Sketch of the evolution of a homonuclear diatomic evolving under its intrinsic rigid rotor Hamiltonian, \(B\hat{\mathbf{\ell}}^2\) with rotational constant \(B\), while undergoing a stroboscopic re-orientation from the \(\z\) to the \(-\z\) axis via a series of ultrafast pulses. \vva{Since the nuclei are identical, alignment along \(-\z\) is the same position as alignment along \(\z\).} %
The first pulse aligns the molecule along the desired \(\z\)-axis, initializing it in what is a close approximation to the corresponding position state.
Subsequent Hamiltonian evolution induces phases, \(\exp(-iB\ell(\ell+1)t)\), on the molecule's rotational states \(|^{\ell}_m\ket\), but all phases disappear at times \(t\) that are multiples of the \vva{revival time \(T_{\text{rev}}=\pi/B\), since \(\ell(\ell+1)\) is even}.
\vva{Shortly after the first revival, when the molecule is again aligned,}
a second pulse, with slightly rotated polarization relative to the first, is incident on the molecule. %
Subsequent pulses can be applied \vva{shortly after later revivals}, with each pulse incrementally edging the position state closer to that aligned along the \(-\z\) axis\vva{; pulses at exact multiples of \(T_{\text{rev}}\) would instead act as a single pulse, since free evolution over \(T_{\text{rev}}\) is trivial and impulsive pulses commute}.
In the impulsive limit, %
where the pulse length is much shorter than \(T_{\text{rev}}\), relative phases are imprinted that cause the molecular wavepacket to rephase at the new, rotated polarization. 
As long as the tilt between polarizations of successive laser pulses is not too large, a sufficiently discretized set of pulses should incrementally re-orient the molecule from \(\z\) to \(-\z\). 
}
\end{figure}
\subsection{Experimental signatures}
\label{subsec:expt}

An Abelian position-state monodromy is
an overall (undetectable) phase of a nuclear spin isomer's position state, but such a phase can be converted into a relative (and thus detectable) phase if the isomer is in superposition with some reference state.
We discuss how to observe \vva{Abelian monodromy via} Aharonov-Bohm-type interference \vva{and how monodromy may be probed in molecular crystals}.

%

%
%
\prg{Molecular interferometry}
One way to convert an Abelian monodromy into a relative phase is to interfere a rotated version of the molecule with a reference, un-rotated version of itself.
This can be done in a double-slit-type experiment: an incoming nuclear spin isomer is split into two ``arms'' and aligned in some fixed position state, a lab-based rotation is applied to one of the two arms, and the two arms are recombined in order to probe their interference.

Initializing in a separable nuclear spin isomer \(\sp\) and nuclear-spin state \(\chi\), denoting the first (second) arm by an overline (underline), and assuming a lab-based symmetry rotation implementing the transformation \(\rg\), the two arms recombine as
\begin{equation}
    |\overline{\sp,\chi}\ket+\l(\rg)|\underline{\sp,\chi}\ket~,
\end{equation}
where \(\l(\rg)\) is the resulting root-of-unity monodromy.

For the simplest example, we can align a homonuclear diatomic (\vva{e.g.,} \ce{H2} or \ce{N2}) along the \(\z\) axis and \vva{rotate it by \(\pi\)} around any equatorial axis.
\vva{Because the two nuclei are identical, alignment along \(-\z\) is the same position as alignment along \(\z\).
This rotation thus traces a closed loop in the position space \(\nicefrac{\so 3}{\D{\infty}}\cong\rptwo\) that cannot be contracted to a point.}
Upon this rotation, the \vva{isomer with even (odd) \(\ell\), e.g., para (ortho) \ce{H2},} yields a positive (negative) monodromy, i.e., \(\l(\rg)=\pm1\), respectively.%
\ifthenelse{\equal{\PeelOffMain}{0}}{
More generally, a \(\C N\)-symmetric molecule can yield an \(N\)th root-of-unity monodromy, but we focus on ortho and para hydrogen from now on for simplicity (see Sec.~\ref{sec:toy-model} and Example~\ref{ex:deuterium-a2-monodromy}).
}{
More generally, a \(\C N\)-symmetric molecule can yield an \(N\)th root-of-unity monodromy, but we focus on ortho and para hydrogen from now on for simplicity.
}%
One way to observe a position-state monodromy is to align the molecule in the lab frame and drag the internuclear axis around by an angle of $\pi$.
However, such adiabatic dragging is difficult to realize in practice due to simultaneous requirements of rotational adiabaticity and high field intensities. %
We propose to induce lab-frame rotations via a stroboscopic sequence of ultrafast laser pulses \cite{Stapelfeldt_colloquium_2003,Seideman_dynamics_2001,Dooley_direct_2003,Spanner_field-free_2004,Lee_coherent_2006}, similar to a quantum kicked-rotor model in which the kick orientation is incremented at each time step (see Fig.~\ref{fig:stroboscopic}).

In practice, the two ``arms'' of the interferometer can represent separated beams in real space that can be obtained via matter-wave diffraction \vva{without a quantum-degenerate gas~\cite{chapman_optics_1995,cronin_optics_2009}}.
\ifthenelse{\equal{\PeelOffMain}{0}}{
We discuss details of and issues with this scenario in Sec.~\ref{sec:expt}.
}{
}

\prg{\vva{Molecular crystals}}
Molecules can arrange themselves in crystalline patterns at particular temperatures and pressures, but, the rotational degrees of freedom of each molecule in such patterns are only \textit{partially} restricted.
Molecular crystals of hydrogen \cite{Harris_orientational_1985}\cite[pg.~563]{Hochli_orientational_1990}, methane \cite[pg.~582]{Hochli_orientational_1990}, and fullerene \cite{David_crystal_1991,Heiney_orientational_1991,David_structural_1992} have been synthesized and studied.

As a molecular crystal is cooled, it forms an \vva{orientationally ordered phase or an} orientational glass \cite{Hochli_orientational_1990}, in which \vva{molecules spontaneously adopt one of a handful of discrete orientations} \cite{devonshire_rotation_1936}.
The particular orientation depends on neighboring orientations. Thus, molecules \vva{such as methane and fullerene} are constrained to approximate position states without the application of intense electric fields.
\vva{Methane and other light molecules still tunnel between symmetry-related orientations.
The resulting tunneling splittings are resolved by inelastic neutron scattering at liquid-helium temperatures, where thermal broadening by phonons is small~\cite{press_ch4_1975,prager_rotational_1997}.} 
Re-orientation may provide a way to probe monodromy, which would be non-Abelian in the case of methane or isotopic fullerene crystals.

\vva{We expect gas-phase homonuclear diatomics to be the nearest-term platform for observing Abelian monodromy, since their laser alignment and rotation are well developed~\cite{Stapelfeldt_colloquium_2003,koch_quantum_2019}.}

\section{This work in context}
\label{sec:prior_work}

Many of the technical tools we use have been utilized in spectroscopy, but in a different Hamiltonian-centric point of view and often with additional (i.e., electronic and/or vibrational) degrees of freedom.

\prg{Momentum states}
\nw{
Our framework for momentum
states is based on what has long been known in the molecular spectroscopy literature \cite{ezra_symmetry_1982,Papousek_molecular_1982,bunker_molecular_1998,bunker_fundamentals_2004,brown_rotational_2003,biedenharn_angular_2010,Herzberg_spectra_2013,Herzberg_infrared_1987,Herzberg_electronic_1966,yurchenko2023computational}.
There exist several prescriptions to determine admissible states, with each prescription considering slightly different symmetries \cite[pg.~54]{ezra_symmetry_1982}.
These prescriptions, along with a more modern take utilizing Schur-Weyl duality \cite{Schmiedt_unifying_2016,Schmiedt_molecular_2017}, focus mainly on determining the statistical weights \(\mst\) of the nuclear-spin factor.
They also tend not to focus on general features, instead working out each molecule in full detail.
By exhaustively listing all admissible combinations of irreps satisfying condition~\eqref{eq:branching}, we identify two general features --- the dimensions of \(\l\) and \(\t\) are always equal, and the multiplicity of the \(\s\) irrep is always one --- and make contact with the theory of induced representations to pave the way for a complete position-momentum phase space.
}

Induced representations have seen several applications in chemistry.
Induced representations of symmetry groups are used in the aforementioned previous work to describe the interplay of symmetry and particular Hamiltonian perturbations
\cite{altmann_induced_1977,Ehlers_induced_1979,harter_simple_1977,harter_orbital_1977,harter_bands_1977,harter_frame_1978,Harter_theory_1979,Harter_theory_1981,drake_molecular_2006}.
Our induced representation is of the \textit{entire} proper-rotation group, acts on the entire admissible state space, and is Hamiltonian-independent.

\nw{
Entangled isomers' states for various symmetric molecules have been studied in pioneering work by Harter, Patterson, and coworkers \cite{harter_simple_1977,harter_orbital_1977,harter_bands_1977,harter_frame_1978,Harter_theory_1979,Harter_theory_1981,drake_molecular_2006}, where entangled momentum basis states are referred to as ``Pauli states'' \cite{Harter_theory_1979}\cite[pg.~622]{biedenharn_angular_2010}.
The explicit form of these states is typically not important for spectroscopic studies because their degeneracy ``\textit{cannot be split by any higher-order interactions}'' \cite[pg.~92]{Herzberg_electronic_1966}.
This may be another reason they have not been investigated further, and we are not aware of general statements about their complete rotation-spin entanglement.
}

\prg{Position states}
Molecular position states are mentioned at the initial step of prescriptions for finding admissible momentum states in the aforementioned work %
(e.g., \cite{harter_frame_1978,drake_molecular_2006}), with a book by Harter \cite{harter_principles_1993} explaining position states for asymmetric and linear molecules in a way that is arguably the closest to our symmetry-centric point of view.

Pendular states \cite{Friedrich_spatial_1991,rost_pendular_1992,Slenczka_pendular_1994,Friedrich_alignment_1995} --- obtained by orienting the molecule along an axis defined by an electric field --- are close approximations to position states.
In the strong-field regime, the position-state space is restricted to a local region of states centered at the position of the pendular state.
This restriction
makes it impossible to notice the topological effects associated with the global position-state space.

Our position states are derived by starting with the Zak basis \cite{zak_finite_1967,justel_zak_2018,albert_robust_2020,culf_group_2022} for \(\so 3\) and entangling Zak basis states with nuclear degrees of freedom according to the prescription stemming from condition~\eqref{eq:branching}.

\prg{Symmetric tops}
We ``mod out'' symmetry on the level of the state space, but one can also do so at the Hamiltonian level.
The canonical Hamiltonian of a 3D rigid body is defined by the body's moment-of-inertia matrix \(I\).
Non-linear molecules are classified as asymmetric (generic diagonal \(I\)), symmetric (doubly degenerate \(I\)), or spherical (\(I\propto\id\), the identity) tops \cite{Papousek_molecular_1982}.

A generic diagonal \(I\) is invariant under rotations forming the dihedral group \(\D2\), which are represented by diagonal matrices with \(\pm 1\) entries and determinant one \cite{wormer_rigid_nodate}.
Thus, the coset space \(\nicefrac{\so 3}{\D 2}\) labels distinct rotationally related asymmetric top Hamiltonians; the relevance of the topology of this space has been noted before \cite{Sudarshan_configuration_1988,Horvathy_inequivalent_1989}.
Similarly, the coset space \(\nicefrac{\so 3}{\D \infty}\) labels distinct Hamiltonians of symmetric tops.

The moment-of-inertia Hamiltonian classification is too coarse to determine the position-state space.
For example, spherical tops admit only one possible moment-of-inertia Hamiltonian (up to constant pre-factors), but can have various state spaces as determined by their symmetry.
Similarly, asymmetric tops need not be asymmetric molecules; e.g., water is an asymmetric top but has the \textit{symmetric} state space \(\nicefrac{\so 3}{\C 2}\).

\prg{Monodromy in molecules}
An Abelian holonomy that is also a monodromy --- a ``topological'' Berry phase --- can occur for eigenstates of particular rovibrational Hamiltonians \cite{mead_determination_1979,mead_geometric_1992,yarkony_diabolical_1996,Resta_manifestations_2000}.%
\ifthenelse{\equal{\PeelOffMain}{0}}{
The Hamiltonian degeneracy occurs at a conical intersection (see Example~\ref{ex:conical}), yielding a $-1$ Berry phase that is independent of the fine-grained details of the state's path.
}{
The Hamiltonian degeneracy occurs at a conical intersection, yielding a $-1$ Berry phase that is independent of the fine-grained details of the state's path.
}%
Non-Abelian holonomy is postulated to occur \cite{Moody_realizations_1986,Li_induced_1987,Bohm_berry_1992,Koizumi_geometric_1995}, but there is no monodromy because the parameter space is simply connected.

Others have studied monodromy effects \cite{Sadovskii_monodromy_1999,Faure_topological_2000,child_quantum_1999,kozin_monodromy_2003,Efstathiou_metamorphoses_2005,Kruglikov_geometric_2009,child_semiclassical_2014}, critical phenomena \cite{Pavlichenkov_phase_1982,Pavlichenkov_rotation_1985,Pavlichenkov_critical_1988}, and other topological features \cite{Zhilinskii_symmetry_2001,Schmidt_topology_2014} of spectra of quantum and semiclassical molecular Hamiltonians, many of which include rotational degrees of freedom.

In contrast to past work, our rotation-spin monodromy effects are intrinsic features of the state space and are independent of any Hamiltonian.
They occur for configuration spaces which are coset spaces of \(\so3\) --- all non-simply-connected except \(\nicefrac{\so 3}{\C\infty} = \SS^2\) --- and which can yield non-Abelian monodromy.

\prg{Chemical bonding}
It is well known that some nuclear spin isomers, such as ortho hydrogen, do not contain states of zero angular momentum.
In other words, a molecule initialized in such an isomer never stops spinning.%
\ifthenelse{\equal{\PeelOffMain}{0}}{
Even more surprising is that some molecules, such as boron trifluoride, do not admit \textit{any} isomer with zero angular momentum (see Example \ref{ex:BF3}).
}{
Even more surprising is that some molecules, such as boron trifluoride, do not admit \textit{any} isomer with zero angular momentum.
}%
The inability to stop spinning can affect how well a nuclear spin isomer bonds during a chemical reaction \cite{Klein_directly_2017}.
This idea has been analyzed from a physics perspective for cyclically-symmetric molecules in Ref.~\cite{fisher_quantum_2018}, and our work can be used to extend their predictions to molecules with arbitrary symmetry.

The work~\cite{fisher_quantum_2018} also identified the root-of-unity monodromy (``branch cut'') associated with perrotationally symmetric isomers with cyclic symmetry.
Our work provides a way to realize that and more general monodromy using lab-based symmetry rotations.

\prg{Holonomy~calculations}
The holonomy of induced-representation spaces is relevant to quantization and gauge theory  \cite{Laidlaw_feynman_1971,Dowker_quantum_1972,Schulman_techniques_2005,Horvathy_Prequantisation_2024,isham_topological_1984,Landsman_geometry_1991,Tanimura_reduction_2000,Sudarshan_configuration_1988,Milnor_existence_1958}.
\ifthenelse{\equal{\PeelOffMain}{0}}{
Since discrete symmetry groups were not directly considered, we \vva{compute the monodromy explicitly} for the spaces of interest (see Sec.~\ref{sec:holonomy-position}).
}{
Since discrete symmetry groups were not directly considered in the literature, we \vva{compute the monodromy explicitly} for the spaces of interest.
}%

\vva{We prove that the Berry connection (that generates the holonomy) is zero for every representation of \(\so 3\) induced from an irrep of a finite subgroup.
We use the heat kernel on \(\so 3\) for finite subgroups and calculate the connection exactly for continuous subgroups.}
\vva{These results complement} a proof of flatness for \(\D\infty\)-symmetric \vva{(\(m=0\)) spin} states, developed in an illuminating paper by \vva{Robbins and Berry} \cite{Robbins_geometric_1994}, as well as flatness results for fixed-\(\ell\) ``anti-coherent'' \vva{states} \cite{zimba_anticoherent_2006,Crann_spherical_2010,Aguilar_when_2020}.

\ifthenelse{\equal{\PeelOffMain}{0}}{
\vva{Previous examples \cite{Vinet_invariant_1988,Giler_geometrical_1989,Levay_modified_1994,levay_canonical_1996,Levay_berry_1995,giller_structure_1993} and the question posed in Ref.~\cite{levay_canonical_1996} motivated our conjecture \cite{Albert_flat_2024} characterizing flat position-state Berry connections.
We prove it for representations of \(\so 3\) induced from proper closed subgroups and show that it fails for general induced representation spaces (see Sec.~\ref{subsec:fundamental-group}).}
}{
\vva{Previous examples \cite{Vinet_invariant_1988,Giler_geometrical_1989,Levay_modified_1994,levay_canonical_1996,Levay_berry_1995,giller_structure_1993} and the question posed in Ref.~\cite{levay_canonical_1996} motivated our conjecture \cite{Albert_flat_2024} characterizing flat position-state Berry connections.
We prove it for representations of \(\so 3\) induced from proper closed subgroups and show that it fails for general induced representation spaces.}
}
\section{Outlook\label{sec:discussion}}

\nw{
Following earlier topological classifications of solid-state band insulators (e.g., \cite{bradlyn_topological_2017}) and nematic liquid crystals \cite{mermin_topological_1979,liu2016generalized}, we classify molecular rotation-spin state spaces, develop a position-momentum duality,
quantify their entanglement patterns, and identify a set of robust ``topological'' features.

\prg{Molecular phase space}
Quantum applications in conventional state spaces, such as qubits and harmonic oscillators, hinge on a position-momentum duality.
Re-invigorating the duality between rotational position and momentum that has been under-represented
in the molecular world, 
we relate established techniques from molecular spectroscopy~\cite{Rasetti_incoherent_1929,Wilson_statistical_1935,Wilson_symmetry_1935,longuet-higgins_symmetry_1963,Van_Vleck_coupling_1951,hougen_interpretation_1971,louck_eckart_1976,harter_orbital_1977,Watson_aspects_1977,berger_classification_1977,ezra_symmetry_1982,Papousek_molecular_1982,bunker_molecular_1998,bunker_fundamentals_2004,brown_rotational_2003,biedenharn_angular_2010,Herzberg_spectra_2013,Herzberg_infrared_1987,Herzberg_electronic_1966,yurchenko2023computational} for finding angular momentum states to those from quantization theory~\cite{isham_topological_1984,Landsman_geometry_1991,Tanimura_reduction_2000,levay_canonical_1996} for finding angular position states.
Our work lays the foundation for using the \textit{entirety} of molecular state space for quantum computing, simulation, and sensing, as well as for finely controlled chemical reactions, frequency conversion, and yet-to-be-discovered future technologies.
}

In technical terms, we show that
the rotation-spin states of a nuclear spin isomer host an induced representation
\cite{mackey_induced_1952,mackey1968induced,coleman_induced_1968,Inui_group_1990,carter_lectures_1995,chirikjian_engineering_2000,folland_course_2016} of the
group of lab-based rotations, induced by an irreducible
representation, or irrep, of the molecule's symmetry subgroup.
While all classical rigid-body state spaces correspond to trivial inducing irreps, spin-statistics restrictions yield non-trivial irreps.
\vva{Nuclear} spin isomers with such state spaces have no classical analogue and admit two interesting and potentially useful features.

\nw{
\prg{Complete entanglement}
The first feature is that symmetry and
spin-statistics force
the rotational and nuclear-spin degrees of freedom to \textit{always} be entangled.
This entanglement is present in the entire molecular
isomers' state space, and is \textit{independent} of and \textit{unaffected} by any symmetric Hamiltonian.
While the numerical spectroscopy community has long had to deal with this effect, \vva{we are not aware of prior statements that symmetry entangles two distinct degrees of freedom in every state of a physical system}.

We systematically quantify the strength of this entanglement in terms of properties of the molecule's symmetry group.
Molecules straddle the spectrum between the quantum and the classical,
and our entanglement classification \vva{quantifies one aspect of} the ``quantumness'' of \vva{a} given molecule.
We find that molecules with non-Abelian symmetry groups skew toward the quantum end of this spectrum.
\vva{Similarly, for qudits with a global on-site symmetry, the maximally mixed state of the symmetry-invariant sector is separable for Abelian symmetries and entangled for non-Abelian ones~\cite{moharramipour_symmetry_2024}.}

Complete entanglement may seem surprising when interpreted as a fragile resource, but it is, in fact, quite generic in Hamiltonian eigenstates \cite{tichy2011essential} and many-body state space in general \cite{gross_most_2009,bremner_are_2009}.
We anticipate that
\vva{our} symmetry-induced intra-isomer entanglement will be 
similar in spirit to fermionic entanglement \cite{Bartlett_reference_2007} --- a useful resource \cite{morris2020entanglement}.
The ability to \textit{switch} (\textit{a.k.a.} transition or interconvert) between isomers is possible in many molecules (e.g., in hydrogen~\cite{buntkowsky2006mechanism} or methane \cite{itano1980avoided}).
When the isomers have different entanglement patterns, this ability may be relevant to chemical reactions \cite{fisher_quantum_2018} and, more speculatively, may yield exotic ways to process quantum information~\cite{fisher_quantum_2015,Freedman_quantum_2019}.
Our explicit state spaces should help reduce the complexity cost of numerical studies (e.g., those using tensor networks \cite{bruognolo2021beginner}) where inter-isomer transitions can occur.
}

\prg{Dissociation}

The complete symmetry-induced rotation-spin entanglement within an isomer cannot be broken by any operation that maintains the molecular symmetry and keeps the molecule intact.
Detecting this entanglement may thus require either a breaking of the molecular symmetry or, more drastically, a breaking (dissociation) of the molecule itself.

Dissociating the molecule should transfer any initial linear and angular momenta into the linear momenta of the individual nuclei.
Any inter-nuclear (e.g., singlet or triplet) entanglement may be maintained after dissociation.
Similarly, complete rotation-spin entanglement may also leave a mark on the relative motion of the scattered nuclei.%
\ifthenelse{\equal{\PeelOffMain}{0}}{
If so, the outgoing nuclear state should be isomer dependent, exhibiting a different entanglement pattern in, say, the entangled \vva{\(\e^{\ast}\)} isomer of dihedrally symmetric boron trifluoride (see Example~\ref{ex:BF3}) than in its separable \(\a_2^{\ast}\) isomer (see Example~\ref{ex:BF3-separable}).
}{
If so, the outgoing nuclear state should be isomer dependent, exhibiting a different entanglement pattern in, say, the entangled \vva{\(\e^{\ast}\)} isomer of dihedrally symmetric boron trifluoride than in its separable \(\a_2^{\ast}\) isomer.
}%

\prg{Fibers and their topological operations}
The second consequence of our result is that \textit{each} position state of certain
nuclear spin isomers hosts a ``hidden'' (cf.~\cite{Zeppenfeld_robust_2023}) qudit or ``fiber'' degree of freedom, and that this fiber can be used to store and protect quantum information that is arguably more natural than engineered molecular encodings
\cite{albert_robust_2020,Jain_AE_2023}.

Quantifying the robustness
of fiber operations using the framework of holonomy (read: Berry phase) \cite{berry_quantal_1984,wilczek_appearance_1984,simon_holonomy_1983,Zanardi_holonomic_1999}, we \vva{prove} that position-state holonomy depends
only on global or ``topological'' properties of paths in position space.
The same dependence is observed in the braiding
of Abelian and non-Abelian anyons \cite{Leinaas_theory_1977,wilczek_quantum_1982,Wu_general_1984,arovas_fractional_1984,nayak_non-abelian_2008,Goldin_particle_1980,Wen_non-abelian_1991,Moore_nonabelions_1991,Levay_berry_1995,nayak_2n-quasihole_1996,read_non-abelian_2009}  and in families of fault-tolerant quantum
gates \cite{gottesman_fibre_2017}.
Such topological holonomy, or \textit{monodromy}, has hitherto been predicted in only engineered and/or exotic systems, while in this case it occurs naturally in the seemingly ``classical'' position states of nuclear spin isomers.
The simplest monodromy is already present in ortho hydrogen, and we outline ways to observe it in the lab.

\prg{Future directions}

\nw{
We restrict to molecules whose orbital, electronic spin, and vibrational degrees of freedom transform trivially under symmetry rotations.
Removing this restriction
should yield an expanded classification based on multi-factor generalizations of Eq.~\eqref{eq:branching}.
This may yield other induced representations not possible in the current framework and generalize interesting ro-vibrational effects \cite{guichardet1984rotation,tachibana1986complete}.

Only proper rotations feature in our rigid-body-based analysis,
but improper rotations
become important when
nuclei are allowed to re-configure themselves via tunneling in ways that are not possible via proper rotations \cite{ezra_symmetry_1982,Papousek_molecular_1982,bunker_molecular_1998,bunker_fundamentals_2004,brown_rotational_2003,biedenharn_angular_2010,Herzberg_spectra_2013,Herzberg_infrared_1987,Herzberg_electronic_1966,yurchenko2023computational}.
It could be interesting to study chiral molecules that can switch chirality via tunneling (e.g., \cite{schmiedt_collective_2016,Schmiedt_molecular_2017,Zeppenfeld_robust_2023}) or non-rigid molecular dimers, trimers, and other clusters.
Our treatment may also be relevant to magnetic rigid-rotor nanoparticles, which feature a similar rotation-spin state space \cite{OKeeffe_quantum_2012,Rusconi_magnetic_2016,Rusconi_linear_2017,ma_torque-free_2021,Rusconi_spin-controlled_2022}.

The fiber is well hidden and delocalized across both angular-momentum and nuclear-spin space, and it will be crucial to determine ways to map quantum information to and from this space.
It will also be of interest to develop uncertainty relations \cite{erb_uncertainty_2010,hayashi2016fourier,wigderson2021uncertainty}, a Wigner-Weyl-type formalism \cite{bizarro1994weyl,mukunda2004wigner}, coherent states \cite{bhaumik1975coherent,janssen1977coherent,gulshani1979generalized,de1989coherent,kowalski2000quantum,gazeau2023covariant}, and extensions of semi-classical methods \cite{heller1981frozen,drake_molecular_2006} to general molecular phase spaces.
Connections between asymmetric molecules and the Dirac equation may also \vva{descend} to symmetric cases \cite{baez2025second}.
Having laid down a complete molecular phase space, we leave these and other exciting quantum applications to future work.
}

\begin{acknowledgments}

V.V.A.\@ acknowledges 
Jonathan Tennyson,
Timur V.\ Tscherbul,
and
Sergei N.\ Yurchenko
for advice on placing these results in context, and Brandon R. Brown for advice on scientific writing.
We also thank
B.\ Andrei Bernevig,
Barry Bradlyn,
Chin-wen Chou,
Leah G.\ Dodson,
Matthew P.\ A.\ Fisher,
Gerald T.\ Fraser,
Joseph T.\ Iosue,
Artur F.\ Izmaylov,
William G.\ Harter,
Joseph T.\ Hodges,
Ted Jacobson,
Shubham P.\ Jain,
Anton Kapustin,
Andrei F.\ Kazakov,
Aleksei Khudorozhkov,
Christiane Koch,
Johannes Kofler,
Richard Kueng,
Prem Kumar,
Dietrich Leibfried,
Monika Leibscher,
David Long,
Svetlana Malinovskaya,
Leo Radzihovsky,
Nathan Schine,
Stafford W.\ Sheehan,
Benjamin Stickler,
Ian Teixeira,
Christopher David White,
and
Jun Ye
for helpful discussions.

E.K.\@ acknowledges support from NSF QLCI grant OMA-2120757.
M.L.\@ acknowledges support by the European Research Council (ERC) Starting Grant No.\ 801770 (ANGULON).
All molecules were drawn in \textsc{Mathematica 13} \cite{wolfram_research_inc_mathematica_2023}.
The following public-domain photos from \href{http://freesvg.org}{freesvg.org} were used for Fig.~\ref{fig:header}: \href{https://freesvg.org/dumbbell-image}{stock dumbbell},
\href{https://freesvg.org/round-pencil-39258}{round pencil},
\href{https://freesvg.org/plane-vector-image}{plane}, and
\href{https://freesvg.org/soccer-ball-clip-art-1595859124}{soccer ball}.
VVA thanks Ryhor Kandratsenia and Olga Albert for providing daycare support throughout this work.

\vva{Parts of this manuscript were prepared with the assistance of ChatGPT 5.6, developed by OpenAI, and Claude Code, developed by Anthropic.
These agents were used to help identify when our conjecture~\cite{Albert_flat_2024} is true and edit the writing of some sections, in accordance with the authors' instructions.
All content, claims, and conclusions have been reviewed and verified by the authors to ensure accuracy and originality.
Certain products, commercial and otherwise, are mentioned in this publication.
These mentions are for informational purposes only, and do not imply recommendation or endorsement by NIST.}

\end{acknowledgments}
\section{Outline of the remaining manuscript}

The rest of the paper is devoted to technical derivations, but an attempt is made to make said derivations relatable.
In addition to providing nearly 50 example demonstrations of the various general ideas with specific molecules, we demonstrate our monodromy result using a simple toy model --- the hydrogen molecule confined to rotate in two dimensions --- in Sec.~\ref{sec:toy-model}.

In Sec.~\ref{sec:asymmetric}, we review what is known about the position and momentum state space of an asymmetric molecule and calculate the holonomy of its position states. 
In Sec.~\ref{sec:symmetric}, we formulate the momentum state space of symmetric molecules, proving the induced representation theorem stated above.
In Sec.~\ref{sec:position}, we formulate position states of symmetric molecules and the Fourier transform that converts between position and momentum.
In Sec.~\ref{sec:holonomy-position}, we calculate the holonomy of molecular position states \vva{and prove} that it depends only on topological features of the position configuration space.
In Sec.~\ref{sec:fiber-codes}, we quantify the strength of the fiber as a quantum error-correcting code.
In Sec.~\ref{sec:expt}, we discuss how to realize the monodromy of homonuclear molecules via various interferometric approaches.

\section{Toy model: hydrogen molecule in two dimensions}
\label{sec:toy-model}

We demonstrate a simple example of the nuclear spin isomer classification and monodromy effects using diatomic molecules, pinned at their center of mass and confined to rotate in the $\x\y$ plane.
All rotations in this section are restricted to be around the $\z$ axis.

An example of an asymmetric diatomic in two dimensions is \ce{HCl}.
Its state space is \vva{the} same as that of the planar rotor (\textit{a.k.a.}\ a particle on a circle) \cite{albert_general_2017}.
Rotor momentum states \(|\ell\ket\) are labeled by an integer $\ell$; their sign determines whether the rotor is spinning clockwise or counterclockwise.
The rotor's momentum operator \(\hat{\ell}\) satisfies $\hat{\ell}|\ell\ket=\ell|\ell\ket$.
Rotor position states,
\begin{equation}\label{eq:position-2D}
|\phi\ket={\textstyle \frac{1}{\sqrt{2\pi}}}\sum_{\ell\in\Z}e^{i\phi\ell}|\ell\ket\,,
\end{equation}
are labeled \vva{by} an angle $\phi$, which determines how the rotor's orientation
deviates from some reference orientation.

On the other hand, the state space of perrotationally symmetric molecules is a subspace of the composite space consisting of a rotational planar-rotor factor and a nuclear-spin factor.
The subspace is defined by a constraint on both factors that stems from symmetry and spin statistics.

For example, the two spin-half nuclei of \ce{H2} are indistinguishable in the molecule's frame,
meaning that the molecule is symmetric under molecule-based rotations by \(\pi\).
A $\pi$-rotation in the molecule's frame rotates the molecule while also exchanging
nuclei.
This operation corresponds to a \textit{perrotation},
\begin{equation}
\lpr_{\pi}=e^{i\pi\hat{\ell}}\ot\textsc{swap}\quad\quad\text{(molecule-based rotation)},\label{eq:perrotation-2D}
\end{equation}
where the operator on the first factor acts on the rotational state
space,
and the operator on the second nuclear-spin factor performs a swap
of the two spins.

The joint rotation-spin constraint is
\begin{subequations}
\begin{align}
\lpr_{\pi}\psi_{\mol}(\phi_{1},\phi_{2},s_{1},s_{2}) & =\psi_{\mol}(\phi_{2},\phi_{1},s_{2},s_{1})\\
 & =-\psi_{\mol}(\phi_{1},\phi_{2},s_{1},s_{2})\,,
\end{align}
\end{subequations}
for any admissible wavefunction $\psi_{\mol}$ of the two nuclear
positions $\phi_{1,2}$ and spin-functions $s_{1,2}$. The minus
sign, stemming from the spin statistics of the two fermionic hydrogen nuclei,
can come from the rotational or the nuclear-spin factor,
resulting in two molecular \textit{isomers} --- para and ``
ortho'' hydrogen.

Basis states of the para isomer consist of tensor products
of rotor states with \textit{even} angular momentum and the unique
anti-symmetric nuclear-spin \textit{singlet} state, $\left|\up\down\right\rangle -\left|\down\up\right\rangle $.
Basis states of the ortho isomer consist of tensor products
of states with \textit{odd} angular momentum and any nuclear-spin
state in the span of the three symmetric \textit{triplet} states,
$\left|\up\up\right\rangle $, $\left|\down\down\right\rangle $,
and $\left|\up\down\right\rangle +\left|\down\up\right\rangle $.
The required spin-statistics phase comes from the nuclear-spin (rotational)
factor for the para (ortho) isomer.

The two isomers' spaces can also be defined using rotor position states \(|\phi\ket\)  \eqref{eq:position-2D}.
If one of the nuclei is at position $\phi_{1}=\phi$, we automatically
know the position of the other nucleus, $\phi_{2}=\phi+\pi$ modulo \(2\pi\), so position
states of this \vva{dumbbell}-like molecule are labeled by a single angle. The $\pi$-perrotation symmetry
implies that states $|\phi\ket$ have to be identified with their
antipodes, $\left|\phi+\pi\right\rangle $. %
Identification is done by superposing
the two states for each angle $\phi\in[0,\pi)$. There are two ways to do so --- using
either a $+1$ or a $-1$ relative phase --- corresponding to the position states of the para and ortho
isomer, respectively.

Omitting the accompanying nuclear-spin factors for brevity, admissible rotational states can be expressed
in two different ways for each isomer,
\begin{align}
|\psi_{\text{para}}\ket & =\sum_{\ell\text{ even}}c_{\ell}^{\text{para}}|\ell\ket=\int_{0}^{\pi}\dd\phi\,c^{\text{para}}(\phi)\left(|\phi\ket+\left|\phi+\pi\right\rangle \right)\nonumber \\
|\psi_{\text{ortho}}\ket & =\sum_{\ell\text{ odd}}c_{\ell}^{\text{ortho}}|\ell\ket=\int_{0}^{\pi}\dd\phi\,c^{\text{ortho}}(\phi)\left(|\phi\ket-\left|\phi+\pi\right\rangle \right),\label{eq:isomer-2D}
\end{align}
for some momentum-state coefficients $c_{\ell}$ and position-state
functions $c(\phi)$.

Monodromy occurs when we examine how the two isomers' states transform
under ``passive'' lab-based symmetry rotations.
Such rotations occur in the lab frame, there is no permutation action on the spins.
The $\pi$-rotation,
\begin{equation}
\rr_{\pi}=e^{i\pi\hat{\ell}}\ot\id\quad\quad\text{(lab-based rotation)},\label{eq:rotation-2D}
\end{equation}
rotates each nucleus to a position that coincides with the initial
position of the other nucleus.

Lab-based rotations should, in some physical settings, be generated by the \textit{total}
angular momentum, $\hat{\ell}+\hat{s}_{1}+\hat{s}_2$,
where $\hat{s}_{1,2}$ are momentum generators for the two spins.
These produce extra transversal operations that re-orient the individual nuclei to a new frame, but they cannot produce perrotational permutation factors like the \textsc{swap} operation in Eq.~(\ref{eq:perrotation-2D}).
We ignore such effects in this work for
simplicity, noting that they may have to be taken into account in an experimental setting.

The hydrogen molecule is symmetric under the lab-based rotation from Eq.~\eqref{eq:rotation-2D},
but the two isomers transform in different ways:
\begin{subequations}\begin{align}
\rr_{\pi}|\psi_{\text{para}}\ket & =+|\psi_{\text{para}}\ket\\
\rr_{\pi}|\psi_{\text{ortho}}\ket & =-|\psi_{\text{ortho}}\ket\,.
\end{align}\end{subequations}
This can be shown either by noting that the two isomers are spanned
by momentum states of different parity, or by directly shifting each position
state by $\pi$ in Eq.~(\ref{eq:isomer-2D}).
In the case that the molecule is initialized in a position state, the different signs correspond to the monodromy of each isomer after an adiabatic path from \(\phi\) to \(\phi+\pi\) in position-state space.

The above two isomers can also be obtained by restricting the theorem
from Sec. \ref{sec:summary} to two dimensions. The symmetry group
of \ce{H2} is $\C 2=\Z_{2}$, a subgroup of $\C\infty \cong \so 2$, the
two-dimensional proper-rotation group.
The symmetry group has two irreps, the trivial irrep \(\a\) and the sign irrep \(\b\).
The even momentum states of the para isomer house
the induced representation $\a\up\so{2}$ of the group of lab-based rotations, while the odd-momentum
ortho isomer corresponds to $\b\up\so{2}$.

More generally, a \(\C N = \Z_N\)-symmetric molecule confined to a plane admits one or more of the induced representations \(\l\up\so{2}\), where \(\l\) is some root-of-unity irrep of \(\C N\).
Each representation corresponds to a particular way to quantize a particle on a circle.
The planar rotor from Eq.~\eqref{eq:position-2D}
corresponds to \(\a\up\so{2}\) --- the representation induced by the sole irrep of the trivial group.

The rest of the manuscript addresses the extra complexities of molecules that are free to rotate in \textit{three dimensions}.
In that case, the rotational factors, \(e^{i\pi\hat\ell}\), on the right-hand side of Eqs.~\eqref{eq:perrotation-2D} and \eqref{eq:rotation-2D} no longer coincide.
So while the full rotation group of an asymmetric molecule is \(\so 2\) in two dimensions, it is \(\so 3\times \so3\) in three dimensions because lab- and molecule-based rotations correspond to independent operations even for rigid bodies with no spin.

\begin{figure}[!t]
\centering{}
\includegraphics[width=1.0\columnwidth]{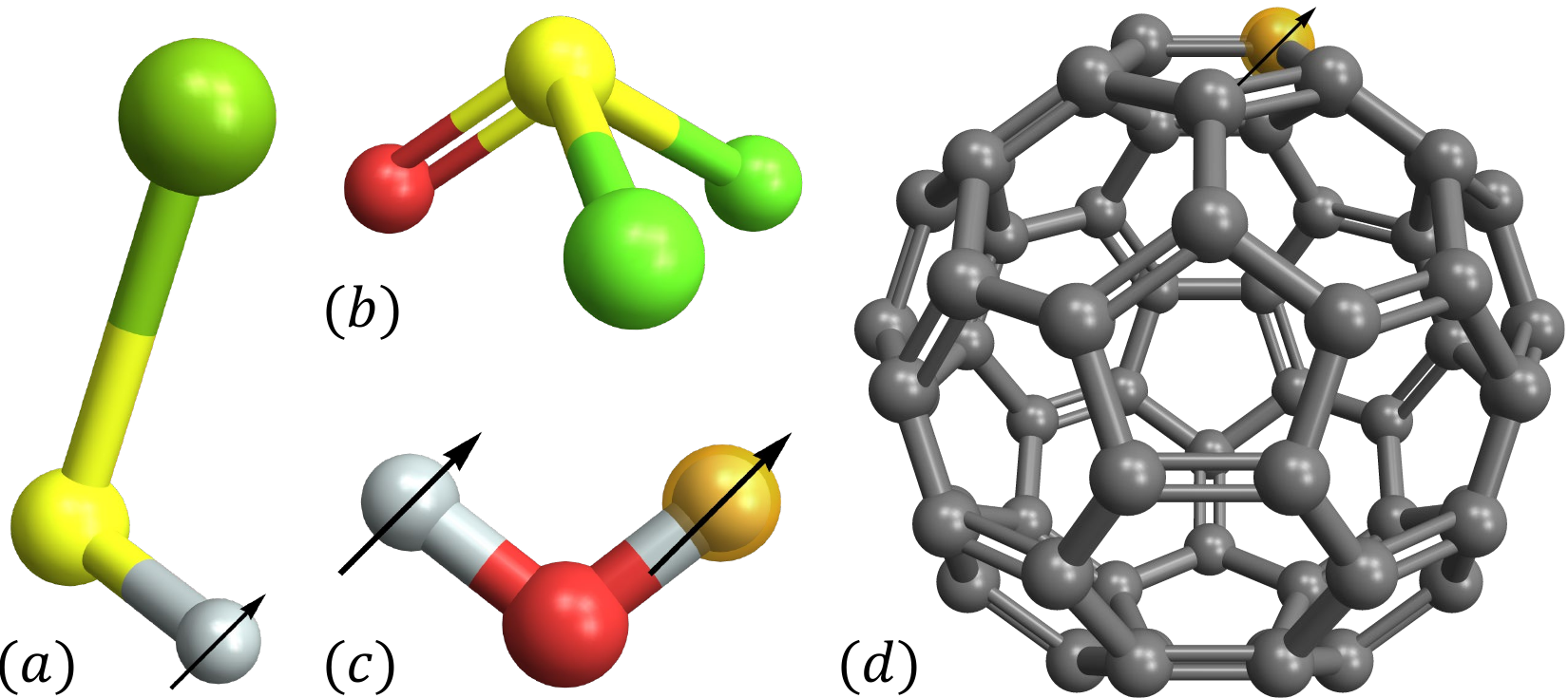}
\caption{\label{fA0:asymmetric}
Examples of asymmetric molecules (a) CaSH, (b) \ce{SOCl2}, (c) \ce{HDO}, and (d) \ce{C59$^{13}$C} from Examples \ref{ex:cash}-\ref{ex:doped-fullerene}.
Any proper rotation maps each molecule into a different position.
}
\end{figure}

\section{Asymmetric molecules}
\label{sec:asymmetric}

A molecule is said to be symmetric under a coordinate transformation when the transformation maps the molecule to the same position that the molecule was in before the transformation was applied.
For example, the \ce{HCl} molecule,
when aligned along some axis, is symmetric under rotations around that axis by any angle.
These rotations are examples of the symmetry transformations or \textit{symmetries} of such a molecule.

Symmetries include rotations that reorient the molecule while keeping the origin of its coordinate frame fixed, or reflections through either planes defined in the frame or the origin itself.
In our symmetry-centric perspective aiming to reveal the \textit{angular} Hilbert space available for particular symmetries (as opposed to the Hamiltonian-centric perspective that focuses on spectra and dynamics of concrete molecules),
we consider only symmetries that can be applied to a rigid molecule by an observer.
All admissible symmetries therefore lie in \(\so 3\), the group of orientation-preserving or \textit{proper} rotations in three dimensions.

We say that a molecule is \textit{asymmetric} if it admits no symmetries other than the identity; see Fig.~\ref{fA0:asymmetric} for examples.
In other words, any nontrivial proper rotation maps the molecule to a position that is different from the molecule's initial position.
From now on, the most abundant nuclear-spin isotope is assumed for each nucleus when no specific isotope is given.

\begin{example}[calcium hydrosulfide]\label{ex:cash}
    Since we consider only proper-rotation symmetries, an asymmetric molecule can still be symmetric under \textit{improper} rotations such as reflections through planes.
    An example of this is the CaSH molecule from Fig.~\ref{fA0:asymmetric}(a). It is not linear \cite{Jarman_high-resolution_1993} and is symmetric under reflection through the plane defined by its three atoms.
    However, it is not symmetric under any proper rotations, so it is considered an asymmetric molecule for our purposes.
\end{example}

\begin{example}[methylamine] \label{ex:methylamine}
    The \ce{CH3NH2} molecule also has reflection symmetry, but no proper rotation leaves the molecule invariant.
    While a proper rotation by \(2\pi/3\) around the axis defined by the \ce{CN} bond \textit{does} cyclically permute the three hydrogen nuclei attached to the carbon, this rotation simultaneously rotates the bent \ce{NH2} end into a different position.
    The final position state of the molecule as a whole is thus different from the initial state under such a rotation.
    In fact, there are \textit{no} proper rotations that permute any subset of nuclei while leaving the rest of the molecule invariant, making this molecule asymmetric.
\end{example}

\begin{example}[semi water]
    While water is symmetric under a proper rotation, substituting one of its identical hydrogen nuclei with deuterium yields semi-heavy water, \ce{HDO} [see Fig.~\ref{fA0:asymmetric}(c)], which is asymmetric.
    Substituting a different spinful isotope for the oxygen in water does not change its symmetry because the oxygen atom is not permuted with another oxygen by any symmetry transformation.
\end{example}

\begin{example}[\ce{C59$^{13}$C} doped fullerene] \label{ex:doped-fullerene}
    Perhaps the most drastic example of the symmetry breaking \cite{drake_molecular_2006} happens when one of the carbons in \ce{C60} fullerene is substituted with carbon-13 [see Fig.~\ref{fA0:asymmetric}(d)].
    Since the carbon-13 is in a different place after any of the original symmetry rotations, the initial icosahedral symmetry of the molecule, which constitutes 60 different proper rotations, breaks down to no symmetry at all.
\end{example}

\subsection{Asymmetric position states}

When applied to a fixed reference position, each rotation \(\gr\in\so 3\) maps the molecule to a position distinct from the reference; otherwise, the molecule would be symmetric.
We assume there are no other possible transformations that the molecule can undergo, so proper rotations \(\gr\in\so 3\) can be used as labels for the distinct position states \(|\gr\ket\) of an asymmetric molecule \cite[Appx.~2]{ezra_symmetry_1982}\cite[Ch.~5]{harter_principles_1993}.
These position states are identical to those of any asymmetric rigid body, e.g., an airplane \cite{Lynch_modern_2017}.

Molecular position states are ``orthonormal'' in the same sense as the position states of a free particle, satisfying
\begin{equation}
\bra\gr|\gr^\pr\ket=\d^{\so 3}(\gr,\gr^\pr)\,,\label{eq:SO3-orthogonality}
\end{equation}
where the Dirac $\d$-function is infinite for rotations $\gr=\gr^\pr$ and zero otherwise.
This relation makes the position states not normalizable --- \(\bra \gr|\gr\ket\) is technically infinite --- but approximate normalizable position states exist which are localized at \(\gr\) and which qualitatively have the same features (see \vva{Sec.~\ref{subsec:approximate-states}}).
We focus on the idealized non-normalizable case since normalization is not relevant to most of our analysis.

Position states are also complete, yielding a decomposition of the identity, \(\id_\rot\), on the rotational state space of an asymmetric molecule,
\begin{align}
\id_{\rot} & =\int_{\so 3}\dd\gr\,|\gr\ket\bra\gr|\,,\label{eq:id-pos-asymmetric}
\end{align}
where $\dd\gr$ is the Haar measure on $\so 3$ with $\int\dd\gr=8\pi^{2}$.

The molecule's reference position, denoted by \(|\gr=\re\ket\) with \(\re\) the identity rotation, defines a \textit{lab frame} of reference.
Meanwhile, the orientation of the molecule relative to that lab frame defines a \textit{molecule frame} (\textit{a.k.a.}\ the Eckhart frame \cite{Eckart_studies_1935,judd_angular_1975,louck_eckart_1976}), which is ``bolted'' to the molecule and not allowed to rotate without the molecule rotating in the same way.

\begin{figure}[!t]
\centering{}
\includegraphics[width=1.0\columnwidth]{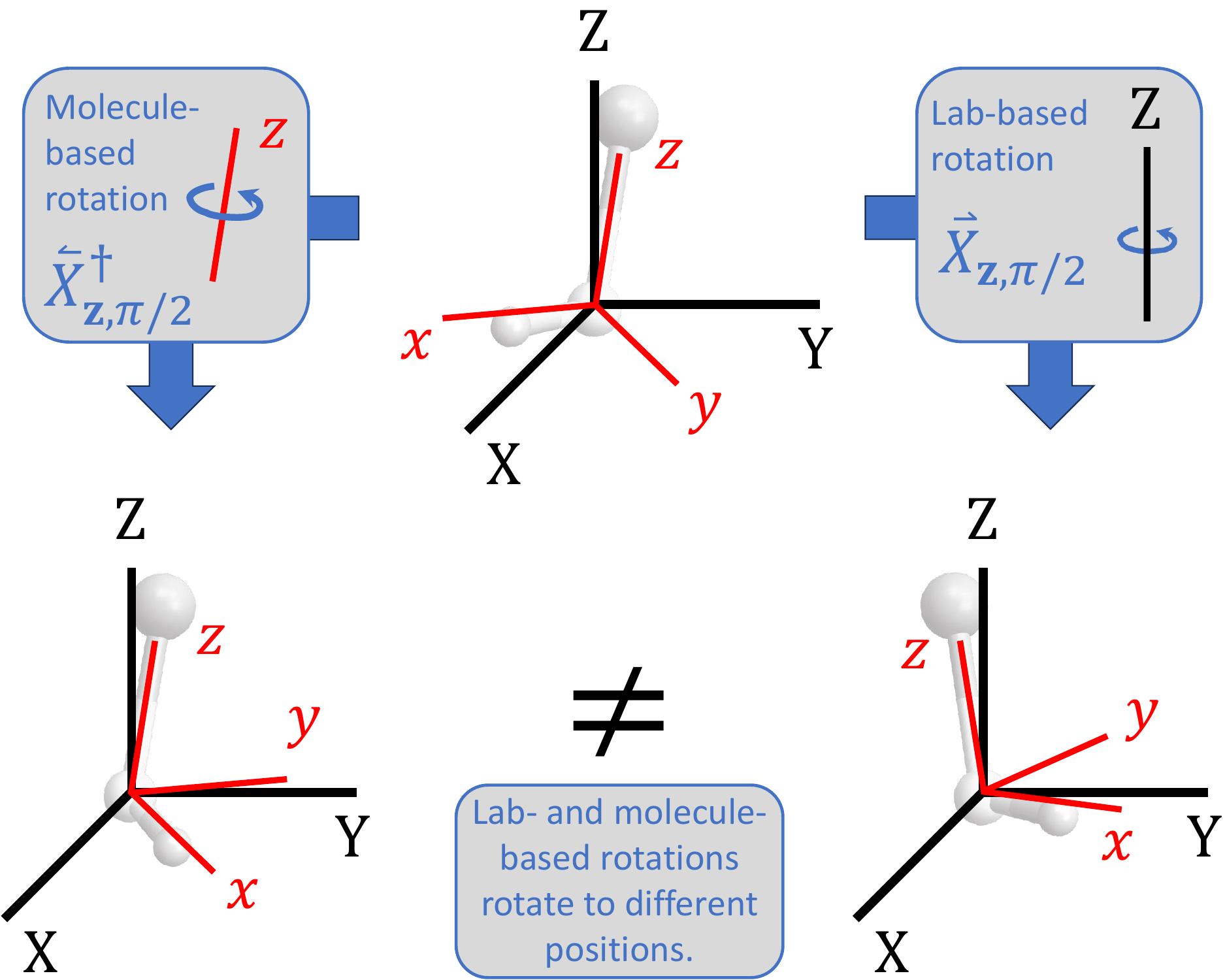}
\caption{\label{fA0:rotations}
Position of an asymmetric molecule is characterized by the relative orientation of the molecule frame (\(\color{red} xyz\) axes) w.r.t.~the lab frame (XYZ axes).
Each frame can be re-oriented independently, yielding two types of rotations.
Molecule-based
and lab-based rotations
rotate around axes defined in the molecule and lab frames, respectively (cf.~\cite[Fig.~3.9]{Lynch_modern_2017}).
The resulting position state after the two types of rotations is generally \textit{not} the same
[see Eq.~\eqref{eq:transformation-lab-based-full}], and the full group of asymmetric rigid-body rotations, \(\so 3\times \so3\), consists of products of the two rotation types. %
}
\end{figure}

The rotation \(\gr\) labeling the molecule's position characterizes the orientation of the molecule frame relative to the lab frame.
Each frame can be re-oriented, and both types of re-orientations can be interpreted as rotations performed on the molecule (see Fig.~\ref{fA0:rotations}).

\begin{subequations}
\label{eq:transformation-lab-based-full}
Re-orienting the molecule frame corresponds to a rotation performed on the molecule around an axis defined in the molecule frame.
Such \textit{molecule-based rotations}, denoted by \(\lr_\rg\) for any \(\rg\in\so3\), act by post-multiplying the rotation matrix labeling the molecular position,
\begin{equation}
    \lr_{\rg}{}|\gr\ket =|\gr\rg^{-1}\ket\,.\label{eq:transformation-body-frame}
\end{equation}

Re-orienting the lab frame is done by \textit{lab-based rotations}, denoted by \(\rr_\rg\).
These act by pre-multiplication,
\begin{equation}
    \rr_{\rg}|\gr\ket =|\rg\gr\ket~.\label{eq:transformation-lab-based}
\end{equation}
\end{subequations}
Since the only piece of information encoded in molecular position is the \textit{relative} orientation of the molecule w.r.t.~the lab frame, we can fix the lab frame in place and think of lab-based rotations as ``active'' rotations of the molecule around an axis defined in the lab frame.

When the molecule is initially in its reference position \(\left|\gr=\re\right\ket\), a lab-based rotation \(\rg\) maps the molecule to the same position as does the molecule-based rotation \(\rg^{-1}\).
This equivalence
corresponds to
\begin{equation}\label{eq:equivalence}
    \rr_{\rg}|\re\ket=\lr_{\rg^{-1}}|\re\ket\equiv|\rg\ket~,
\end{equation}
where it is necessary to take the inverse of \(\rg\) since the two types of rotations change the relative orientation between the two frames in opposite ways.

In general, lab- and molecule-based rotations map an initial position \(\gr\) to two \textit{different} positions,
\(\rg\gr\neq\gr\rg\), because the group \(\so3\) is \textit{non-Abelian}.
Since their actions on position states are not equivalent,  the two types of rotations correspond to distinct and independent operators on the rotational state space.
The full group of asymmetric rigid-body rotations is \(\so 3 \times \so 3\), realized by products of the lab- and molecule-based rotations.

It is not possible to use one type of rotation to induce the effect of the other in a \textit{state-independent} way.
To see this, let \(\gr\) be the initial molecular position. Applying a lab-based rotation \(\rg\) yields the final position \(\rg\gr\).
Alternatively, one can apply a molecule-based rotation \(\rh\) that yields the same final position, \(\rg\gr = \gr \rh^{-1}\), by solving for \(\rh\).
However, the solution, \( \rh = \gr^{-1} \rg^{-1} \gr\), depends on the initial state because proper rotations do not commute.

The lab frame and its corresponding \(\so 3\) factor break down when the molecule is placed in an environment that is less symmetric than free space because symmetry-related positions are no longer distinguishable; this occurs, e.g., when the molecule is placed in a crystal lattice \cite[pg. 372]{harter_principles_1993}.
The molecule frame and its corresponding \(\so 3\) factor break down when the molecule itself is symmetric under some proper rotations.
Maintaining a perfectly symmetric \textit{external} environment, we demonstrate, in subsequent sections, several features imposed by a molecule's \textit{internal} symmetry.

\subsection{Asymmetric rotational states}

Lab- and molecule-based rotations are each generated by their own angular momentum operators, with each set yielding a basis for a copy of the Lie algebra of \(\so3\).
Due to the equivalence from Eq.~\eqref{eq:equivalence}, these two sets are coupled in the sense that they share the same total integer angular momentum \(\ell\geq 0\).

A molecule
can admit different values of the total momentum's projection onto the lab- and molecule-frame axes.
As a consequence of the Peter-Weyl theorem \cite{folland_course_2016,carter_lectures_1995,arovas_lecture_nodate}, the identity on the rotational state space, \(\id_\rot\) from Eq.~\eqref{eq:id-pos-asymmetric}, can alternatively be decomposed into a direct sum indexed by the total momentum \(\ell\),
\begin{equation}\label{eq:id-rot}
    \id_{\rot}  =\bigoplus_{\ell\geq0}\id_{2\ell+1}\ot \id_{2\ell+1}~,
\end{equation}
with blocks of dimension \((2\ell+1)^2\) that further split into two tensor factors
spanned by states of quantized values of the momentum components into the lab- and molecule-frame axes, respectively.
Above, \(\id_d\) for natural number \(d\) denotes the \(d\)-dimensional identity matrix.

The choice of basis for the two \(\id_{2\ell+1}\) factors in Eq.~\eqref{eq:id-rot} is arbitrary, but both are typically expressed in terms of \textit{rotational states} \(|^\ell_m,\om\ket\), where \(m\) and \(\om\) denote the \(\z\)-axis projection of the momentum into the respective frames.
We relegate \(m\) to a subscript because we will mostly focus on \(\om\).
The decomposition~\eqref{eq:id-rot} becomes
\begin{equation}
\id_{\rot}=\sum_{\ell\geq0}\sum_{|m|\leq\ell}\sum_{|\om|\leq\ell}|_{m}^{\ell},\om\ket\bra_{m}^{\ell},\om|~.
\label{eq:id-mom-asymmetric-1}
\end{equation}

Lab-based (molecule-based) rotations act only on the first (second) factor in Eq.~\eqref{eq:id-rot}, decomposing in terms of rotational states as \cite{harter_principles_1993}\cite[footnote 145]{albert_robust_2020}
\begin{subequations}
    \label{eq:rot}
    \begin{align}
        \rr_{\rg}&=\bigoplus_{\ell\geq0}D^{\ell}(\rg)\ot \id_{2\ell+1} \label{eq:active-rot} \\
        \lr_{\rg}&=\bigoplus_{\ell\geq0}\id_{2\ell+1}\ot D^{\ell\star}(\rg)\,,\label{eq:passive-rot}
    \end{align}
\end{subequations}
where $D^{\ell}$ is a matrix representing \(\so3\) rotations.

Rotational states \(|^\ell_m,\om\ket\) can be expressed as superpositions of position states \(|\gr\ket\), and \vva{vice versa},
\begin{subequations}
\begin{align}
    |_{m}^{\ell},\om\ket&={\textstyle \sqrt{\frac{2\ell+1}{8\pi^{2}}}}\int_{\so 3}\dd\gr\,D_{m\om}^{\ell\star}(\gr)|\gr\ket\\
    |\gr\ket&=\sum_{\ell\geq0}{\textstyle \sqrt{\frac{2\ell+1}{8\pi^{2}}}}\sum_{|m|,|\om|\leq\ell}D_{m\om}^{\ell}(\gr)|_{m}^{\ell},\om\ket\,.\label{eq:asymmetric-position}
\end{align}
\end{subequations}
The overlap between elements of these Fourier-dual angular position and momentum states,
\begin{equation}
\bra_{m}^{\ell},\om|\gr\ket={\textstyle \sqrt{\frac{2\ell+1}{8\pi^{2}}}}D_{m\om}^{\ell}(\gr)={\textstyle \sqrt{\frac{2\ell+1}{8\pi^{2}}}}\bra m|D^{\ell}(\gr)|\om\ket\,,\label{eq:wigner-ME}
\end{equation}
is proportional to a Wigner \(D\)-matrix element, i.e., a matrix element of the rotation matrix \(D^\ell\) in the $\z$-axis momentum component eigenbasis \cite[Ch. 4]{varshalovich_quantum_1988}\cite{zare_angular_1991}.

The breaking down of the molecule frame due to symmetry results in
restricting the molecule-frame momentum space \(\{|\om\ket\}\) to particular subspaces.
We perform these restrictions in Sec.~\ref{sec:symmetric}, obtaining sets of momentum and position states for symmetric molecules.
\subsection{Incorporating nuclear spin}

Asymmetric molecules have no symmetries, meaning that any rotations that permute nuclei will necessarily map the molecule to a distinct orientation (see Example~\ref{ex:methylamine}).
There are thus no restrictions on the state space due to spin statistics, so the nuclei are not required to be in any particular (e.g., single or triplet) state.
We denote the identity on the state space of any nuclear spins as \(\id_{\nat}\).

The combined rotational-nuclear space for an asymmetric molecule is a tensor product of the rotational and nuclear-spin spaces,
\begin{subequations}\label{eq:identity}
\begin{align}
\id_{\mol} & =\id_{\rot}\ot \id_{\nat}\\
 & =\int_{\so 3}\dd\gr\,|\gr\ket\bra\gr| \ot \id_{\nat}~,\label{eq:identity2}\\
 & =\sum_{\ell\geq0}\sum_{|m|\leq\ell}\sum_{|\om|\leq\ell}|_{m}^{\ell},\om\ket\bra_{m}^{\ell},\om|\ot \id_{\nat}~,\label{eq:id-mom-asymmetric}
\end{align}
\end{subequations}
where we plug in the two position-state and rotational-state decompositions of the rotational factor from Eqs.~\eqref{eq:id-pos-asymmetric} and \eqref{eq:id-mom-asymmetric-1}, respectively.
If the nuclear spin of each atom participating in a given molecule is zero, then the \(\id_\nat\) factor is dropped, and only the rotational factor remains.

Given a basis \(|\chi\ket\) for the nuclear-spin factor \(\id_\nat\), a basis for the entire state space of an asymmetric molecule consists of states
\begin{equation}\label{eq:basis-separable}
     |^\ell_m,\om\ket_{\rot}\ot|\chi\ket_{\nat}~,
\end{equation}
where we have explicitly split each basis state into a factor coming from the rotational space and a factor coming from the nuclear-spin space.
Each of these states is rotation-spin \textit{separable}, i.e., is in tensor-product form with respect to the rotation-spin factorization.
Rotation-spin non-separable or \textit{entangled} states can then be expressed as superpositions of more than one of these states.

While there exist separable states in the state space of asymmetric molecules, the combination of symmetry and spin-statistics can make it impossible for separable states to exist at all in state spaces of symmetric nuclear spin isomers.
In later sections, we show that, after imposing symmetry and spin-statistics constraints, rotation-spin basis states are dressed by the nuclear spin in a way that makes \textit{each} basis state entangled [see Eq.~\eqref{eq:entangled-state}].
Since such states form an orthonormal basis, superposing them can only increase their degree of entanglement and thus cannot yield a separable state.

The nuclei carry angular momentum \(S\) in the form of nuclear spin, which can be defined with respect to the molecule frame (\(\overrightharpoon{S}\)) or the lab frame (\(\overleftharpoon{S}\)) \cite[Table I]{Rusconi_magnetic_2016}. 
When incorporating spin in certain physical settings, lab-based and molecule-based rotations are  extended to rotate both the rotational degrees of freedom and the nuclear spins.
In other words, while the original rotations in Eq.~\eqref{eq:rot}, \( \rr_{\rg}\) and \( \lr_{\rg}\), are generated by rotational angular momenta, \(\overrightharpoon{J}\) and \(\overleftharpoon{J}\), composite rotations are generated by the total angular momenta, \(\overrightharpoon{J}+\overrightharpoon{S}\) and \(\overleftharpoon{J}+\overleftharpoon{S}\), in the lab and molecule frames, respectively.
The extra nuclear-spin terms induce tensor-product rotations of the form \(U^{\otimes M}\) acting on any subset of \(M\) identical spins, where \(U\) is an \(\su 2\) rotation representing \(\rg\).
These operations are distinct from the nuclear-spin permutations highlighted in this work (due to Schur-Weyl duality \cite{harrow_applications_2005,Schmiedt_unifying_2016}) and so are neglected from now on for simplicity, with the caveat that they may have to be accounted for in an experimental setting.

\subsection{Holonomy of asymmetric position states}\label{sec:asymmetric-holonomy}

We are interested in what happens when a molecule is initialized in a particular position state \(|\gr(0)\ket\) and then adiabatically traverses a \textit{closed path} --- $\{\gr(t)\,,\,t\in[0,1]\}$ with \(\gr(1)\) identified with \(\gr(0)\) --- in its position space.
This path can, for example, be induced by applying lab-based rotations \(\rr_{\gr(t)}\), which rotate the molecule to a different position state according to Eq.~\eqref{eq:transformation-lab-based}.

The full molecular position state in Eq.~\eqref{eq:identity2} includes a nuclear-spin factor.
This factor tags along and is unaffected during the position-state path.
As such, we omit the nuclear states below for simplicity.

Upon traversing a closed path in its position-state space \(\so 3\), the molecular state undergoes a \textit{holonomy} \cite{simon_holonomy_1983,Zanardi_holonomic_1999,avron_adiabatic_2012},
\begin{equation}
|\gr(0)\ket\to U_{\hol}|\gr(0)\ket\,.
\end{equation}
The holonomy operator is, in this case, a scalar whose argument is called the geometric or \textit{Berry phase} \cite{berry_quantal_1984}.
Note that a Hamiltonian is not required for determining the holonomy, which can be determined solely from the set of states and their parameter space.

The holonomy is invariant under changes in phase convention of the ``instantaneous'' states, $|\gr(t)\ket\to e^{i\varphi(t)}|\gr(t)\ket$,
and different phase conventions (sometimes called ``gauges'') and path parameterizations provide different ways to calculate the same holonomy.

We parameterize closed paths by the same set of coordinates for all points except, potentially, the last point \(\gr(t=1)\).
To account for cases when the last point requires a different set of coordinates, we express
the holonomy as a product of two terms \cite[Eq.~(1.2)]{read_non-abelian_2009},
\begin{equation}\label{eq:holonomy-abelian}
U_{\hol}=U_{\mon}\exp\left(i\int_{\gr(0)}^{\gr(1)}\dd\gr\,A(\gr)\right)\,.
\end{equation}

The second term on the right-hand side of Eq.~\eqref{eq:holonomy-abelian} is an integral of the \textit{Berry connection},
\begin{equation}\label{eq:connection-abelian}
A(\gr)=i\bra\gr|\p\gr\ket/\bra\gr|\gr\ket\,,
\end{equation}
defined for all points $\gr=\gr(t)$ in the path but the last one (i.e., for \(t < 1\)), and with the partial derivative being with respect to \(t\).
This connection depends on ``local'' features of the parameter space, such as the curvature.
We will show that \(A(\gr(t))=0\) for all \(0\leq t<1\), which means the connection is \textit{locally flat}.

The first term on the right-hand side of Eq.~\eqref{eq:holonomy-abelian} --- the \textit{monodromy} $U_{\mon}$ --- can occur when $\gr(1)$ is labeled with a \textit{different} set of coordinates than $\gr(0)$.
In order to resolve the labeling mismatch, the final state \(|\gr(1)\ket\) has to be re-expressed in terms of the initial state \(|\gr(0)\ket\).
A non-trivial $U_{\mon}$ is precisely the transformation relating the two states in the case when they are not equal; this happens when \(\gr(t)\) parameterizes a non-contractible path.
As such, the monodromy piece depends on ``global'' or ``topological'' features of the parameter space.

The celebrated Aharonov-Bohm phase \cite{Ehrenberg_refractive_1949,Aharonov_significance_1959} is due to a monodromy
\cite{Vasselli_background_2019}, and so is the \(\pi\)-phase due to circling a conical intersection \cite{mead_geometric_1992}.
In such Hamiltonian-based cases, a non-simply-connected parameter space is induced by the structure of the eigenstates.

\begin{example}[conical intersection toy model]\label{ex:conical}
    Consider evolving a family of two-level system states \cite{alexandradinata_pedagogical_nodate},
    \begin{equation}
        |\phi\ket=\begin{pmatrix}\phantom{-}\sin\phi/2\\
        -\cos\phi/2
        \end{pmatrix}\quad\quad\text{for}\quad\quad 0\leq\phi<2\pi~,
    \end{equation}
    along the path \(\phi(t) = 2\pi t\). The Berry connection is locally flat for this case, \(\bra\phi|\p\phi\ket=0\), a fact that we can immediately observe because all states in the path are real-valued \cite{simon_holonomy_1983}. However, the state at the end point, \(\phi(1)=2\pi\), still has to be re-expressed in terms of that at the initial point, \(\phi(0)=0\), in order to conform to the chosen parameterization.
    The two states differ by a phase, \(\left|\phi=2\pi\right\rangle =-\left|\phi=0\right\rangle\), and re-expressing reveals a monodromy of \(-1\).
    This simple example holds more generally for any \(2\times2\) symmetric matrix \cite{Berry_Anticipations_1990}.
\end{example}

We now show that the Berry connection in Eq.~\eqref{eq:connection-abelian} is locally flat for asymmetric molecular position states \(|\gr\ket\).
Plugging in the expression in terms of rotational states in Eq.~\eqref{eq:asymmetric-position}, using $D_{m\om}^{\ell\star}(\gr)=D_{\om m}^{\ell}(\gr^{-1})$, and re-expressing the result as a trace of a product within the angular momentum space yields
\begin{subequations}
\begin{align}
\bra\gr|\p\gr\ket&=\sum_{\ell\geq0}\sum_{|m|,|\om|\leq\ell}{\textstyle \frac{2\ell+1}{8\pi^{2}}}D_{m\om}^{\ell\star}(\gr)\p D_{m\om}^{\ell}(\gr)\\&=\sum_{\ell\geq0}{\textstyle \frac{2\ell+1}{8\pi^{2}}}\tr\left(D^{\ell}(\gr^{-1})\,\p D^{\ell}(\gr)\right)\,.\label{eq:current}
\end{align}
\end{subequations}

The quantity inside the trace is a representation \vva{of} the molecule's angular velocity matrix relative to the molecule frame, \(\gr^{-1}\p\gr\) \cite[Prop. 3.9]{Lynch_modern_2017}.
Angular velocity is valued in the tangent space of the space \(\so 3\) of position-state labels, otherwise known as the \(\so 3\) or angular momentum Lie algebra \cite{varshalovich_quantum_1988}.
This Lie algebra is generated by the angular momentum operators, all of which are traceless. Therefore, the angular velocity matrix is also traceless, \(\tr(D^{\ell}(\gr^{-1})\,\p D^{\ell}(\gr)) = 0\) for all \(\ell\).
Since \vva{each term of} the numerator \(\bra\gr|\p\gr\ket\) is zero and the denominator \(\bra\gr|\gr\ket\) is positive \vva{after the regularization of Sec.~\ref{subsec:approximate-states}}, the connection \eqref{eq:connection-abelian} is zero \vva{(a special case of Theorem~\ref{thm:flat})}.

Since the space \(\so 3\) is not simply connected \cite{arovas_lecture_nodate}, there could be a nontrivial monodromy piece.
However, \(U_{\mon}\) turns out to be the identity for this particular case.
For example, if the original coordinates parameterizing position states are Euler angles $(\alpha,\beta,\gamma)$ with $0\leq\alpha<2\pi$ (\textit{excluding} \(2\pi\)), but the path is parameterized by $(2\pi t,0,0)$ (\textit{including} \(2\pi\)), then the final point $(2\pi,0,0)$ has to be mapped back into the initial point $(0,0,0)$.
The states corresponding to the two points are proportional (up to the monodromy) and, in this case, are exactly equal,
\begin{equation}
    |\gr=(2\pi,0,0)\ket=|\gr=(0,0,0)\ket~,
\end{equation}
yielding a \(U_{\mon}\) of identity.
We prove the general case using tools from the symmetric molecule formulation (see Example~\ref{ex:rot-symmetric-holonomy}).

Together, the above results show that there is no holonomy for closed loops in the position state space of an asymmetric molecule and, by extension, any asymmetric rigid body.
This should not be surprising since other ``classical'' states of fixed position do not have Berry phases either, such as position states of a free particle on a line or a planar rotor.
We will see in Sec.~\ref{sec:holonomy-position} that this is no longer the case for position states of many symmetric molecules with non-zero nuclear spin.

\begin{table}
\begin{centering}
\begin{tabular}{lccl}
\toprule
\multirow{2}{*}{symmetry~} & \multicolumn{2}{c}{examples} & \multirow{2}{*}{~~~~~~~~~~irreps}\tabularnewline
 & rotational & perrotational & \tabularnewline
\midrule
$\C{1}$ & \ce{CaSH} & none & $\a$ \tabularnewline
$\C{2N\phantom{+1}}$ & \ce{SO2} & \ce{H2O} & $\a$, $\b$,\! $^{1}\e_{i\leq N-1}$,\! $^{2}\e_{i\leq N-1}$\tabularnewline
$\C{2N+1}$ & \ce{HCo(CO)4} & \ce{CaOCH3} & $\a$, $^{1}\e_{i\leq N}$,$^{2}\e_{i\leq N}$\tabularnewline
$\C{\infty}\cong\so{2}$ & \ce{HCl} & none & $\a$, $\lambda \in \Z_{\neq 0}$\tabularnewline
\midrule
$\D{2\phantom{N+1}}$ & \ce{C2S4} & \ce{C2H4} & $\a$, $\b_{1}$, $\b_{2}$, $\b_{3}$ \tabularnewline
$\D{2N\phantom{+1}}$ & \ce{S8} & \ce{XeF4} & $\a_{1}$, $\b_{1}$, $\a_{2}$, $\b_{2}$, $\e_{i\leq N-1}$\tabularnewline
$\D{2N+1}$ & \ce{SO3} & \ce{BF3} & $\a_{1}$, $\a_{2}$, $\e_{i\leq N}$\tabularnewline
$\D{\infty}\cong \OO{2}$ & \ce{S2} & \ce{H2} & $\a_{1}$, $\a_{2}$, $\lambda \in \Z_{>0}$\tabularnewline
\midrule
$\T$ & \ce{XeO4} & \ce{CH4} & $\a$, $^{1}\e$, $^{2}\e$, $\ti$\tabularnewline
$\O$ & \ce{Mo(CO)6} & \ce{SF6} & $\a_{1}$, $\a_{2}$, $\e$, $\ti_{1}$, $\ti_{2}$\tabularnewline
$\phantom{.}\I$ & \ce{C60} & \ce{C20H20} & $\a$, $\ti_{1}$, $\ti_{2}$, $\g$, $\h$\tabularnewline
\bottomrule
\end{tabular}
\par\end{centering}
\caption{
List of rigid-body symmetry groups, i.e., the possible subgroups of \(\so 3\): the cyclic ($\C{M}=\Z_M$ for \(M \geq 1\)), dihedral ($\D{M}$ for \(M \geq 2\)), continuous (\(\C\infty\cong\so2\), \(\D\infty\cong\OO2\)), and exceptional ($\T$, $\O$, $\I$) groups.
An example of a rotationally and perrotationally symmetric molecule is provided when possible.
Asymmetric molecules can be thought of as rotationally symmetric molecules whose symmetry group is the trivial group \(\C1\).
As such, symmetry alone is not sufficient to determine whether a molecule is rotationally or perrotationally symmetric: one needs to also know about the spin of any identical nuclei that are permutable via a symmetry rotation.
The last column lists group irreps, labeled by Mulliken symbols \cite{altmann_point-group_1994,ceulemans_group_2013} (see also \cite{Mulliken_electronic_1933,elcoro_double_2017}).
We use complex (instead of real) irreps because we do not consider time-reversal symmetry.
\label{tab:groups}}
\end{table}
%

\section{Symmetric rotational states}
\label{sec:symmetric}

In this work, we define a symmetric molecule as one whose position is invariant under a set of proper-rotation symmetry operations.
This set forms the molecule's
\begin{equation*}
    \text{\textit{rigid-body symmetry group},}~\G\subset\so3,
\end{equation*}
a subgroup of the group \(\so 3\) of proper (i.e, orientation-preserving) rotations.
This symmetry group is a subgroup of the \textit{full} symmetry group of the ``static molecular model'' \cite[Eq.~(2.21)]{ezra_symmetry_1982} which can include reflections, improper rotations, and parity symmetry.
Since we concern ourselves only with symmetry transformations that are realizable by rigid-body rotations, we consider only \(\G\) and refer to it as simply ``the symmetry group'' throughout the text.

Subgroups of \(\so 3\) can be discrete, meaning that they have a finite set of elements, or continuous, meaning that some of their elements are parameterized by a continuous variable (see Table \ref{tab:groups}).

The two continuous symmetry groups are isomorphic to the group of planar rotations, \(\C\infty \cong \so 2\), and the group of planar rotations and reflections, \(\D\infty\cong \OO 2\).
The \ce{HCl} molecule, along with any heteronuclear diatomic, is invariant under rotations by any angle around its principal axis, meaning that its symmetry group is \(\C\infty\).
The \ce{H2} molecule, along with any homonuclear diatomic, is invariant under those same rotations and also \(\pi\)-rotations around axes perpendicular to its principal axis, all of which exchange its two constituent atoms; its symmetry group is \(\D\infty\).

The possible discrete symmetry groups include the trivial group \(\C{1}\) (corresponding to no symmetry at all),
the cyclic groups \(\C N\), and the dihedral groups \(\D N\) (both for \(N\geq 2\)).
All asymmetric molecules, studied in Sec.~\ref{sec:asymmetric}, have the trivial symmetry group.
The \(\C N\) group can be thought of as a discretized version of \(\C\infty\), where rotations are only by multiples of \(2\pi/N\).
Examples of \(\C 3\)-symmetric molecule is ammonia (\ce{NH3}), whose atoms \textit{do not} lie in a plane.
The \(\D N\) group is a similarly discretized version of \(\D\infty\).
Boron trifluoride (\ce{BF3}) or sulphur trioxide (\ce{SO3}), whose four atoms \textit{do} lie in a plane, both have \(\D 3\) symmetry.

Rounding out the list of symmetry groups are the three ``exceptional'' subgroups --- the tetrahedral \(\T\), octahedral \(\O\), and icosahedral \(\I\) group.
Methane is a canonical example of tetrahedral symmetry --- the symmetry group of a tetrahedron.
Examples of octahedral and icosahedral symmetry include sulfur hexafluoride (\ce{SF6}) and fullere (\ce{C60}), respectively.
To reiterate, a (rigid-body) symmetry group consists of only proper rotations collected from the full symmetry group of a molecule, which is \(\O_{\text{h}}\) and \(\I_{\text{h}}\), respectively, in the latter two cases.

There are cases where symmetry rotations permute identical nuclei, \textit{and} those nuclei have non-zero nuclear spin.
A simple example is water, which admits a proper rotation that exchanges the two hydrogen atoms while leaving the oxygen atom intact.
In such cases, in addition to restricting the state space to only \(\G\)-symmetric states, we also have to take into account any Bose or Fermi spin statistics realized by the nuclei.

We split symmetric molecules into two categories, depending on whether or not symmetries permute identical spinful nuclei.
If all identical nuclei that are permuted under symmetry rotations have zero nuclear spin, then we call the molecule \textit{rotationally symmetric}.
Identical nuclei in rotationally symmetric molecules can be treated as points, and such molecules are special cases of symmetric rigid bodies.
While technically correct, ascribing the invariance of such molecules to the statistics of any ``spin-zero nuclei'' is as necessary as attaching a zero spin to the end of a \vva{dumbbell}.

If any of the identical permuted nuclei have non-zero nuclear spin, then we call the molecule \textit{perrotationally symmetric}.
Nuclei in perrotationally symmetric molecules have a non-trivial nuclear-spin space whose factors are permuted by molecule-based symmetry rotations, requiring a treatment deviating from that of symmetric rigid bodies.

For example,
dihydrogen \ce{H2} is \(\D\infty\)-symmetric, but its two atoms have spin one-half nuclei.
A molecule-based rotation that exchanges the identical nuclei in \ce{H2} results in a \(-1\) Fermi spin-statistics factor, making the molecule perrotationally symmetric.
On the other hand, disulfur \ce{S2} is also \(\D\infty\)-symmetric, but has spinless nuclei.
There is no spin-statistics factor to worry about, so \ce{S2} is rotationally symmetric and can be treated in the same way as a dumbbell.

A similar distinction is observed between rotationally \(\D 3\)-symmetric sulfur trioxide (\ce{SO3}) and perrotationally \(\D 3\)-symmetric boron trifluoride (\ce{BF3}), and between rotationally \(\T\)-symmetric xenon tetroxide (\ce{XeO4}) and perrotationally \(\T\)-symmetric methane (\ce{CH4}).
More examples are provided in Table~\ref{tab:groups}.
To reiterate, shape and symmetry are not enough to determine whether a molecule is rotationally or perrotationally symmetric; one needs to also know about the spin of any identical nuclei.

\subsection{Rotationally symmetric molecules}\label{sec:symmetry}

We now consider cases where molecule-based symmetry-group rotations \textit{do not} permute identical spinful nuclei, meaning that the rotational state space of such symmetric molecules is identical to that of a rigid body with the same symmetry.

Rotationally \(\G\)-symmetric molecular states span a subspace of the asymmetric molecular state space, defined by \(\id_\mol\) \eqref{eq:identity} and spanned by the basis in Eq.~\eqref{eq:basis-separable}.
This subspace is defined by the restriction that
\begin{equation}
\lr_{\rg}|\psi_{\mol}\ket=|\psi_{\mol}\ket~,\quad\quad\forall\rg\in\G\,,\label{eq:symm-restriction}
\end{equation}
for any state \(|\psi_\mol\ket\) of the symmetric molecule and any molecule-based rotation \(\lr\).

In representation-theoretic terms, Eq.~\eqref{eq:symm-restriction} implies that the molecule transforms under the trivial irreducible representation, or irrep, of $\G$.
In this irrep, denoted by $\irt$ (and sometimes by \(\a_1\)), all elements \(\rg\) are mapped into one, i.e., $\irt(\rg)=1$.
We need to keep only states in the decomposition of $\id_{\mol}$ (\ref{eq:identity}) that transform as this irrep.
We can determine this space by evaluating molecule-based rotations at elements of the symmetry group, decomposing the result into all the irreps of $\G$, and keeping only copies of the trivial irrep.

Restricting $\rg$ to rotations in the symmetry group, each $D$-matrix for a given \(\ell\), while forming an irreducible representating of \(\so 3\), forms a \textit{reducible} representation of \(\G\).
This representation is unitarily equivalent to a block matrix whose blocks are labeled by any \(\G\)-irreps $\ir$ that are present in the \(D\)-matrix.
Each block is characterized by two numbers, the dimension, \(\dim\ir\), of the irrep and the number of copies of the irrep --- the frequency or \textit{multiplicity} \(\ml_\ell \ir\) --- present in the block.
The irrep dimension is \(\ell\)-independent, while the multiplicity can depend on the angular momentum.

We can ``distribute out'' the irrep and unitarily transform each block into tensor-product form, with the first factor housing the irrep, and the second factor housing the multiplicity space.
This yields the \textit{isotypic decomposition} (\textit{a.k.a.}~canonical decomposition) \cite{harrow_applications_2005,folland_course_2016},
\begin{equation}
D^{\ell\star}(\rg)\cong\bigoplus_{\ir\uparrow\ell}\ir(\rg)\ot \id_{\ml_{\ell}\ir}\,,\label{eq:isotypic-wigner-1}
\end{equation}
where \(\ir(\rg)\) is the matrix representation of the group element \(\rg\), and \(\id_{\ml_{\ell}\ir}\) is the identity matrix.
Not all \(\G\)-irreps participate in a given rotation matrix, and the shorthand notation \(\ir\up\ell\) denotes the subset of irreps that are present in the above decomposition; such irreps are determined by ``branching'' or ``subduction'' rules.

\begin{example}[\(\G=\C\infty\) isotypic decomposition]\label{ex:cinfty}
    For example, rotation matrices \(D^\ell\), when restricted to the group \(\G=\C\infty\) of \(\z\)-axis rotations, are already diagonal w.r.t.\ the \(\C{\infty}\)-irrep basis,
\begin{equation}\label{eq:abelian-decomp}
    D^{\ell\star}(\z,\phi)=\sum_{|\om|\leq\ell} e^{i\om\phi}|\om\ket \bra\om|~,
\end{equation}
where \(\phi\) denotes the angle of the \(\z\)-axis rotation.\footnote{\label{fn:notation-abuse}
We abuse notation by substituting a rotation's axis-angle \((\vv,\phi)\) or Euler-angle \((\alpha,\beta,\gamma)\) parameterizations \cite{chirikjian_engineering_2000} for the rotation  itself as the argument of the rotation matrix \(D^{\ell}\).
}

The set of distinct \(\C\infty\)-irreps is in one-to-one correspondence with the integers \(\lambda\in\Z\) (with \(\lambda=0\) the trivial irrep \(\a\)), and each one-dimensional block houses the irrep \(\ir_{\lambda=\om} (\phi) = e^{i\om\phi}\).
The set of participating irreps is thus
\begin{equation}
    \ir\up\ell = \{\ir_{\lambda=\om}~\text{such that}~|\om|\leq\ell\}~.
\end{equation}
The multiplicity space is trivial in this case since each participating irrep is featured only once for each \(\ell\).
\end{example}

Plugging the decomposition \eqref{eq:isotypic-wigner-1} into Eq.~(\ref{eq:passive-rot}) yields
\begin{equation}
\lr_{\rg}\cong\bigoplus_{\ell\geq0}\id_{2\ell+1}\ot\bigoplus_{\ir\uparrow\ell}\ir(\rg)\ot \id_{\ml_{\ell}\ir}\ot \id_{\nat}\,.\label{eq:rot-sym}
\end{equation}
The next step is to apply the symmetry restriction (\ref{eq:symm-restriction}),
which means keeping only the blocks corresponding to trivial irreps, $\ir=\irt$.
We do this with the help of a Kronecker \(\d\)-function: $\d_{\ir,\irt}=1$ when $\ir=\irt$ and zero otherwise.
The trivial irrep is one-dimensional, $\dim\irt=1$, so the first factor in the decomposition \eqref{eq:rot-sym} reduces to a scalar factor, \(\irt(\rg)=1\).
The second multiplicity factor remains nontrivial because each $\ell$ can, in general, contain more than one copy of the trivial irrep.

All of the trivial-irrep blocks can be combined into a projection onto the state space of a \(\G\)-symmetric molecule,
\begin{align}
\id_{\mol}^{\G} & =\bigoplus_{\ell\geq0}\id_{2\ell+1}\ot\bigoplus_{\ir\uparrow\ell}\d_{\ir,\irt}\id_{\ml_{\ell}\irt}\ot \id_{\nat}\,.\label{eq:id-precursor}
\end{align}
Expressing the first two factors in terms of a basis, Eq.~\eqref{eq:id-precursor} becomes
\begin{equation}
\id_{\mol}^{\G}=\sum_{\ell\downarrow\irt}\sum_{|m|\leq\ell}\sum_{\k=1}^{\ml_{\ell}\irt}|_{m\k}^{\ell}\ket\bra_{m\k}^{\ell}|\ot \id_{\nat}~,\label{eq:id-mom-symmetric}
\end{equation}
where $m$ and $\k$ index the two factors for each $\ell$ whose \(D\)-matrix contains \(\irt\) in its decomposition.
The sum over \(\ell\) becomes restricted to the subset of angular momenta which subduce to the trivial irrep. In (Frobenius \cite{folland_course_2016}) reciprocity to the notation used in Eq.~\eqref{eq:isotypic-wigner-1}, we denote this subset by the shorthand \(\ell\down\irt\).

The symmetric rotational states \(|_{m\k}^{\ell}\ket\) are the angular momentum states of a $\G$-symmetric rigid body.
The key difference from the asymmetric case is the second factor in Eq.~\eqref{eq:id-precursor}, spanned by the basis \(\{|\k\ket\}\) of states that transform as the trivial \(\G\)-irrep.
Hamiltonian eigenstates and, more generally, any molecular states can be expressed as superpositions of the symmetric rotational states.
\begin{figure}[!t]
\centering{}
\includegraphics[width=1.0\columnwidth]{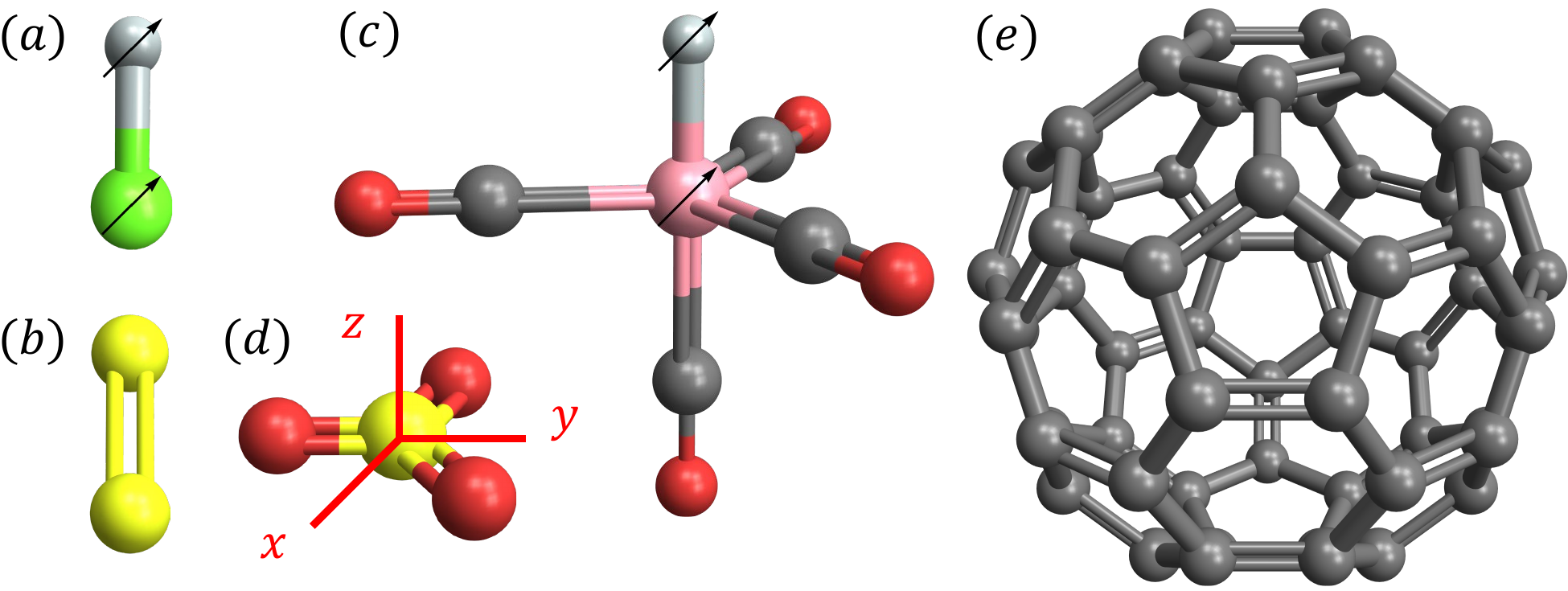}
\caption{\label{fA1:symmetric} Rotationally symmetric molecules (a) \ce{HCl}, (b) \ce{S2}, (c) \ce{HCo(CO)4}, (d) \ce{SO3}, and (e) \ce{C60} from Examples \ref{ex:heteronuclear}-\ref{ex:fullerene}.
Proper rotations in a symmetry group \(\G\) leave the positions of these molecules invariant.
\ce{SO3} (d) is shown in the reference position state chosen in Example~\ref{ex:sulfur-trioxide}.
}
\end{figure}

\begin{example}[calcium monohydrosulfide]\label{ex:asymmetric}
The procedure for determining the state space of a symmetric molecule can also be applied to asymmetric molecules, such as the bent triatomic \ce{CaSH} \cite{Sheridan_high-resolution_2007,augenbraun_molecular_2020} [see Fig.~\ref{fA0:asymmetric}(a)].
The symmetry group of such molecules is trivial, containing only the identity element, \(\G=\C1=\bra \re \ket\).
This group has only one irrep, the trivial irrep \(\a\).
Each rotation matrix $D^\ell$, when restricted to \(\G\), evaluates to the identity matrix whose dimension, $2\ell+1 = \ml_\ell \a$, indexes the multiplicity of the trivial irrep for that angular momentum.
We wind up with a re-expression of the asymmetric rotational states from Eq.~\eqref{eq:id-mom-asymmetric}, $|_{m\k}^{\ell}\ket=|_{m}^{\ell},\om=\k-\ell-1\ket$ for all \(\ell,m\) and for \(1\leq\k\leq2\ell+1\).
\end{example}

\begin{example}[hydrogen chloride]\label{ex:heteronuclear}
Aligning the \ce{HCl} molecule along the \(\z\)-axis, we observe that any rotation around the \(\z\)-axis leaves it invariant.
There are no other rotations that do this, so \(\G=\C{\infty}\).

Group elements \(\rg\) of \(\C{\infty}\) are rotations by angle \(\phi\in[-\pi,\pi)\), and their corresponding irreps, \(e^{i\lambda\phi}\), are indexed by integers \(\lambda\).
The case \(\lambda=0\) corresponds to the trivial irrep.

Recalling Example~\ref{ex:cinfty}, the \(\z\)-axis projection eigenstate state $|\om\ket$ houses a copy of the \(\lambda=\om\) irrep.
The trivial irrep is featured only once, at \(\om=0\), so there is no need for a $\k$ index.
The \(\C \infty\)-symmetric rotational states are the subset of the asymmetric rotational states \(\{|_{m}^{\ell},\om\ket\}\) from Eq.~\eqref{eq:id-mom-asymmetric} with zero \(\om\), $|_{m}^{\ell}\ket\equiv|_{m}^{\ell},\om=0\ket$ for all \(\ell,m\).
This basis corresponds to the spherical harmonics \cite{harter_principles_1993}.

The states \(\{|^\ell_m\ket\}\) span the rotational state space of any diatomic whose atoms are distinct, even in cases when its nuclei admit non-zero nuclear spin.
They also describe the state space of any \(\C\infty\)-symmetric linear \textit{polyatomic} molecule, including \ce{N-N-O} or \ce{C-C-N}.
\end{example}

\begin{example}[disulfur]\label{ex:disulfur}
    The \ce{S2} molecule is a symmetric molecule with symmetry group \(\G=\D\infty\).
    Aligning this dumbbell-like molecule along the \(\z\) axis, the rotations correspond to \(\z\)-axis rotations by arbitrary angles and form the subgroup \(\C\infty\), while ``reflections'' correspond to \(\pi\)-rotations around any equatorial axis.
    Symmetry-group rotations \textit{do} exchange identical sulfur nuclei, but these nuclei do not have any nuclear spin, so this molecule is rotationally symmetric.

    The \(\D\infty\) group admits a trivial irrep \(\a_{1}\), and a ``sign'' irrep \(\a_{2}\), which represents \(\z\)-axis rotations by \(+1\) and equatorial-axis rotations by \(-1\).
    The group also admits a countably infinite set of two-dimensional irreps, indexed by \(\lambda\in\Z\), for which \(\z\)-axis rotations by angle \(\phi\) are represented by \(\exp(i \lambda \phi \sigma_z)\), and the \(\y\)-axis ``reflection'' is represented by \(\sigma_x\) (given the usual Pauli matrices \(\sigma_{x,z}\)).

    Since \(\C\infty\) is a subgroup of \(\D\infty\), we can first apply the analysis of the previous example and consider which of the admissible states from that case, \(\{|^\ell_m,\om=0\ket\}\), transform as the trivial irrep of the larger group.

    Since we have already determined how the \(\z\)-axis rotations act via Eq.~\eqref{eq:abelian-decomp}, we are left to determine the remaining equatorial rotations.
    Each such rotation is a product of a \(\z\)-axis rotation and a ``fiducial'' \(\y\)-axis \(\pi\)-rotation.
    The latter is simply represented for each \(\ell\),$^{\ref{fn:notation-abuse}}$
    \begin{equation}\label{eq:dihedral-reflection}
        D^{\ell\star}(\y,\pi)=\sum_{|\om|\leq\ell}\left(-1\right)^{\ell+\om}\left|-\om\right\rangle \bra\om|~,
    \end{equation}
    implying that the trivial irrep occurs only in \(\om=0\) states with \textit{even} momenta, with the remaining \textit{odd} angular momentum states realizing the sign irrep.
    The basis \(\{|^\ell_m,\om=0\ket\}\) for even \(\ell\) corresponds to the even-momentum spherical harmonics.

    The above basis spans the rotational state space of any \(\D\infty\)-symmetric linear molecule with spinless identical nuclei, e.g., \ce{C-N-C} or the closed-shell O$_{2}^{2-}$.

\end{example}

\begin{example}[cobalt tetracarbonyl hydride]\label{ex:cobalt}
The \ce{HCo(CO)4} molecule is symmetric under rotations by $\pm2\pi/3$, which form the group $\G=\C 3$ \cite{Veillard_cobalt_1990}.
Symmetry-group rotations \textit{do} permute identical C and O nuclei, but none of these nuclei have any nuclear spin, so this molecule is rotationally symmetric.

The group \(\C 3\) has three elements, denoted by angles \(\phi\in\{0,2\pi/3,4\pi/3\}\),
and three irreps, with the trivial one denoted as $\a$.
Picking the rotations to be around the (principal) \(\z\)-axis, we can utilize the decomposition from Eq.~\eqref{eq:abelian-decomp} and determine that the trivial irrep occurs whenever \(\om\) is a multiple of three.
For example, when \(\ell=4\), there are three states, \(\left|\k=1\right\rangle \equiv\left|\om=-3\right\rangle \), \(\left|\k=2\right\rangle \equiv\left|\om=0\right\rangle \), and \(\left|\k=3\right\rangle \equiv\left|\om=3\right\rangle \).

More generally, for \(\C N\)-symmetric molecules that are aligned so that the symmetry group consists of \(\z\)-axis rotations, the basis \(\{|^\ell_{m\k}\ket\}\) is a subset of the asymmetric rotational states \(\{|_{m}^{\ell},\om\ket\}\) from Eq.~\eqref{eq:id-mom-asymmetric} for which \(\om\) is zero modulo \(N\).
The basis states are
\begin{equation}\label{eq:CN-basis}
    |_{m\k}^{\ell}\ket\equiv\left|_{m}^{\ell},\om=N(\k-1-n)\right\ket
\end{equation}
for all \(\ell,m\), and for \(\k\) ranging from one to the multiplicity \(\ml_\ell \irt = 2n+1\), where \(n=\lfloor\ell/N\rfloor\).
\end{example}

\begin{example}[sulfur trioxide]\label{ex:sulfur-trioxide}
The symmetry group of the planar \ce{SO3} molecule is \(\G=\D 3\) \cite{Meyer_centrifugally_1991,Maki_determination_2009}, the symmetry group of an equilateral triangle and the permutation group of the triangle's three vertices.
This group is generated by two elements, the cyclic permutation \((\perm{123})\), which maps vertex \(\perm 1\to\perm 2\), \(\perm 2\to \perm 3\), and \(\perm 3\to \perm 1\), and the swap permutation \((\perm{23})\), which swaps vertices \(\perm 2\) and \(\perm 3\).
The permutation resulting from applying both of these generating permutations depends on the order in which the two are applied, making \(\D 3\) the smallest \textit{non-Abelian} symmetry group.

Symmetry-group rotations leave \ce{SO3} invariant while also permuting its three indistinguishable oxygen nuclei.
These nuclei can be thought of as the three vertices acted on by the group's permutation representation.
Since the nuclei are spinless, we can treat the molecule as rotationally symmetric and project to the group's trivial irrep \(\a_1\).

We place the molecule in the \(\x\y\)-plane such that the cyclic nuclear permutation is performed by a \(\z\)-axis rotation by \(2\pi/3\).
We further orient the molecule within the \(\x\y\)-plane such that the swap of the 2nd and 3rd nuclei is performed by a \(\y\)-axis rotation by \(\pi\) [see Fig.~\ref{fA1:symmetric}(d)].
Using the axis-angle representation of rotations \cite{chirikjian_engineering_2000}, the two generating elements of the group correspond to \((\z,2\pi/3)\) and \((\y,\pi)\), respectively.

The actions of the two generating rotations are described in Eqs.~\eqref{eq:abelian-decomp} and \eqref{eq:dihedral-reflection}, respectively.
The asymmetric molecular states need to be superposed to yield states that transform according to the trivial irrep, yielding symmetric rotational states (cf. Wang functions \cite{wormer_rigid_nodate,Watson_aspects_1977,Papousek_molecular_1982})
\begin{equation}\label{eq:sulfur-trioxide-states}
    |_{m\k}^{\ell}\ket={\textstyle \frac{1}{\sqrt{2}}}\big(\left|_{m}^{\ell},\om=3(\k-1)\right\rangle -
    \left(-1\right)^{\ell+\k}\left|_{m}^{\ell},\om=3(1-\k)\right\rangle \big)~,
\end{equation}
where \(1\leq \k\leq \mult(\ell)\), and where the multiplicity is listed in \cite[Table 23.10]{altmann_point-group_1994}.
One can verify that \(D^{\ell\star}(\z,2\pi/3)|^\ell_{m\k}\ket=D^{\ell\star}(\y,\pi)|^\ell_{m\k}\ket=|^\ell_{m\k}\ket\), as desired.
\end{example}

\begin{example}[C$_{60}$ fullerene]\label{ex:fullerene}
The proper rotational symmetry group of the fullerene \cite{Changala_rovibrational_2019,liu_collision-induced_2022,liu_ergodicity_2023} is the icosahedral group, \(\G=\I\). Carbon-12 has no nuclear spin, so the molecule is rotationally symmetric.

The trivial irrep of \(\I\), denoted by \(\a\), is present only for certain angular momenta.
For example, the smallest nonzero momentum which houses this irrep is \(\ell=6\), and the set of momenta that contain a trivial irrep, along with their multiplicities, are tabulated in \cite[Table 74.10]{ceulemans_group_2013}.
Basis states \(\{|\k\ket\}\) are not proportional to the \(\z\)-axis basis states \(|\om\ket\), and their explicit expressions \cite{qiu_icosahedral_2002} are not particularly illuminating.
\end{example}

\begin{figure}[!t]
\centering{}
\includegraphics[width=1.0\columnwidth]{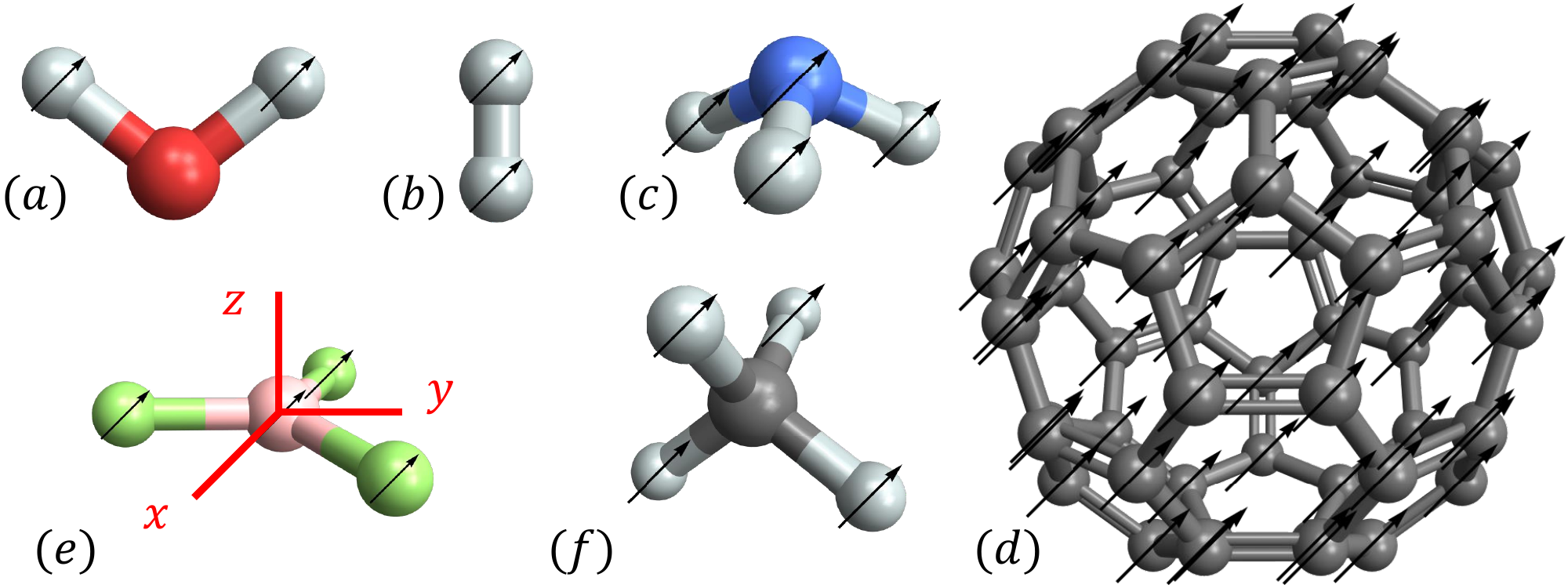}
\caption{\label{fA2:perrotation-symmetric}
Perrotationally symmetric molecules (a) \ce{H2O}, (b) \ce{D2}, (c) \ce{NH3}, (d) \ce{$^{13}$C60}, (e) \ce{BF3}, and (f) \ce{CH4} from Examples \ref{ex:water-singlet}-\ref{ex:fullerene-h}.
Proper rotations in a symmetry group \(\G\) leave the positions of these molecules invariant but can permute identical spinful nuclei.
}
\end{figure}
\subsection{Perrotationally symmetric molecules}\label{sec:spin}

Some symmetry transformations leave a molecule invariant while also permuting a set of indistinguishable molecular nuclei of nonzero spin \cite{Wilson_statistical_1935,Wilson_symmetry_1935,longuet-higgins_symmetry_1963,hougen_interpretation_1971}.
A simple example is water, which admits a proper rotation that exchanges the two spin-half hydrogen atoms while leaving the oxygen atom intact.

In order to enforce the required nuclear-spin statistics,
the definition (\ref{eq:passive-rot}) of molecule-frame rotations can be modified for the case of molecule-based symmetry-group rotations.
Each symmetry-group element corresponds to a rotation \(\lr\) \eqref{eq:passive-rot} tensored with a unitary matrix, \(\Perm\), that permutes the tensor factors in the nuclear spin space according to how the rotation permutes the nuclei.
We denote the full \textit{perrotation} operation \cite{brester_Kristallsymmetrie_1923,wigner_uber_1930,gilles_internal_1972}\cite[Eq.~(2.25)]{ezra_symmetry_1982} by a different overhead arrow,
\begin{equation}
\lpr_{\rg}=\lr_{\rg}\ot\Perm(\rg)~,\quad\forall\rg\in\G~.\label{eq:perrotations}
\end{equation}
Lab-based symmetry-group rotations, \(\rr\), are not appended with nuclear-spin permutations as the nuclei are merely moved from one position to another.

Not all molecule-frame symmetries permute identical spinful nuclei.
Those that do not have an identity permutation acting on the spin factor, and are still included in the symmetry group \(\G\) because they are necessary for enforcing rigid-body symmetry constraints.

Application of rotations \(\rg\) that also permute nuclei needs to yield a \(\pm1\) factor determined by the bosonic or fermionic spin statistics of permuted nuclei.
This sign is realized by a particular one-dimensional \(\G\)-irrep, which we call \(\s\).
Taken together, symmetry and spin-statistics require that
\begin{equation}
\lpr_{\rg}|\psi_{\mol}\ket=\s(\rg)|\psi_{\mol}\ket~,\quad\quad\forall\rg\in\G\,.\label{eq:symm-restriction-1}
\end{equation}
The rotational state space of a symmetric molecule with nontrivial nuclear permutations is the subspace of the space defined by \(\id_\mol\ot \id_{\nat}\) \eqref{eq:identity} that satisfies the restriction from Eq.~\eqref{eq:symm-restriction-1}.

In cases where no spinful nuclei are permuted by any symmetry-group rotations, the above reduces to the symmetry-only constraint in Eq.~\eqref{eq:symm-restriction}, for which \(\s\) is the trivial irrep.
In the general case, our task is to find all states in the decomposition of $\id_{\mol}\ot \id_{\nat}$ that transform as the irrep \(\s\).
We do so by decomposing perrotations \eqref{eq:perrotations} into irreps of $\G$ and keeping any copies of \(\s\).

The permutation piece ``\(\Perm\)'' of the perrotations decomposes in the same way as each rotation matrix \(D^\ell\) does in Eq.~\eqref{eq:isotypic-wigner-1}.
We index the set of distinct \(\G\)-irreps present in the ``\(\Perm\)'' representation of \(\G\) by \(\Lambda\), with \(\ml \Lambda\) denoting the multiplicity of each irrep,
\begin{equation}
\Perm(\rg)=\bigoplus_{\Lambda}\Lambda(\rg)\ot \id_{\ml\Lambda}\,.
\end{equation}
The nuclear-spin space is independent of angular momentum, so all above parameters are \(\ell\)-independent.

Plugging both the above \(\Perm\)-matrix decomposition and the $\ir$-decomposition of the rotation \(\lr\) from Eq.~\eqref{eq:rot-sym} into the perrotation \eqref{eq:perrotations} yields
\begin{equation}
\lpr_{\rg}=\bigoplus_{\ell\geq0}\id_{2\ell+1}\ot\bigoplus_{\ir\uparrow\ell}\bigoplus_{\Lambda}[\ir\ot\Lambda](\rg)\ot \id_{\ml_{\ell}\ir}\ot \id_{\ml\Lambda}\,,\label{eq:rot-nuc-decomp}
\end{equation}
where we have collected the rotational and nuclear irrep spaces into
the collective representation $[\ir\ot\Lambda](\rg)=\ir(\rg)\ot\Lambda(\rg)$.
The joint symmetry and spin statistics restriction from Eq.~(\ref{eq:symm-restriction-1}) implies that we need to further decompose the tensor-product representation $\ir\otimes\Lambda$ into irreps and keep any copies of $\s$ in that decomposition.

There are multiple ways to obtain $\s$, each one falling out from the tensor product of a particular $\ir$ and $\Lambda$.
Each distinct pair \((\ir=\l,\Lambda=\t)\) of admissible irreps yields a subspace of rotational states called a \textit{nuclear spin isomer}, or isomer for short.
All admissible combinations can be extracted from direct-product tables \cite[Appx. E]{ceulemans_group_2013} and are tabulated in Table \ref{tab:isomer-table}.

Given a symmetry group, there turn out to be only two possible values for \(\s\) for each group, either trivial or not.
Additionally, fixing \(\s\) and \(\l\) uniquely determines \(\t\).
As such, each isomer \(\sp\) in Table \ref{tab:isomer-table} can be unambiguously and succinctly denoted by its rotational irrep only, either as \(\sp = \l\) or as \(\sp=\l^{\ast}\).
The non-primed labels are for isomers for which \(\s\) is trivial, while the labels with an asterisk \vva{mark} a nontrivial \(\s\).

From now on, we focus on a single isomer, noting that the full set of states of a given molecule corresponds to the span of all of the molecule's isomers.

For each isomer $\sp$, corresponding to the irrep triple $[\l\ot\t]\downarrow\s$, the dimensions of \(\l\) and \(\t\) turn out to be always the same.
We define the following shorthand,
\begin{subequations}\label{eq:dims}
    \begin{align}
    \dm & \equiv\dim\l=\dim\t\\
    \mult(\ell) & \equiv\ml_{\ell}\l\\
    \mst & \equiv\ml\t\,.
    \end{align}
\end{subequations}

The identity on the state space of \(\sp\) decomposes in a similar way as the identity for a symmetric molecule in Eq.~\eqref{eq:id-mom-symmetric} in terms of momentum states \(|^\ell_{m\k}\ket\),
\begin{equation}
\textnormal{$\id$}_{\mol}^{\G,\sp}=\sum_{\ell\downarrow\l}\sum_{|m|\leq\ell}\sum_{\k=1}^{\mult(\ell)}|_{m\k}^{\ell}\ket\bra_{m\k}^{\ell}|\ot\sum_{\chi=1}^{\mst}|\chi\ket\bra\chi|~.\label{eq:id-mom-rotnuc}
\end{equation}
The indices $\ell$ and $m$ correspond to the standard lab-frame total angular momentum and its $\z$-axis
projection, respectively.
The first sum is only over those angular momenta which contain \(\l\) in the decomposition of the rotation matrices \(D^\ell\) via Eq.~\eqref{eq:isotypic-wigner-1}; we denote this set by \(\ell\down\l\).
The third index $\k$ labels the multiplicity space of $\l$ for each $\ell$.
The fourth $\chi$ index labels the multiplicity subspace of $\t$ inside the nuclear-spin space, as only states in this subspace are allowed to accompany the isomer's rotational states due to the spin-statistics requirement.
The multiplicity \(\mst\) is known as the statistical weight \cite{biedenharn_angular_2010}.

The key difference from the symmetric case is that the definition, and corresponding degree of rotation-spin entanglement, of each basis state $|_{m\k}^{\ell}\ket$ depends heavily on the dimension of the \(\l\) irrep, specifically, on whether $\dm=1$ or $\dm > 1$.
\begin{table*}[!htb]
    \begin{minipage}{.5\linewidth}
      \centering
        \begin{tabular}{cccccc}
        \toprule
        ~~~symmetry~~~ & ~~isomer \(\sp\)~~ & ~$\l$~ & ~$\t$~ & ~$\s$~ & $~~~\dm~~~$\tabularnewline
        \midrule
        \multirow{7}{*}{$\C{2N}$} & $\a$ & $\a$ & $\a$ & $\a$ & $1$\tabularnewline
         & $\b$ & $\b$ & $\b$ & $\a$ & $1$\tabularnewline
         & ${}^1\e_{i}$ & ${}^1\e_{i}$ & ${}^2\e_{i}$ & $\a$ & $1$\tabularnewline
          & ${}^2\e_{i}$ & ${}^2\e_{i}$ & ${}^1\e_{i}$ & $\a$ & $1$\tabularnewline
        \cmidrule{2-6} \cmidrule{3-6} \cmidrule{4-6} \cmidrule{5-6} \cmidrule{6-6}
         & $\phantom{{}^j}\a^{\ast}$ & $\a$ & $\b$ & $\b$ & $1$\tabularnewline
         & $\phantom{{}^j}\b^{\ast}$ & $\b$ & $\a$ & $\b$ & $1$\tabularnewline
         & ${}^j\e_{i}^{\ast}$ & ${}^j\e_{i}$ & ${}^j\e_{N-i}$ & $\b$ & $1$\tabularnewline
        \midrule
        \multirow{3}{*}{$\C{2N+1}$} & $\a$ & $\a$ & $\a$ & $\a$ & $1$\tabularnewline
         & ${}^1\e_{i}$ & ${}^1\e_{i}$ & ${}^2\e_{i}$ & $\a$ & $1$\tabularnewline
         & ${}^2\e_{i}$ & ${}^2\e_{i}$ & ${}^1\e_{i}$ & $\a$ & $1$\tabularnewline
        \midrule
        \multirow{4}{*}{$\T$} & $\a$ & $\a$ & $\a$ & $\a$ & $1$\tabularnewline
        & ${}^1 \e\phantom{_{i}}$ & ${}^1 \e\phantom{_{i}}$ & ${}^2 \e\phantom{_{i}}$ & $\a$ & $1$\tabularnewline
        & ${}^2 \e\phantom{_{i}}$ & ${}^2 \e\phantom{_{i}}$ & ${}^1 \e\phantom{_{i}}$ & $\a$ & $1$\tabularnewline
         & $\ti$ & $\ti$ & $\ti$ & $\a$ & $3$\tabularnewline
        \midrule
        \multirow{6}{*}{$\O$} & $\a_{i}$ & $\a_{i}$ & $\a_{i}$ & $\a_{1}$ & $1$\tabularnewline
         & $\e$ & $\e$ & $\e$ & $\a_{1}$ & $2$\tabularnewline
         & $\ti_{i}$ & $\ti_{i}$ & $\ti_{i}$ & $\a_{1}$ & $3$\tabularnewline
        \cmidrule{2-6} \cmidrule{3-6} \cmidrule{4-6} \cmidrule{5-6} \cmidrule{6-6}
         & $\a_{i}^{\ast}$ & $\a_{i}$ & $\a_{3-i}$ & $\a_{2}$ & $1$\tabularnewline
         & $\e^{\ast}$ & $\e$ & $\e$ & $\a_{2}$ & $2$\tabularnewline
         & $\ti_{i}^{\ast}$ & $\ti_{i}$ & $\ti_{3-i}$ & $\a_{2}$ & $3$\tabularnewline
        \midrule
        \multirow{4}{*}{$\I$} & $\a$ & $\a$ & $\a$ & $\a$ & 1\tabularnewline
         & $\ti_{i}$ & $\ti_{i}$ & $\ti_{i}$ & $\a$ & $3$\tabularnewline
         & $\g$ & $\g$ & $\g$ & $\a$ & $4$\tabularnewline
         & $\h$ & $\h$ & $\h$ & $\a$ & $5$\tabularnewline
        \bottomrule
        \end{tabular}
    \end{minipage}%
    \begin{minipage}{.5\linewidth}
      \centering
        \begin{tabular}{cccccc}
        \toprule
        ~~~symmetry~~~ & ~~isomer \(\sp\)~~ & ~$\l$~ & ~$\t$~ & ~$\s$~ & $~~~\dm~~~$\tabularnewline
        \midrule
        \multirow{2}{*}{$\D 2$} & $\a$ & $\a$ & $\a$ & $\a$ & $1$\tabularnewline
         & $\b_{i}$ & $\b_{i}$ & $\b_{i}$ & $\a$ & $1$\tabularnewline
        \midrule
        \multirow{6}{*}{$\D{4N}$} & $\a_{i}$ & $\a_{i}$ & $\a_{i}$ & $\a_{1}$ & $1$\tabularnewline
         & $\b_{i}$ & $\b_{i}$ & $\b_{i}$ & $\a_{1}$ & $1$\tabularnewline
         & $\e_{i}$ & $\e_{i}$ & $\e_{i}$ & $\a_{1}$ & $2$\tabularnewline
        \cmidrule{2-6} \cmidrule{3-6} \cmidrule{4-6} \cmidrule{5-6} \cmidrule{6-6}
         & $\a_{i}^{\ast}$ & $\a_{i}$ & $\b_{3-i}$ & $\b_{2}$ & $1$\tabularnewline
         & $\b_{i}^{\ast}$ & $\b_{i}$ & $\a_{3-i}$ & $\b_{2}$ & $1$\tabularnewline
         & $\e_{i}^{\ast}$ & $\e_{i}$ & $\e_{2N-i}$ & $\b_{2}$ & $2$\tabularnewline
        \midrule
        \multirow{2}{*}{$\D{4N+1}$} & $\a_{i}$ & $\a_{i}$ & $\a_{i}$ & $\a_{1}$ & $1$\tabularnewline
         & $\e_{i}$ & $\e_{i}$ & $\e_{i}$ & $\a_{1}$ & $2$\tabularnewline
        \midrule
        \multirow{6}{*}{$\D{4N+2}$} & $\a_{i}$ & $\a_{i}$ & $\a_{i}$ & $\a_{1}$ & $1$\tabularnewline
         & $\b_{i}$ & $\b_{i}$ & $\b_{i}$ & $\a_{1}$ & $1$\tabularnewline
         & $\e_{i}$ & $\e_{i}$ & $\e_{i}$ & $\a_{1}$ & $2$\tabularnewline
        \cmidrule{2-6} \cmidrule{3-6} \cmidrule{4-6} \cmidrule{5-6} \cmidrule{6-6}
         & $\a_{i}^{\ast}$ & $\a_{i}$ & $\b_{i}$ & $\b_{1}$ & $1$\tabularnewline
         & $\b_{i}^{\ast}$ & $\b_{i}$ & $\a_{i}$ & $\b_{1}$ & $1$\tabularnewline
         & $\e_{i}^{\ast}$ & $\e_{i}$ & $\e_{2N+1-i}$ & $\b_{1}$ & $2$\tabularnewline
        \midrule
        \multirow{4}{*}{$\D{4N+3}$} & $\a_{i}$ & $\a_{i}$ & $\a_{i}$ & $\a_{1}$ & $1$\tabularnewline
         & $\e_{i}$ & $\e_{i}$ & $\e_{i}$ & $\a_{1}$ & $2$\tabularnewline
        \cmidrule{2-6} \cmidrule{3-6} \cmidrule{4-6} \cmidrule{5-6} \cmidrule{6-6}
         & $\a_{i}^{\ast}$ & $\a_{i}$ & $\a_{3-i}$ & $\a_{2}$ & $1$\tabularnewline
         & $\e_{i}^{\ast}$ & $\e_{i}$ & $\e_{i}$ & $\a_{2}$ & $2$\tabularnewline
        \midrule
        \multirow{3}{*}{$\D{\infty}$} & $\a_{i}$ & $\a_{i}$ & $\a_{i}$ & $\a_{1}$ & $1$\tabularnewline
        \cmidrule{2-6} \cmidrule{3-6} \cmidrule{4-6} \cmidrule{5-6} \cmidrule{6-6}
         & $\a_{1}^{\ast}$ & $\a_{1}$ & $\a_{2}$ & $\a_{2}$ & $1$\tabularnewline
         & $\a_{2}^{\ast}$ & $\a_{2}$ & $\a_{1}$ & $\a_{2}$ & $1$\tabularnewline
        \bottomrule
        \end{tabular}
    \end{minipage}
    \caption{Table of all possible isomers of perrotationally symmetric nuclear spin isomers.
    Given a symmetry group, the isomer corresponding to the triple \(\l\ot\t\down\s\) is unambiguously labeled by its rotational irrep (see Tab.~\ref{tab:groups}) for trivial \(\s\), with an ``\(\ast\)'' present in the superscript in case \(\s\) is non-trivial.
    The irrep dimension, $\dm\equiv\dim\l$, serves as the Schmidt rank, which quantifies the amount of entanglement of the isomer's rotational basis states.
    \label{tab:isomer-table}
    }
\end{table*}
\subsubsection{$\protect\dm=1$ separable isomers\label{subsec:separable-isomer}}

In this case, both irreps $\l$ and $\t$ are one-dimensional, and their tensor product directly
yields the correct spin statistics, $\l\otimes\t=\s$.
To identify the state space, we can separately project the rotational and nuclear-spin space onto subspaces transforming as the two respective irreps.

Utilizing the Kronecker \(\d\)-function from Eq.~\eqref{eq:id-precursor} for each space, we obtain
\begin{equation}\label{eq:id-precursor-perrotational}
\id_{\mol}^{\G,\sp}=\bigoplus_{\ell\geq0}\id_{2\ell+1}\ot\bigoplus_{\ir\uparrow\ell}\bigoplus_{\Lambda}\d_{\ir,\l}\id_{\mult(\ell)}\ot\d_{\Lambda,\t}\id_{\mst}\,.
\end{equation}
Since both irreps are one-dimensional, the two factors \(\ir\ot\Lambda\) in Eq.~\eqref{eq:rot-nuc-decomp} reduce to a scalar, leaving three remaining factors --- the lab-frame \(\z\)-axis projection factor and the two multiplicity spaces --- and thereby confirming Eq.~\eqref{eq:id-mom-rotnuc}.

In correspondence with the three factors in Eq.~\eqref{eq:id-precursor-perrotational} and generalizing Eq.~\eqref{eq:basis-separable}, the basis states are then
\begin{equation}\label{eq:separable-state}
|_{m\k}^{\ell}\ket_{\rot}|\chi\ket_{\nat}\,,
\end{equation}
where we explicitly split each basis state into a factor coming from the rotational space and a factor coming from the nuclear space.
Each basis state is expressible in tensor-product form, so each basis state is separable w.r.t.\ the rotation-spin decomposition.
Possible entangled states include superpositions of such states for a given isomer or of states belonging to different isomers.

\begin{example}[water, \(\a^{\ast}\) para isomer]\label{ex:water-singlet}
Water is a symmetric molecule with proper-rotation symmetry group \(\G=\C 2\).
Its only element corresponds to a \(\pi\)-rotation around some axis \(\v\) such that the two hydrogen atoms and the oxygen are left intact.
This symmetry group has two irreps, the trivial irrep \(\a\) and the sign irrep \(\b\), for which the rotation is represented by \(-1\).

The molecule's sole perrotation from Eq.~\eqref{eq:perrotations} with a non-identity permutation component is$^{\ref{fn:notation-abuse}}$
\begin{equation}
   \lpr_{\v,\pi}=\bigoplus_{\ell\geq0}\id_{2\ell+1}\ot D^{\ell\star}(\v,\pi)\ot \nat(\perm{12})\,,\label{eq:water-perrotation}
\end{equation}
where \(\Perm(\v,\pi)=\nat(\perm{12})\) denotes the swap operation acting on the two nuclear-spin factors.

Since hydrogen nuclei are spin-\(1/2\), i.e., fermionic, the nuclear exchange yields a \(-1\) sign.
Admissible molecular states should satisfy Eq.~\eqref{eq:symm-restriction-1} with \(\s = \b\), the sign irrep of the symmetry group.
According to Table \ref{tab:isomer-table}, one way to achieve this is to pick \(\l=\a\) and \(\t=\b\), corresponding to the \(\a^{\ast}\) isomer of water.

Since the rotational side transforms as the trivial irrep, identifying the basis \(\{|\k\ket\}\) is done in the same way as for symmetric molecules in Sec.~\ref{sec:symmetry}.
We align the molecule so that the symmetry-group rotation is around the \(\z\)-axis, which allows us to apply Eq.~\eqref{eq:CN-basis} for \(N=2\).
In other words, the \(\C 2\)-symmetric rotational states are the subset of the asymmetric states \(\{|_{m}^{\ell},\om\ket\}\) for even \(\om\), with multiplicity \(\mult(\ell) = 2\lfloor\ell/2\rfloor+1\).

The nuclear-spin states \(\{|\chi\ket\}\) that are paired up with the above rotational states span the multiplicity space of the \(\t=\b\) irrep.
The \(-1\) factor upon exchange comes from this irrep, meaning that this space consists of all anti-symmetric states.

Each hydrogen nucleus is spanned by spin up and down states, \(\left|\up\right\ket\) and \(\left|\down\right\ket\), respectively.
The composite four-dimensional space admits one anti-symmetric ``singlet'' state, \((\left|\up\down\right\ket-\left|\down\up\right\ket)/\sqrt{2}\), and there is no need for the index \(\chi\) since nuclear-spin \vva{multiplicity} \(\mst=1\).
Isomers whose nuclei are in a singlet state are called para isomer.
The singlet state is tensored with an \vva{arbitrary} nuclear-spin state of the oxygen.

The fact that only some nuclear-spin states pair up with the trivial-irrep rotational states highlights a difference from the symmetric rigid-body case from Sec.~\ref{sec:symmetry}.
In that case, the overall irrep \(\s\) is also the trivial irrep, but \textit{all} nuclear states are allowed since symmetry rotations do not permute identical spinful nuclei.
\end{example}

\begin{example}[water, \(\b^{\ast}\) ortho isomer]\label{ex:triplet-water}
According to Table \ref{tab:isomer-table}, there is a second way to obtain the \(\s=\b\) irrep for water, namely, by picking \(\l=\b\) and \(\t=\a\).
This choice corresponds to the \(\b^{\ast}\) isomer of water.

For this isomer, the rotational side transforms as the non-trivial irrep \(\l=\b\).
The \(-1\) factor upon exchange comes from this irrep.
Observing Eq.~\eqref{eq:abelian-decomp}, this irrep occurs in all asymmetric states \(\{|_{m}^{\ell},\om\ket\}\) for which \(\om\) is odd.
The lowest angular momentum of this isomer is thus \(\ell = 1\).

The hydrogen nuclear-spin states \(\{|\chi\ket\}\) that are paired up with the above rotational states span the multiplicity space of \(\t=\a\), i.e., the trivial irrep.
This space, with multiplicity \(\mst=3\), consists of all nuclear-spin triplet states and is spanned by \(\left|\chi=1\right\ket\equiv\left|\up\up\right\ket\), \(\left|\chi=2\right\ket\equiv\left|\down\down\right\ket\), and \(\left|\chi=3\right\ket\equiv(\left|\up\down\right\ket+\left|\down\up\right\ket)/\sqrt{2}\).
Isomers whose nuclei are in triplet states are called ortho isomers.
The triplet hydrogen-nuclei states are tensored with an \vva{arbitrary} nuclear-spin state of the oxygen.

\end{example}

\begin{example}[deuterated hydrogen, \(\a_{1}\) ortho isomer]\label{ex:deuterium-a1}
A simple isomer with bosonic nuclei is D$_2$, where ``D'' stands for deuterium.
This is a symmetric molecule with the same symmetry group as disulfur from Example~\ref{ex:disulfur}, \(\G=\D\infty\).
However, the equatorial-axis rotations exchange the two spin-\(1\) deuterium nuclei, resulting in a \(+1\) spin-statistics factor that is realized by the trivial irrep.
Admissible molecular states should satisfy Eq.~\eqref{eq:symm-restriction-1} with \(\s = \a_{1}\).
According to Table \ref{tab:isomer-table}, one way to achieve this is to pick \(\l=\t=\a_{1}\), corresponding to the \(\a_{1}\) isomer of D$_2$.

Since the rotational side transforms as the trivial irrep, identifying the basis \(\{|\k\ket\}\) is done in the same way as for symmetric molecules in Sec.~\ref{sec:symmetry}.
Per Example~\ref{ex:disulfur},
asymmetric states \(\{|_{m}^{\ell},\om=0\ket\}\) for even \(\ell\) transform as the trivial irrep of \(\D\infty\).

The nuclear-spin states also transform as the trivial irrep, and so are in the same triplet subspace as in Example~\ref{ex:triplet-water}, with multiplicity \(\mst = 3\).
\end{example}

\begin{example}[deuterated hydrogen, \(\a_{2}\) para isomer]\label{ex:deuterium-a2}
The D$_2$ molecule admits another para isomer --- the \(\a_{2}\) isomer from Table \ref{tab:isomer-table} --- which corresponds to picking the 1D irreps \(\l=\t=\a_{2}\).
A basis for this isomer consists of tensor products of rotational states \(|^\ell_m,\om=0\ket\) with \(\ell\) \textit{odd} and singlet nuclear states (with multiplicity \(\mst = 1\)).

None of the two-dimensional \(\D\infty\)-irreps feature in either the ortho or para isomer of this molecule.
This trend extends to all \(\D\infty\)-symmetric molecules, including ordinary hydrogen, because no combination of two-dimensional irreps yields the irreps \(\a_{1,2}\) that describe the spin statistics.

In the case of ordinary \ce{H2}, the spin statistics are fermionic (\(\s=\a_2\)), and there are two isomers --- \(\a_1^\ast\) and \(\a_2^\ast\).
These admit the same two respective sets of rotational states as the \(\a_1\) and \(\a_2\) deuterium isomer.
However, the ortho-para jargon is switched: the \(\a_1^\ast\) admits a nuclear-spin singlet and so is called para-hydrogen, while the \(\a_2^\ast\) hydrogen isomer is ortho.
\end{example}

\begin{example}[ammonia, \(^2 \e\) isomer]\label{ex:ammonia}
The \ce{NH3} molecule is not planar, admitting the cyclic \(\C 3\) symmetry.
We ignore any tunneling effects for clarity \cite{bunker_molecular_1998},
noting that this example applies to any linear molecule that has been appended with three hydrogens in a \(\C3\) symmetric way, e.g., calcium monomethoxide (\ce{CaOCH3})  \cite{Mitra_direct_2020}.

Rotations in this group are by angles \(0\), \(2\pi/3\), and \(4\pi/3\), which wind up cyclically permuting the three hydrogen atoms.
This group has three irreps, the trivial one \(\a\), and two irreps \(^{j=1,2}\e\) that are complex conjugate to each other and for which \(^j\e(2\pi/3)=\exp(i\frac{2\pi}{3}j)\).

A cyclic permutation of nuclei is equivalent to performing two exchanges, yielding a \(+1\) spin-statistics factor.
This corresponds to the trivial irrep \(\s=\a\).
According to Table \ref{tab:isomer-table}, one way to achieve this is to pick \(\l={}^2\e\) and \(\t={}^1\e\), corresponding to the \({}^2\e\) isomer.

Aligning the molecule such that symmetry-group rotations are around the \(\z\) axis and observing Eq.~\eqref{eq:abelian-decomp}, we see that the \(|\om\ket\) states of this isomer are all those for which \(\om\equiv 2\) modulo 3.
The first momentum that harbors such states is \(\ell=1\).
More generally, for \(\C N\)-symmetric molecules, the momenta of the \(^2 \e_i\) isomer satisfy \(\ell\geq i\).
\end{example}

\begin{example}[boron trifluoride, \(\a_2^{\ast}\) isomer]\label{ex:BF3-separable}
The \ce{BF3} molecule has the same dihedral (\(\D 3\)) symmetry \cite{Kuchitsu_structure_1998} as sulfur trioxide from Example~\ref{ex:sulfur-trioxide}.
Symmetry-group rotations leave both molecules invariant while also permuting their three indistinguishable nuclei.
In contrast to \ce{SO3}, the fluorine nuclei of \ce{BF3} are spinful, making the latter molecule perrotationally symmetric.

The \(\D 3\) group --- the permutation group of three objects --- has three irreps.
The two one-dimensional irreps are the trivial irrep \(\a_1\) and the ``sign'' irrep \(\a_2\), in which all two-object swaps are mapped to \(-1\) while remaining elements are mapped to \(+1\).
The remaining two-dimensional irrep is called \(\e\), for which we let \(\sigma = e^{-i\frac{2\pi}{3}}\) and pick a basis \(\{|\nu=1\ket,|\nu=2\ket\}\) such that
\begin{equation}\label{eq:D3-2D-irrep}
    \e(\perm{123})=\begin{pmatrix}\sigma & 0\\
0 & \sigma^{\star}
\end{pmatrix}\quad\text{and}\quad\e(\perm{23})=\begin{pmatrix}0 & 1\\
1 & 0
\end{pmatrix}\,.
\end{equation}

The spin statistics of the spin-half \vva{fluorine} nuclei of \ce{BF3} require a \(-1\) to be produced every time a symmetry-group rotation swaps any two nuclei.
This corresponds to the sign \(\D 3\)-irrep, meaning that \(\s=\a_2\).
According to Table \ref{tab:isomer-table}, one way to achieve this is to pick \(\l=\a_2\) and \(\t=\a_1\).
This corresponds to the \(\a_2^{\ast}\) isomer of \ce{BF3}.

We orient the molecule within the \(\x\y\)-plane in the same way as we did with \ce{SO3}.
The cyclic nuclear permutation \((\perm{123})\) is performed by a \(\z\)-axis rotation by \(2\pi/3\), with corresponding perrotation$^{\ref{fn:notation-abuse}}$
\begin{equation}
   \lpr_{\z,2\pi/3}=\bigoplus_{\ell\geq0}\id_{2\ell+1}\ot D^{\ell\star}(\z,2\pi/3)\ot \nat(\perm{123})\,,\label{eq:D3-perrotation}
\end{equation}
where \(\Perm(\z,2\pi/3)=\nat(\perm{123})\) denotes the matrix that cyclically permutes the three nuclear-spin tensor factors.
The swap \((\perm{23})\) of the 2nd and 3rd nuclei is performed by a \(\y\)-axis rotation by \(\pi\) [see Fig.~\ref{fA2:perrotation-symmetric}(d)].

The rotations corresponding to the two generating permutations, \((\perm{123})\) and \((\perm{23})\), act on the asymmetric rotational states \eqref{eq:id-mom-asymmetric} as
\begin{subequations}\label{eq:wigners}
\begin{align}
    D^{\ell\star}(\z,2\pi/3)|_{m}^{\ell},\om\ket &=\sigma^{\om}|_{m}^{\ell},\om\ket \\
    D^{\ell\star}(\y,\pi)|_{m}^{\ell},\om\ket	   &=\left(-1\right)^{\ell+\om}|_{m}^{\ell},-\om\ket~,
\end{align}
\end{subequations}
respectively.
The asymmetric molecular states \(|\om\ket\) need to be superposed to yield states that transform according to the sign irrep [cf.\ the trivial-irrep states in Eq.~\eqref{eq:sulfur-trioxide-states}].
The resulting rotational states of the isomer are
\begin{equation}\label{eq:bf3-states}
    |_{m\k}^{\ell}\ket={\textstyle \frac{1}{\sqrt{2}}}\big(\left|_{m}^{\ell},\om=3(\k-1)\right\rangle +\left(-1\right)^{\ell+\k}\left|_{m}^{\ell},\om=3(1-\k)\right\rangle \big)~,
\end{equation}
where \(1\leq \k\leq \mult(\ell)\), and where the multiplicity is listed in \cite[Table 23.10]{altmann_point-group_1994}.

The \(2^3=8\)-dimensional space of the three fluorine nuclei decomposes into four copies of the trivial irrep \(\a_1\), with the entire admissible subspace spanned by the states \(\left|\up\up\up\right\ket\) and \(\left|\up\down\down\right\ket+\left|\down\up\up\right\ket+\left|\down\down\up\right\ket\), along with their two counterparts obtained by flipping all spins.
The quadruplet fluorine-nuclei states are then tensored with an \vva{arbitrary} nuclear-spin state of the boron to form the isomer's nuclear-spin states.

Interestingly, the nuclear states do not decompose into any copies of the sign irrep \(\a_2\), thereby eliminating the possibility of the \(\a_2=\a_1\ot\a_2\) isomer for this molecule.
This is known as having ``missing levels'' in the spectroscopy literature \cite{Harter_theory_1979,bunker_molecular_1998,bunker_fundamentals_2004}.
Since the \(\a_2\) isomer houses the \(\ell=0\) state, which transforms according to \(\l=\a_1\), symmetry and spin statistics make sure that \ce{BF3} never has zero angular momentum.
\end{example}

\begin{example}[$^{13}$\ce{C60} fullerene, \(\a\) isomer]
Fullerenes made up of carbon-12 have no nuclear spin and can thus be treated as rotationally symmetric molecules with icosahedral ($\G=\I$) symmetry; we study this case in Example~\ref{ex:fullerene}.
On the other hand, isotopic $^{13}$\ce{C60} fullerenes are perrotationally symmetric since carbon-13 has a nuclear spin of \(1/2\).

Each symmetry-group rotation acts in a way that permutes an even subset of the $60$ carbon-13 nuclei, resulting in a \(+1\) spin-statistics factor that is realized by the trivial icosahedral irrep, \(\s=\a\).
According to Table \ref{tab:isomer-table}, one way to achieve this is to pick \(\l=\t=\a\), corresponding to the \(\a\) isomer.

The rotational states transform as the trivial irrep, \(\l=\a\), so identifying the basis \(\{|\k\ket\}\) is done in the same way as in Example~\ref{ex:fullerene}.

The nuclear-spin space is restricted to only states \(\{|\chi\ket\}\) that transform under the trivial irrep.
The full space is of dimension \(2^{60}\), and finding a basis for the states is not the most tractable of tasks.
We can, however, numerically determine the nuclear multiplicity of \(\mst = \numprint{19215358678900736}\) (see Table \ref{tab:buckeyballstatweight}),
confirming earlier studies \cite{Balasubramanian_applications_1985,harter_nuclear_1992,harter_erratum_1992,Balasubramanian_comment_1992,Bunker_spherical_1999}.

The other isomers of $^{13}$\ce{C60} use any of the four other \(\I\)-irreps, \(\{\ti_1,\ti_2,\g,\h\}\), for both \(\l\) and \(\t\).
All such isomers have \(\dm>1\).
\end{example}

\subsubsection{$\protect\dm\protect>1$ entangled isomers}

In general, the resulting tensor-product representation $\ir\ot\Lambda$ from Eq.~\eqref{eq:rot-nuc-decomp} with \(\ir=\l\) and \(\Lambda=\t\) is of dimension $\dm^{2}$, per the notation defined in Eq.~\eqref{eq:dims}.
In cases where the dimension \(\dm>1\), this tensor-product irrep has to be \textit{restricted} to any copies of the one-dimensional $\s$ irrep that yield the correct spin statistics defined in Eq.~\eqref{eq:symm-restriction-1}.
For the \(\su 2\) Lie group, performing this restriction is analogous to coupling angular momentum vectors and then restricting to a particular total angular momentum sector \cite{georgi_lie_2021}.

Let \(\{|\n\ket_{\rot}\}_{\n=1}^{\dm}\) and \(\{|\n\ket_{\nat}\}_{\n=1}^{\dm}\) define bases for the irrep spaces of \(\l\) and \(\t\), respectively.
The task is to find any states in the composite space, spanned by \(\{|\n\ket_{\rot}|\n^\pr\ket_{\nat}\}\), that transform according to the scalar irrep \(\s\).
In all cases (see Table \ref{tab:isomer-table}), there turns out to be only one such state,
which we denote by \(|\s\ket\) and which can be written as
\begin{align}\label{eq:state}
|\s\ket & ={\frac{1}{\sqrt{\dm}}}\sum_{\n=1}^{\dm}s_{\nu}|\n\ket_{\rot}|\nu\ket_{\nat}\,,
\end{align}
whose Clebsch-Gordan coupling coefficients $s_{\nu}=\pm1$ depend on \(\l\), \(\t\), and \(\s\).

When \(\l=\t\) and \(\s\) are trivial, we can prove the above result by writing down the projection onto the above state,
\begin{equation}
    \varPi_{\mol}\equiv |\s\ket\bra\s|~,
\end{equation}
as a sum over symmetry group elements and applying Schur orthogonality \textit{a.k.a.}\ the ``Grand Orthogonality Theorem'' \cite[Sec. 2.4]{arovas_lecture_nodate}.
More generally, the above result can be obtained by consulting Clebsch-Gordan tables \cite[Appx. F]{ceulemans_group_2013}.

Incorporating the above into the decomposition from Eq.~\eqref{eq:rot-nuc-decomp}, we first select the isomer corresponding to \(\ir=\l\) and \(\Lambda=\t\) and then project the tensor-product irrep into the above state,
\begin{equation}
    [\ir\ot\Lambda](\rg)\to\d_{\ir,\l}\d_{\Lambda,\t}\s(\rg)\varPi_{\mol}~.
\end{equation}
Backing out the identity on the resulting space, we have
\begin{align}\label{eq:identity-precursor}
     \!\!\!\id_{\mol}^{\G,\sp} & =\bigoplus_{\ell\geq0}\id_{2\ell+1}\ot\bigoplus_{\ir\uparrow\ell}\bigoplus_{\Lambda}\varPi_{\mol}\ot\d_{\ir,\l}\id_{\mult(\ell)}\ot\d_{\Lambda,\t}\id_{\mst},
\end{align}
where we have split the rotational and nuclear-spin multiplicity factors and plugged in Eqs.~\eqref{eq:dims}.

Re-expressing the above identity factor as Eq.~\eqref{eq:id-mom-rotnuc} yields basis states
\begin{equation}\label{eq:entangled-state}
|_{m\k}^{\ell}\ket|\chi\ket\equiv{\frac{1}{\sqrt{\dm}}}\sum_{\n=1}^{\dm}s_{\nu}|_{m\k}^{\ell},\nu\ket_{\rot}|\nu,\chi\ket_{\nat}\,,
\end{equation}
where we have incorporated the state \(|\s\ket\) \eqref{eq:state} and explicitly split each state in the superposition on the right-hand side into a factor coming from the rotational space and a factor coming from the nuclear space.
This notation is different from that in Eq.~\eqref{eq:basis}, where we explicitly write the fixed entangled state; we continue to absorb it henceforth for notational simplicity.

When \(\dm=1\), the internal \(\nu\) irrep index disappears, and Eq.~\eqref{eq:entangled-state} reduces to the separable case from Eq.~\eqref{eq:separable-state}.
When \(\dm\geq2\), the basis state for each \(\ell\), \(m\), \(\k\), and \(\chi\) is a \textit{completely} rotation-spin entangled state with \(\dm\) components, i.e., Schmidt rank \(\dm\) \cite{Ekert_entangled_1995}.
This entanglement is enforced by the combination of symmetry and spin statistics, manifest in the restriction from Eq.~\eqref{eq:symm-restriction-1}, and is impossible to separate without breaking the symmetry/spin-statistics requirement or transitioning to another isomer.
We can pick superpositions of the above states that yield a different basis with different labels, but there is no way to remove the sum over \(\nu\) in this way because it is its own separate factor.

\begin{example}[boron trifluoride, \(\e^{\ast}\) isomer]\label{ex:BF3}
Recalling Example~\ref{ex:BF3-separable}, we know that the spin statistics of the spin-\(1/2\) \vva{fluorine} nuclei of \ce{BF3} require a \(-1\) to be produced every time a symmetry-group rotation swaps any two nuclei.
This corresponds to the sign \(\D 3\)-irrep, meaning that \(\s=\a_2\).
According to Table \ref{tab:isomer-table}, the way to achieve this using multi-dimensional irreps is to pick \(\l=\t=\e\) and restrict their product to \(\a_2\).
This corresponds to the \(\e^{\ast}\) entangled isomer of \ce{BF3}.

By comparing Eq.~\eqref{eq:wigners} to Eq.~\eqref{eq:D3-2D-irrep}, we observe that a pair of states \(\{|^\ell_m,\pm|\om|\ket\}\) form a basis for the \(\e\) irrep whenever \(|\om|\) is 1 modulo 3.
Taking care of our chosen order of the irrep basis, we also observe that \(\{|^\ell_m,\mp|\om|\ket\}\) form a basis for the \(\e\) irrep whenever \(|\om|\) is 2 modulo 3.
The multiplicity \(\mult(\ell)\) for each angular momentum \(\ell\) is tabulated in \cite[Table 23.10]{altmann_point-group_1994}.

For example, for \(|m|\leq\ell=1\), we have only \(\mult(1)=1\) copy of \(\e\), so there is no need for the multiplicity index \(\k\).
We can then define the following basis,
\begin{subequations}
\begin{align}
    \left|_{m}^{\ell=1},\n=1\right\rangle _{\rot}&	=|_{m}^{\ell=1},\om=1\ket \\
    \left|_{m}^{\ell=1},\n=2\right\rangle _{\rot}&	=|_{m}^{\ell=1},\om=-1\ket~,
\end{align}
\end{subequations}
and verify that the two rotations from Eq.~\eqref{eq:wigners}, when expressed in this basis, reduce to the two matrices from Eq.~\eqref{eq:D3-2D-irrep}.

On the nuclear side, we omit the factor due to the boron spin for simplicity since that state is unrestricted by spin statistics.
The \(2^3=8\) dimensional spin space of the three spin-half \vva{fluorine} nuclei houses two copies of \(\e\), so the multiplicity \(\mst = 2\).
The first copy is spanned by the basis states
\begin{align}\label{eq:spin-irrep}
    \left|\nu=1,\chi=1\right\rangle _{\nat}&=\left(\left|\up\down\down\right\rangle +\sigma^{\star}\left|\down\up\down\right\rangle +\sigma\left|\down\down\up\right\rangle \right)/\sqrt{3}\nonumber\\
    \left|\nu=2,\chi=1\right\rangle _{\nat}&=\left(\left|\up\down\down\right\rangle +\sigma\left|\down\up\down\right\rangle +\sigma^{\star}\left|\down\down\up\right\rangle \right)/\sqrt{3}~,
\end{align}
where one can verify that permuting the nuclei is equivalent to applying a corresponding product of powers of the matrices from Eq.~\eqref{eq:D3-2D-irrep}.
The basis for the second copy, \(\{\left|\nu,\chi=2\right\ket_{\nat}\}_{\nu=1,2}\), can be obtained from Eq.~\eqref{eq:spin-irrep} by flipping all spins.

To construct the state \(|\s=\a_2\ket\) from Eq.~\eqref{eq:state}, we consult Clebsch-Gordan tables for \(\D 3\) \cite[23.11, 3rd table]{altmann_point-group_1994}.
This yields the entangled ``rotation-spin singlet'' state of Schmidt rank \(2\),
\begin{equation}
    |\a_{2}\ket=\left(|\nu=1\ket_{\rot}|\nu=2\ket_{\nat}-|\nu=2\ket_{\rot}|\nu=1\ket_{\nat}\right)/\sqrt{2}~.
\end{equation}
Plugging this into Eq.~\eqref{eq:entangled-state} and using the bases for the ``\(\rot\)'' and ``\(\nat\)'' \(\e\)-irreps defined above yields the basis \(\{|^\ell_{m\k},\chi\ket\}\) for this isomer.
For example,
\begin{equation}
    |_{m}^{1},\chi\ket=\left(\left|_{m}^{1},1\right\rangle _{\rot}\left|2,\chi\right\rangle _{\nat}-\left|_{m}^{1},2\right\rangle _{\rot}\left|1,\chi\right\rangle _{\nat}\right)/\sqrt{2}
\end{equation}
for \(|m|\leq \ell = 1\), \(\chi\in\{1,2\}\), and without a \(\k\) index since the multiplicity \(\mult(\ell=1)=1\).
One can verify that these yield the correct spin statistics and transform as the \(\s=\a_2\) irrep under symmetry-group perrotations, such as the one from Eq.~\eqref{eq:D3-perrotation}.

\end{example}

\begin{example}[methane, \(\ti \) ortho isomer]\label{ex:methane-e}
    Methane \ce{CH4} has tetrahedral symmetry \( \T \) \cite{Herzberg_electronic_1966,van1973spin,berger_classification_1977,hougen2001methane}, with perrotations exchanging the four hydrogen atoms in the same way as the four corners of a tetrahedron are permuted by tetrahedral-group rotations.
    Using \cref{tab:isomer-table}, we see there are four isomers of \( \T \): $\a$, ${}^1 \e\phantom{_{i}}$, ${}^2 \e\phantom{_{i}}$, and $\ti$. Each of these is separable except for $\ti$, which is entangled with $\dm = 3$.

    The rotational state factor is not easy to track analytically, but they have been studied in related work by Harter and Patterson \cite{harter_simple_1977,harter_orbital_1977,harter_bands_1977,Harter_theory_1979,Harter_theory_1981,drake_molecular_2006}.

    The \(2^4=16\) dimensional spin space of the \vva{four} spin-half hydrogen nuclei houses three copies of \(\ti \), so the multiplicity \(\mst = 3\).
    The first copy is spanned by the basis states
    \begin{subequations}
    \begin{align}
        \left|\nu=1,\chi=1\right\rangle _{\nat}&=\vva*{\left(\left| \down \down \up \down \right\rangle + \left| \down \down \down \up \right\rangle - \left| \up \down \down \down \right\rangle - \left| \down \up \down \down \right\rangle \right)/2}     \\
        \left|\nu=2,\chi=1\right\rangle _{\nat}&=\vva*{\left(\left| \down \up \down \down \right\rangle + \left| \down \down \down \up \right\rangle - \left| \up \down \down \down \right\rangle - \left| \down \down \up \down \right\rangle \right)/2}     \\
        \left|\nu=3,\chi=1\right\rangle _{\nat}&=\vva*{\left(\left| \down \up \down \down \right\rangle + \left| \down \down \up \down \right\rangle - \left| \up \down \down \down \right\rangle - \left| \down \down \down \up \right\rangle \right)/2}~.
    \end{align}
    The second copy is spanned by the basis states
    \begin{align}
        \left|\nu=1,\chi=2\right\rangle _{\nat}&=\left(\left| \down \down \up \up \right\rangle - \left| \up \up \down \down \right\rangle \right)/\sqrt{2}     \\
        \left|\nu=2,\chi=2\right\rangle _{\nat}&=\left(\left| \down \up  \down \up \right\rangle - \left| \up \down \up \down \right\rangle \right)/\sqrt{2}     \\
        \left|\nu=3,\chi=2\right\rangle _{\nat}&=\left(\left| \down \up \up \down \right\rangle - \left| \up \down \down \up \right\rangle \right)/\sqrt{2}~.
    \end{align}
    The third copy is spanned by the basis states
    \begin{align}
        \left|\nu=1,\chi=3\right\rangle _{\nat}&=\vva*{\left(\left| \down \up \up \up \right\rangle + \left| \up \down \up \up \right\rangle - \left| \up \up \down \up \right\rangle - \left| \up \up \up \down \right\rangle \right)/2}     \\
        \left|\nu=2,\chi=3\right\rangle _{\nat}&=\vva*{\left(\left| \down \up \up \up \right\rangle + \left| \up \up \down \up \right\rangle - \left| \up \down \up \up \right\rangle - \left| \up \up \up \down \right\rangle \right)/2}     \\
        \left|\nu=3,\chi=3\right\rangle _{\nat}&=\vva*{\left(\left| \down \up \up \up \right\rangle + \left| \up \up \up \down \right\rangle - \left| \up \down \up \up \right\rangle - \left| \up \up \down \up \right\rangle \right)/2}~.
    \end{align}
    \end{subequations}
    Notice that $\chi=i$ corresponds to a superposition of those states with exactly $i$ $\up$-spins and $(4-i)$ $\down$-spins.
    \vva{All three copies are orthonormal and transform under the same \(\ti\) matrices when the nuclei are permuted.}

    As with \ce{BF3}, admissible states include the factor \(|\s\ket\), which consists of the above states summed over \(\m\) for each \(\ell\), \(m\), \(\k\), and \(\chi\).
    Exact expressions are no more illuminating than the general case.
\end{example}

\subsubsection{Relative fraction of entanglement}

A particular perrotationally \(\G\)-symmetric molecule admits a set \(\{\sp\}\) of isomers, each with dimension \(\dm^{\sp}\), rotational multiplicities \(\mult^{\sp}(\ell)\), and statistical weight \(\mst^{\sp}\).
The identity of the entire molecular state space is a sum over the projections onto all of the molecule's isomers,
\begin{equation}
      \id_{\mol}=\sum_{\sp}\id_{\mol}^{\G,\sp}~,
\end{equation}
where each term on the right-hand side is determined by the prescription from earlier in this subsection.

Each isomer's state space is infinite-dimensional.
However, by counting the total number of basis states for the entangled cases (\(\dm^{\sp}>1\)) and dividing by the total number of states, we can determine the \textit{relative entangled-state fraction},
\begin{subequations}\label{eq:fent}
\begin{align}
    \fent&=\vva*{\lim_{\ell_{\max}\to\infty}}\frac{\vva*{\sum_{\ell\leq\ell_{\max}}(2\ell+1)}\sum_{\sp:\dm^{\sp}>1}\mult^{\sp}(\ell)\mst^{\sp}}{\vva*{\sum_{\ell\leq\ell_{\max}}(2\ell+1)}\sum_{\sp}\mult^{\sp}(\ell)\mst^{\sp}}\\&=\frac{\sum_{\sp:\dm^{\sp}>1}\dm^{\sp}\mst^{\sp}}{\sum_{\sp}\dm^{\sp}\mst^{\sp}}~.
\end{align}
\end{subequations}
\vva{The factor \(2\ell+1\) counts the lab-frame projections \(m\) of each momentum \(\ell\).}
The second equality is the result of calculating the same relative fraction in position space (see Sec.~\ref{sec:position}), which allows us to obtain a closed-form expression.
\vva{Without the limit, the first expression gives the relative fraction for the momentum-constrained subspace \(\ell\leq\ell_{\max}\).}

\begin{example}[boron trifluoride]\label{ex:BF3:fraction}
Combining all isomers of \ce{BF3} and calculating the relative entangled-state fraction reveals that exactly \textit{half} of all basis states are entangled.
This value, along with those of other molecules with dihedral perrotation symmetry, is listed in Table \ref{tab:fractions}.

\begin{table}[htp]
    \centering
    \renewcommand{\arraystretch}{1.2}
    \begin{tabular}{cccccc} \toprule
        symmetry & ~~molecule~~ & \multicolumn{4}{c}{entangled-state fraction, $\fent$} \\
        & & $\ell \leq 2$ & $\ell \leq 4$ & $\ell \leq 8$ & $\ell \leq \infty$ \\ \midrule
        \(\D 3\) & $\ce{BF3}$ & \vva*{$0.684$} & \vva*{$0.509$} & \vva*{$0.517$} & $1/2 = 0.5$ \\
         \( \D 4 \) & \ce{XeF4} &
 \vva*{0.348} & \vva*{0.367} & \vva*{0.372} & $3/8 = 0.375$ \\
         \( \D 5 \) & \ce{C5H5-} & \vva*{0.619} & \vva*{0.778} & \vva*{0.742} & $3/4 = 0.75$ \\
         \( \D 6 \) & \ce{C6H6} &
 \vva*{0.723} & \vva*{0.598} & \vva*{0.635} & $5/8 = 0.625$  \\
         \( \D 7\) & \ce{C7H7+} & \vva*{0.780} & \vva*{0.857} & \vva*{0.843} & $27/32 \approx 0.844$ \\
         \( \D 8\) & \ce{C8H8^{2-}} & \vva*{0.742} & \vva*{0.722} & \vva*{0.725} & ~~$93/128 \approx 0.727$~~ \\
         \bottomrule
    \end{tabular}
    \caption{\label{tab:fractions}Fraction of states that are entangled [see Eq.~\eqref{eq:fent}] for planar molecules with dihedral perrotation symmetry.}
\end{table}

\end{example}

\begin{example}[methane]\label{ex:methane-fraction}
    Methane admits four isomers: \(\a\) (meta), \(^1\e\), \(^2\e\) (both para, stemming from the 2D real-valued \(\e\) irrep of the improper group \(\T_d\)), and \(\ti\) (ortho).
    Only the last one is completely entangled.
    We count the entangled-state fraction \eqref{eq:fent} of methane to be \(\fent\approx0.56\), meaning that just over half of methane's rotational states are completely entangled.
\end{example}

\begin{example}[$^{13}$\ce{C60} fullerene]\label{ex:fullerene-h}

The molecule \ce{^{13}C60} has icosahedral $\I$ symmetry. From \cref{tab:isomer-table} we see there are five isomers $\a$, $\ti_1$, $\ti_2$, $\g$, and $\h$. All of these are entangled except $\a$ (which has $\dm = 1$), with Schmidt ranks of $\dm = 3,3,4,5$ respectively.
The isomers $\g$ and $\h$ have the highest degree of entanglement of any isomer because $\I$ is the only group that contains irreps of dimension greater than $3$.

Multiplicities \(\mult(\ell)\) are listed in Table \ref{tab:buckeyballrotmult} for the first few momenta, while statistical weights are listed in Table \ref{tab:buckeyballstatweight}.
Combining these yields \(\fent\approx0.98\), meaning that one is hard-pressed to find a separable state in the entire molecule.

\begin{table}[htp]
    \centering
\begin{tabular}{lccccc}
\toprule
$\ell$~~ & $\a$ & $\ti_{1}$ & $\ti_{2}$ & $\g$ & $\h$\tabularnewline
\midrule
$0$ & $1$ & $0$ & $0$ & $0$ & $0$\tabularnewline
$1$ & $0$ & $1$ & $0$ & $0$ & $0$\tabularnewline
$2$ & $0$ & $0$ & $0$ & $0$ & $1$\tabularnewline
$3$ & $0$ & $0$ & $1$ & $1$ & $0$\tabularnewline
$4$ & $0$ & $0$ & $0$ & $1$ & $1$\tabularnewline
$5$ & $0$ & $1$ & $1$ & $0$ & \vva*{$1$}\tabularnewline
\bottomrule
\end{tabular}~~~~~%
\begin{tabular}{rccccc}
\toprule
$\ell~~$ & $\a$ & $\ti_{1}$ & $\ti_{2}$ & $\g$ & $\h$\tabularnewline
\midrule
$6$~~ & $1$ & $1$ & $0$ & $1$ & $1$\tabularnewline
$7$~~ & $0$ & $1$ & $1$ & $1$ & $1$\tabularnewline
$8$~~ & $0$ & $0$ & $1$ & $1$ & $2$\tabularnewline
$9$~~ & $0$ & $1$ & $1$ & $2$ & $1$\tabularnewline
$10$~~ & $1$ & $1$ & $1$ & $1$ & $2$\tabularnewline
\bottomrule
\end{tabular}
    \caption{Multiplicities $\mult(\ell)$ for each choice of rotational irrep $\l$ for \ce{^{13}C60} from Example~\ref{ex:fullerene-h}.}
    \label{tab:buckeyballrotmult}
\end{table}

\begin{table}[htp]
    \centering
\begin{tabular}{ccc}
\toprule
$\sp_{\phantom{2}}$ & ~~$\dm$~~ & $\mst$\tabularnewline
\midrule
$\a_{\phantom{2}}$ & $1$ & $\numprint{19215358678900736}$\tabularnewline
$\ti_{1}$ & $3$ & $\numprint{57646074961907712}$\tabularnewline
$\ti_{2}$ & $3$ & $\numprint{57646074961907712}$\tabularnewline
$\g_{\phantom{2}}$ & $4$ & $\numprint{76861433640804352}$\tabularnewline
$\h_{\phantom{2}}$ & $5$ & $\numprint{96076792318656512}$\tabularnewline
\bottomrule
\end{tabular}
    \caption{Irrep \(\dm\)imensions and nuclear-spin multiplicities \(\mst\) of the five \(\sp\)pecies of \ce{^{13}C60} \cite{Balasubramanian_applications_1985,harter_nuclear_1992,harter_erratum_1992,Balasubramanian_comment_1992,Bunker_spherical_1999,bunker_molecular_1998}.
    The statistical weights are large because the entire nuclear state space has dimension \(\sum_{\sp} \dm^{\sp} \mst^{\sp} = 2^{60}\).
    They roughly follow the ratio $1:3:3:4:5$.}
    \label{tab:buckeyballstatweight}
\end{table}
\end{example}

\section{Symmetric position states}
\label{sec:position}

In this section, we formulate the state space of general rotationally
and perrotationally symmetric molecules in terms of position states,
culminating in Eqs.~\eqref{eq:entangled-position-states} and \eqref{eq:entangled-momentum-isomer-in-position} for a general induced representation.
The formulation yields the same state space as that derived in Sec.~\ref{sec:symmetric}, and we develop the Fourier transform that relates position states of this section and momentum states from the previous section.

\subsection{Asymmetric molecules}

The position state space of a symmetric molecule can be thought of as a subspace of the asymmetric molecular state space defined in Eq.~\eqref{eq:identity}.
For such purposes, it is more convenient for us to express the factor spanned
by the molecule-frame $\z$-axis projection states \(|\om\ket\)
using a more general, $\G$-adapted basis \(|\nu\kappa\ket\) stemming from
the isotypic decomposition from Eq.~\eqref{eq:isotypic-wigner-1}.
The basis change is
\begin{equation}
    |_{m}^{\ell},\om\ket\quad\to\quad|_{m\k}^{\ell},\nu\ket~,
\end{equation}
where we group indices in a way that is consistent with the previous section.

Recalling Eq.~\eqref{eq:isotypic-wigner-1}, the $\G$-adapted basis $\{|\nu\k\ket\}$ block
diagonalizes the rotation matrices $D^{\ell}$, with each block corresponding
to a distinct $\G$-irrep $\ir$, and the entire set of participating
irreps denoted by $\ir\up\ell$ (see Example~\ref{ex:cinfty}). For each $\ir$,
the index $\nu$ (along with Greek indices \(\mu,\sigma\)) labels the internal irrep space (and is removed when
$\dim\ir=1$), and the index $\k$ goes over the multiplicity space
(and is removed when $\ml_{\ell}\ir=1$). We suppress the $\ir$ index
in the $\{|\nu\k\ket\}$ basis label set because, in all cases of
interest, we will be selecting a particular irrep --- the trivial
irrep $\irt$ for rotationally symmetric molecules, and a general
irrep $\l$ for perrotationally symmetric nuclear spin isomers.

Indices of the original $|_{m}^{\ell},\om\ket$ rotational states
are in one-to-one correspondence with the Wigner $D$-matrix elements
$D_{m\om}^{\ell}$ \eqref{eq:wigner-ME}, i.e., matrix elements of rotations $D^{\ell}$
in the $\z$-axis basis. Generalizing the molecule-frame momentum factor, we define
the corresponding \textit{$\G$-adapted matrix elements} (cf. \cite[Eq.~(2.33)]{harter_frame_1978}),
\begin{subequations}
    \label{eq:adapted-wigner-ME}
\begin{equation}\label{eq:adapted-wigner-ME-1}
\bra_{m\k}^{\ell},\nu|\gr\ket={\textstyle \sqrt{\frac{2\ell+1}{8\pi^{2}}}}D_{m;\nu\k}^{\ell}(\gr)={\textstyle \sqrt{\frac{2\ell+1}{8\pi^{2}}}}\bra m|D^{\ell}(\gr)|\nu\k\ket\,,
\end{equation}
which are proportional to rotation matrix elements in two \textit{different}
bases: the original
$\z$-axis projection basis for the left index, and the $\G$-adapted basis for the right index.

If the irrep is one-dimensional, then there is no internal irrep index, and the matrix elements are
\begin{equation}\label{eq:adapted-wigner-ME-2}
    \bra_{m\k}^{\ell}|\gr\ket={\textstyle \sqrt{\frac{2\ell+1}{8\pi^{2}}}}D_{m;\k}^{\ell}(\gr)\quad\text{(1D irrep)}\quad,
\end{equation}
with the semicolon and Greek index distinguishing them from the ordinary Wigner \(D\)-matrices \(D^\ell_{m\om}\) from Eq.~\eqref{eq:wigner-ME}.
The latter are recovered by picking $\G$ to be the trivial group (see Example~\ref{ex:asymmetric}).
For groups such as \(\C N\), the adapted matrix elements are a subset of the Wigner elements, in which case we default to the Wigner set.
\end{subequations}

\begin{subequations}
\label{eq:asym-adapted-fourier}The above change of basis yields
the following expression for the $\G$-adapted rotational states in
terms of position states,
\begin{align}
|_{m\k}^{\ell},\nu\ket & ={\textstyle \sqrt{\frac{2\ell+1}{8\pi^{2}}}}\int_{\so 3}\dd\gr\,D^{\ell\star}_{m;\nu\k}(\gr)|\gr\ket\,.\label{eq:asym-adapted-fourier-momentum}
\end{align}
The inverse expression,
\begin{equation}
|\gr\ket=\sum_{\ell\geq|m|\geq0}{\textstyle \sqrt{\frac{2\ell+1}{8\pi^{2}}}}\sum_{\ir\up\ell}\sum_{\nu=1}^{\dim\ir}\sum_{\k=1}^{\ml_{\ell}\ir}D_{m;\nu\k}^{\ell}(\gr)|_{m\k}^{\ell},\nu\ket\,,\label{eq:asym-adapted-fourier-position}
\end{equation}
is easily obtained by remembering that the orthogonality and completeness
properties of the Wigner $D$-matrices \cite{varshalovich_quantum_1988,albert_robust_2020} are
maintained under basis changes.
\end{subequations}

These bases provide an alternative decomposition of the rotation-spin state space of an asymmetric molecule, decomposing the identity from Eq.~\eqref{eq:identity} as
\begin{subequations}\label{eq:identity-g-adapted}
\begin{align}
\id_{\mol}&=\int_{\so 3}\dd\gr\,|\gr\ket\bra\gr|\ot\id_{\nat}~,\label{eq:identity2-g-adapted}\\&=\sum_{\ell\geq|m|\geq0}\sum_{\ir\up\ell}\sum_{\nu=1}^{\dim\ir}\sum_{\k=1}^{\ml_{\ell}\ir}|_{m\k}^{\ell},\nu\ket\bra_{m\k}^{\ell},\nu|\ot\id_{\nat}~.\label{eq:id-mom-asymmetric-g-adapted}
\end{align}
\end{subequations}

\subsection{Rotationally symmetric molecules}

The position states of an asymmetric molecule are in one-to-one correspondence
with elements of $\so 3$ since any proper rotation, by definition,
rotates the molecule from some initial position into a \textit{different}
final position.
Rotationally symmetric molecules (see Sec.~\ref{sec:symmetric} for a definition) admit a proper-rotation
subgroup $\G$ that leaves the molecule invariant per the restriction from Eq.
\eqref{eq:symm-restriction}, so not all $\so 3$ rotations are needed to label
distinct orientations of a symmetric molecule.

Re-stating the symmetry restriction in terms of group theory, given
a label $\cs$ for a molecular position state, any labels of the form
$\cs\rg$ for symmetry-group rotations $\rg\in\G$ correspond to the
same position and are therefore redundant. This redundancy implies
that each $\cs$ is representative of its corresponding \textit{left
coset} of $\G$ in $\so 3$,
\begin{equation}
\cs\G=\{\cs\rg\,|\,\rg\in\G\}\,.
\end{equation}

The set of labels for distinct position states of a $\G$-symmetric
molecule thus corresponds to elements of $\nicefrac{\so 3}{\G}$, the space of
left cosets of $\G$ in $\so 3$. Each coset can be represented by
only one of its elements, and particular choices of representatives
are referred to as ``sections'', ``gauges'', or ``transversals''.

Position states $|\cs\ket$ of $\G$-symmetric molecules form a subspace
of the state space of an asymmetric molecule
and can be expressed as equal superpositions of elements of their corresponding
cosets,
\begin{equation}
|\cs\ket\equiv\frac{1}{\sqrt{|\G|}}\sum_{\rg\in\G}|\gr=\cs\rg\ket\,,\label{eq:coset-states}
\end{equation}
where $|\gr\ket$ is a position state of an asymmetric molecule from
Eq.~\eqref{eq:asym-adapted-fourier-position}, and where $|\G|$ is the number of elements in the group.
For continuous groups $\G\in\{\C{\infty},\D{\infty}\}$, the sum turns
into an integral, and $|\G|$ becomes the group volume; we cover such
cases in the examples.

The above coset states satisfy the symmetry constraint from Eq.~\eqref{eq:symm-restriction}
since the application of a molecule-based rotation
merely permutes the elements in the superposition.
For any $\rh\in\G$, we have
\begin{subequations}
  \label{eq:coset-invariance}
\begin{align}
  \lr_{\rh}|\cs\ket&={\textstyle \frac{1}{\sqrt{|\G|}}}{\textstyle  \sum_{\rg}}\left|\gr=\cs\rg\rh^{-1}\right\rangle \\
  &={\textstyle \frac{1}{\sqrt{|\G|}}}{\textstyle \sum_{\rg}}\left|\gr=\cs\rg\right\rangle~.
\end{align}
\end{subequations}
In the first equality, we use the definition of the action of molecule-based
rotations on asymmetric position states, \(\lr_\rg|\gr\ket=|\gr\rg^{-1}\ket\) in Eq.~\eqref{eq:transformation-body-frame}.
In the second, we use the group resummation property,
\begin{equation}
{\textstyle \sum_{\rg}}f(\rg\rh)={\textstyle \sum_{\rg}}f(\rg)\,,\label{eq:resummation}
\end{equation}
for any \(\rh\in\G\) and any function $f$ on the group.

We now express each position state \(|\gr\ket\) in the above coset states as a superposition of $\G$-adapted rotational states using Eq.~(\ref{eq:asym-adapted-fourier-position}).
Rearranging
sums, writing out the $\G$-adapted $D$-matrices (\ref{eq:adapted-wigner-ME}),
and splitting up the product between $\cs$ and $\rg$ yields
\begin{align}
|\cs\ket & =\sum_{\ell\geq|m|\geq0}{\textstyle \sqrt{\frac{2\ell+1}{8\pi^{2}/|\G|}}}\sum_{\ir\up\ell}\sum_{\nu=1}^{\dim\ir}\sum_{\k=1}^{\ml_{\ell}\ir}\label{eq:1D-interim-position-states}\\
 & \,\,\,\,\,\,\,\,\,\,\,\,\,\,\,\,\,\,\,\Big\langle m\Big|D^{\ell}(\cs)\Big[{\textstyle \frac{1}{|\G|}}{\textstyle \sum_{\rg}}D^{\ell}(\rg)\Big]\Big|\nu\k\Big\rangle\,\,|_{m\k}^{\ell},\nu\ket\,.\nonumber
\end{align}
This equation can be further simplified by noticing that the sum
in square brackets is a projection onto all copies of the trivial irrep. In
terms of the $\G$-adapted basis,
\begin{equation}
\frac{1}{|\G|}\sum_{\rg\in\G}D^{\ell}(\rg)|\nu\k\ket=\d_{\ir,\irt}|\k\ket\,,\label{eq:1D-projection}
\end{equation}
where the $\nu$ index is not present since $\dim\irt=1$.

Plugging in the expression for the projection reduces the position-state
expression (\ref{eq:1D-interim-position-states}) in the following
ways. First, the sum over $\ir$ goes away due to $\ir=\irt$. As a result, the sum over $\ell$ is
reduced to the sum over only those momenta which contain at least
one copy of the trivial irrep; we denote this set by $\ell\down\irt$.
Second, the sums over $\nu$ and $\k$ index the irrep and multiplicity
spaces of $\irt$, and the former goes away since the irrep dimension
is one, i.e., $|_{m\k}^{\ell},\nu\ket\to|_{m\k}^{\ell}\ket$.
Altogether, this yields
\begin{equation}
|\cs\ket=\sum_{\ell\down\irt}\sum_{|m|\leq\ell}{\textstyle \sqrt{\frac{2\ell+1}{8\pi^{2}/|\G|}}}\sum_{\k=1}^{\ml_{\ell}\irt}D_{m;\k}^{\ell}(\cs)|_{m\k}^{\ell}\ket\,,\label{eq:sym-position}
\end{equation}
which expresses position states precisely in terms of the rotational
states $\{|_{m\k}^{\ell}\ket\}$ that we obtained in Sec.~\ref{sec:symmetry}.

The symmetric molecular position states (\ref{eq:sym-position})
constitute an orthonormal ``basis'' for the state space of a $\G$-symmetric
rigid body,
\begin{equation}\label{eq:rot-symmetric-ortho}
\bra\cs|\cs^\pr\ket=\d^{\nicefrac{\so 3}{\G}}(\cs,\cs^\pr)\,,
\end{equation}
where the coset-space Dirac $\d$-function is infinite for coset representatives
$\cs=\cs^\pr$ and zero otherwise. Position states resolve the identity
on the symmetric state space as
\begin{align}
\id_{\mol}^{\G} & =\int_{\nicefrac{\so 3}{\G}}\dd\cs|\cs\ket\bra\cs|\ot\id_{\nat}\,,\label{eq:id-pos-symmetric}
\end{align}
dual to the rotational-state identity in Eq.~\eqref{eq:id-mom-symmetric}.

The reverse Fourier transform expressing $|_{m\k}^{\ell}\ket$ in
terms of position states can be derived by using the fact that cosets
partition $\so 3$. Given a fixed set of coset representatives,
each rotation $\gr\in\so 3$ can be written as a product of a particular representative
and symmetry-group element,
\begin{equation}
\gr=\cs\rg\quad\text{for some}\quad\cs\in\nicefrac{\so 3}{\G}\quad\text{and}\quad\rg\in\G\,.\label{eq:coset-partition}
\end{equation}
As a result, any integral over $\so 3$ can be split into an
integral over the coset space and a sum over the symmetry group. Applying
this to Eq.~(\ref{eq:asym-adapted-fourier-momentum}) and specializing to the trivial (one-dimensional) irrep yields
\begin{equation}
|_{m\k}^{\ell}\ket={\textstyle \sqrt{\frac{2\ell+1}{8\pi^{2}}}}\int_{\nicefrac{\so 3}{\G}}\dd\cs\sum_{\rg\in\G}D_{m;\k}^{\ell\star}(\cs\rg)|\gr=\cs\rg\ket\,~,
\end{equation}
where \(\k\) indexes the multiplicity of \(\irt\).

Next, we simplify the $\G$-adapted matrix elements by noticing that
taking the $\k$th element is effectively evaluating $\rg$ in the
trivial $\G$-irrep,
\begin{subequations}
\begin{align}
D_{m;\k}^{\ell\star}(\cs\rg)&=\bra m|D^{\ell\star}(\cs)D^{\ell\star}(\rg)|\k\ket\\&=\bra m|D^{\ell\star}(\cs)\irt(\rg)|\k\ket\\&=\bra m|D^{\ell\star}(\cs)|\k\ket\,.
\end{align}
\end{subequations}
Plugging this in, evaluating $\sum_{\rg\in\G}1=|\G|$, and expressing the sum of position states \(|\gr=\cs\rg\ket\) as a symmetric position state \(|\cs\ket\) from Eq.~\eqref{eq:coset-states} yields
\begin{equation}
|_{m\k}^{\ell}\ket={\textstyle \sqrt{\frac{2\ell+1}{8\pi^{2}/|\G|}}}\int_{\nicefrac{\so 3}{\G}}\dd\cs\,D_{m;\k}^{\ell\star}(\cs)|\cs\ket\,,
\end{equation}
completing the Fourier transform on $\nicefrac{\so 3}{\G}$.

\begin{example}[calcium monohydrosulfide]\label{ex:asymmetric-position}
    The procedure for determining the position state space of a symmetric molecule can also be applied to asymmetric molecules.
    Recalling Example~\ref{ex:asymmetric}, the symmetry group of such molecules is trivial, containing only the identity element, \(\G=\bra \re \ket\).
    The coset space \(\nicefrac{\so 3}{\C{1}} = \so 3\), meaning that the entire group remains as the set of labels of asymmetric molecular position states.

    The trivial irrep is the only irrep, so the sum over $\ir$ in the position-state expression (\ref{eq:asym-adapted-fourier-position}) goes away.
    The trivial irrep is one-dimensional, so there is also no $\nu$ index in that equation.
    The $\k$ index goes from 1 to $2\ell+1=\ml_{\ell}\irt$, reducing Eq.~(\ref{eq:asym-adapted-fourier-position}) to the position states from Eq.~\eqref{eq:asymmetric-position}.
\end{example}

\begin{example}[hydrogen chloride]\label{ex:cinfty-position}
    Position states of any \(\C\infty\)-symmetric molecules from Example~\ref{ex:cinfty} are labeled by points on the ordinary two-sphere \(\SS^2\).
    In our formalism, the sphere is recovered as the coset space \(\nicefrac{\so 3}{\C\infty}\cong\SS^2\) and corresponds to a subset of coset representatives.

    A convenient partition of \(\so3\)-rotations into coset representatives \(\cs\) and symmetry-group elements \(\rg\) per Eq.~\eqref{eq:coset-partition} can be done using the Euler angle parameterization, \(\gr=(\alpha\beta\gamma)=\cs\rg\), when \(\C\infty\) is the group of rotations around the \(\z\)-axis.
    In that case, the \(\gamma\) angle is reserved for labeling the subgroup elements \(\rg\), while the pair \((\alpha\beta)\) is retained for the position states \(\cs\).

    For \(\C\infty\), the position state sum in Eq.~\eqref{eq:coset-states} becomes an integral,
    \begin{equation}\label{eq:continuous-coset-states}
        |\cs\ket\equiv{\textstyle\frac{1}{\sqrt{|\G|}}}\int_{\G}\dd\G~|\gr=\cs\rg\ket\,,
    \end{equation}
    where the group volume is \(\C\infty=\int_0^{2\pi}d\gamma = 2\pi\).
    When \(\C\infty\) is the subgroup of \(\z\)-axis rotations in the Euler-angle parameterization, the above integral reduces to one over \(\gamma\).

    The \(\C\infty\)-adapted matrix elements correspond to the spherical harmonics,
    \begin{equation}
        Y_{m}^{\ell}(\cs)={\textstyle \sqrt{\frac{2\ell+1}{4\pi}}}D_{m0}^{\ell\star}(\cs)~,
    \end{equation}
    and the position and momentum states for this case are
    \begin{subequations}\label{eq:heteronuclear-position}
        \begin{align}
            |\cs\ket&=\sum_{\ell\geq0}\sum_{|m|\leq\ell}Y_{m}^{\ell\star}(\cs)|_{m}^{\ell}\ket\\
            |_{m}^{\ell}\ket
            &=\int_{\SS^{2}}\dd\cs\,Y_{m}^{\ell}(\cs)|\cs\ket~,
        \end{align}
    \end{subequations}
    where \(|^\ell_m\ket = |^\ell_m,\om=0\ket\) are the rotational basis states determined in Example~\ref{ex:cinfty}.
\end{example}

\begin{example}[disulfur]\label{ex:disulfur-position}
    The \ce{S2} molecule, alogn with any spinless homonuclear diatomic, is a \(\D\infty\)-symmetric rigid body (see Example~\ref{ex:disulfur}).
    Its symmetry group includes the rotations from the previous heteronuclear example, along with all \(\pi\)-rotations that permute the two identical (spinless) nuclei.

    Position states of this case are of the form in Eq.~\eqref{eq:continuous-coset-states}, with the integral being over \(\G=\D\infty\), with group volume
    \(|\D\infty|= 4\pi\).

    The coset space labeling \ce{S2} positions is \(\nicefrac{\so 3}{\D \infty}\cong \rptwo\) --- the two-dimensional real projective plane.
    This is also known as the two-sphere with opposite points identified,
    and each point can be identified with a rod in three-dimensional space \cite{mermin_topological_1979}.
    In terms of the Euler-angle parameterization from the previous example, each projective-plane point \((\alpha\beta)\) is a rod with endpoints at \((\alpha\beta)\) and its antipode \((\alpha+\pi,\pi-\beta)\).

    Position and momentum states for this case are,
    \begin{subequations}\label{eq:homonuclear-position}
        \begin{align}
            |\cs\ket&=\sqrt{2}\sum_{\ell\text{ even}}\sum_{|m|\leq\ell}Y_{m}^{\ell\star}(\cs)|_{m}^{\ell}\ket\\|_{m}^{\ell}\ket&=\sqrt{2}\int_{\rptwo}\dd\cs\,Y_{m}^{\ell}(\cs)|\cs\ket~,
        \end{align}
    \end{subequations}
    where the square root is due to a different normalization for \(\G=\D\infty\), i.e., \(8\pi^2/|\D\infty|=2\pi\).
    These are closely related to those of the \(\C\infty\)-symmetric case from Eq.~\eqref{eq:heteronuclear-position}.
    The position states can even be obtained by superposing each \(\C\infty\)-symmetric state \(\cs=(\alpha\beta)\) with its antipode, which restricts the spherical-harmonic sum to only the even-momentum harmonics.
\end{example}

\begin{example}[cobalt tetracarbonyl hydride]\label{ex:cobalt-position}
    Recalling Example~\ref{ex:cobalt}, this molecule is \(\C 3\) symmetric.
    The coset space labeling its positions is \(\nicefrac{\so 3}{\C 3}\), a member of the \textit{lens space} family, \(\mathsf{L}_{2N,1} \equiv \nicefrac{\so 3}{\C N}\).
    We handle the general-\(N\) case directly due to its simple generality.

    Picking the group to consist of \(\z\)-axis rotations, a simple parameterization for the coset space consists of Euler angles \((\alpha\beta\gamma)\), with \(0\leq\gamma<2\pi/N\).
    The \(N\to\infty\) case \vva{reduces to} the \(\C\infty\)-symmetric Example~\ref{ex:cinfty-position} (cf. \cite[Appx. D]{albert_robust_2020}).

    The symmetry group \(\C N\) is discrete, so Eq.~\eqref{eq:coset-states} applies directly.
    Position and momentum states (with the latter worked out in Example~\ref{ex:cobalt}) can be expressed in terms of the standard Wigner \(D\)-matrix elements,
    \begin{subequations}\label{eq:cobalt-position}
        \begin{align}
            |\cs\ket&=\sum_{\ell\geq0}{\textstyle \sqrt{\frac{2\ell+1}{8\pi^{2}/N}}}\sum_{|m|\leq\ell}\sum_{\k=1}^{2n+1}D_{m,N(\k-n-1)}^{\ell}(\cs)|_{m\k}^{\ell}\ket\\|_{m\k}^{\ell}\ket&={\textstyle \sqrt{\frac{2\ell+1}{8\pi^{2}/N}}}\int_{\mathsf{L}_{2N,1}}\dd\cs\,D_{m,N(\k-n-1)}^{\ell\star}(\cs)|\cs\ket~,
        \end{align}
    \end{subequations}
    where \(n=\lfloor \ell/N \rfloor\).
\end{example}

\begin{example}[sulfur trioxide]\label{ex:sulfur-trioxide-position}
    Recalling Example~\ref{ex:sulfur-trioxide}, this molecule is \(\D3\) symmetric.
    The coset space labeling its positions is \(\so{3}/\D{3}\), a member of the \textit{prism space} family, \(\nicefrac{\so 3}{\D N}\).
    The \(N\to\infty\) case of such spaces yields the real projective plane from Example~\ref{ex:disulfur-position}.

    Recalling the form of the rotational states from Eq.~\eqref{eq:sulfur-trioxide-states}, the corresponding \(\D3\)-adapted rotation matrix elements from Eq.~\eqref{eq:adapted-wigner-ME-2}, expressed in terms of the standard Wigner \(D\)-matrix elements, are
    \begin{equation}
        D_{m;\k}^{\ell\star}(\cs)={\textstyle \frac{1}{\sqrt{2}}}\big(D_{m,3(\k-1)}^{\ell\star}(\cs)-(-1)^{\ell+\k}D_{m,3(1-\k)}^{\ell\star}(\cs)\big),
    \end{equation}
    for \(\cs\in \nicefrac{\so 3}{\D 3}\).
    These are then plugged into the Fourier transform on this prism space,
    \begin{subequations}\label{eq:sulfur-trioxide-position}
        \begin{align}
            |\cs\ket&=\sum_{\ell\geq0}{\textstyle \sqrt{\frac{2\ell+1}{4\pi^{2}/3}}}\sum_{|m|\leq\ell}\sum_{\k=1}^{\mult(\ell)}D_{m;\k}^{\ell}(\cs)|_{m\k}^{\ell}\ket\\|_{m\k}^{\ell}\ket&={\textstyle \sqrt{\frac{2\ell+1}{4\pi^{2}/3}}}\int_{\nicefrac{\so 3}{\D 3}}\dd\cs\,D_{m;\k}^{\ell\star}(\cs)|\cs\ket~.
        \end{align}
    \end{subequations}

\end{example}

\begin{example}[\ce{C60} fullerene]
    The position-state space of this icosahedrally symmetric molecule is the \vva{Poincar\'e} homology sphere \(\nicefrac{\so 3}{\I}\) \cite{Kirby_eight_1979,baez_icosahedron_2017}.
    This space was once a candidate model for the shape of an assumed-to-be-periodic universe \cite{luminet_dodecahedral_2003}.
    Position states are difficult to express in terms of icosahedral harmonics, but admit a simple coset-state expression,
    \begin{equation}
        |\cs\ket\equiv{\textstyle\frac{1}{\sqrt{60}}}\sum_{\rg\in\I}|\gr=\cs\rg\ket~.
    \end{equation}

\end{example}

\subsection{Perrotationally symmetric molecules}
\label{sec:spin-position}

Molecular position states of perrotationally symmetric molecules have
to satisfy the joint symmetry and spin-statistics condition from Eq.~\eqref{eq:symm-restriction-1},
under which the collective rotation-spin space transforms according
to the 1D $\G$-irrep $\s$. In Sec.~\ref{sec:spin}, we showed that there are various
combinations of rotational irreps $\l$ and nuclear irreps $\t$ such
that their direct product restricts to $\s$.

Here, we develop position states for a single isomer \(\sp\), which \vva{corresponds} to the triple $\l\ot\t\down\s$.
This requires generalizing
the procedure for rotationally symmetric molecules from the previous
subsection.
The resulting states are still parameterized, in part, by
coset-space labels $\cs\in\nicefrac{\so 3}{\G}$, but the non-Abelian nature of
$\l$ irrep attaches extra internal degrees of freedom to each position
state in the entangled case.

\subsubsection{$\protect\dm=1$ separable isomers}

In the separable case, both irreps $\l$ and $\t$ are one-dimensional, and
their tensor product directly yields the correct spin statistics,
$\l\otimes\t=\s$. To identify the position state space, we have
to formulate position states on the rotational factor that transform
according to $\l$.

The result is obtained by a simple modification of the coset-state
superposition in Eq.~(\ref{eq:coset-states}) to a superposition whose
coefficients are evaluations of the symmetry-group elements in the
desired irrep. For each position-state label $\cs\in\nicefrac{\so 3}{\G}$, the
state is
\begin{equation}
|\cs\ket\equiv{\textstyle \frac{1}{\sqrt{|\G|}}}\sum_{\rg\in\G}\l(\rg)\left|\gr=\cs\rg\right\rangle \,.\label{eq:1D-coset-states}
\end{equation}
These states are Dirac-\(\d\) orthogonal, satisfying Eq.~\eqref{eq:rot-symmetric-ortho} for any one-dimensional \(\l\).
When $\l=\irt$, they reduce to the rotationally symmetric case in
Eq.~(\ref{eq:coset-states}).

The above generalized coset states transform correctly under the rotational part of the perrotation from Eq.~\eqref{eq:perrotations}.
Generalizing Eq.~\eqref{eq:coset-invariance}, we have, for any $\rh\in\G$,
\begin{subequations}
\label{eq:separable-transform-under-rotations}
\begin{align}
\lr_{\rh}|\cs\ket&={\textstyle \frac{1}{\sqrt{|\G|}}}{\textstyle \sum_{\rg}}\l(\rg)\left|\gr=\cs\rg\rh^{-1}\right\rangle \\&={\textstyle \frac{1}{\sqrt{|\G|}}}{\textstyle \sum_{\rg}}\l(\rg\rh)\left|\gr=\cs\rg\right\rangle \\&={\textstyle \frac{1}{\sqrt{|\G|}}}{\textstyle \sum_{\rg}}\l(\rg)\l(\rh)\left|\gr=\cs\rg\right\rangle \\&=\l(\rh)|\cs\ket\,.
\end{align}
In the third equality, we recall that $\l$
is a scalar representing the group and split the product of $\rg$
and $\rh$.
In the fourth, we re-express the result in terms of the original coset states.
\end{subequations}

The generalized coset states (\ref{eq:1D-coset-states}) can be expressed
in terms of the rotational states $\{|_{m\k}^{\ell}\ket\}$ from Eq.
\eqref{eq:separable-state}, where $\k$ indexes the multiplicity of $\l$. The derivation
is a slight generalization of the rotationally symmetric ($\l=\irt$)
derivation in Eqs.~(\ref{eq:1D-interim-position-states}-\ref{eq:sym-position}).
We omit it for brevity, noting that the entangled isomer derivation
done in the next subsection encompasses both.

Position and rotational states for the rotational factor of a separable
isomer are
\begin{subequations}
\label{eq:separable-Fourier-transform}
\begin{align}
|\cs\ket & =\sum_{\ell\down\l}\sum_{|m|\leq\ell}{\textstyle \sqrt{\frac{2\ell+1}{8\pi^{2}/|\G|}}}\sum_{\k=1}^{\mult(\ell)}D_{m;\k}^{\ell}(\cs)|_{m\k}^{\ell}\ket\\
|_{m\k}^{\ell}\ket & ={\textstyle \sqrt{\frac{2\ell+1}{8\pi^{2}/|\G|}}}\int_{\nicefrac{\so 3}{\G}}\dd\cs D_{m;\k}^{\ell\star}(\cs)|\cs\ket\,.
\end{align}
\end{subequations}
Tacking on the nuclear states $\{|\chi\ket\}$, which are required
to transform according to $\t$, the isomer's identity decomposes
as
\begin{align}
\id_{\mol}^{\G,\sp} & =\int_{\nicefrac{\so 3}{\G}}\dd\cs|\cs\ket\bra\cs|\ot\sum_{\chi=1}^{\mst}|\chi\ket\bra\chi|\,,
\end{align}
dual to the rotational-state decomposition from Eq.~\eqref{eq:id-mom-rotnuc}.

\begin{example}[water, $\a^{\ast}$ para isomer]
    Recalling Example~\ref{ex:water-singlet}, \ce{H2O} is \(\C2\)-symmetric, so its position states are parameterized by points in the lens space \(\mathsf{L}_{4,1} \equiv \nicefrac{\so 3}{\C 2}\).
    Using the Euler-angle parameterization from Example~\ref{ex:cobalt-position},
    each coset consists of only two elements, \((\alpha\beta\gamma)\) and \((\alpha,\beta,\gamma+\pi)\).
    Coset representatives are parameterized by $(\alpha\beta\gamma)$ with $\alpha=[0,2\pi)$, $\beta=[0,\pi]$, and $\gamma=[0,\pi)$.

    Since \(\l=\a\) is the trivial irrep for this isomer, we have \(\a(\gamma=0) = \a(\gamma=\pi) = +1\).
    The coset states from \eqref{eq:1D-coset-states} are
    \begin{equation}
    \left|\cs=(\alpha\beta\gamma)\right\rangle ={\textstyle \frac{1}{\sqrt{2}}}\big(\left|\gr=(\alpha,\beta,\gamma)\right\rangle + \left|\gr=(\alpha,\beta,\gamma+\pi)\right\rangle \big)~,
    \end{equation}
    where we abuse notation and use a coset representative's Euler-angle parameterization as a stand-in for the representative \(\cs\) itself.
    Applying the symmetry-group rotation \(\lr_{00\pi}\) to this state yields \(+1\), confirming that the state transforms according to the trivial irrep.

    The Fourier transform between these states and the rotational states is that from Eq.~\eqref{eq:homonuclear-position} for \(N=2\).
    Per Example~\ref{ex:water-singlet}, a basis for the isomer consists of tensor products of the above states with the singlet state $(\left|\up\down\right\rangle -\left|\down\up\right\rangle )/\sqrt{2}$ of the two hydrogen nuclei and an arbitrary state of the oxygen nucleus.
\end{example}

\begin{example}[water, $\b^{\ast}$ ortho isomer]\label{ex:triplet-water-position}
    The other isomers of water, considered in Example~\ref{ex:triplet-water}, admit the \(\l=\b\) irrep, for which \(\b(\gamma=0) = +1\) and \(\b(\gamma=\pi) = -1\).
    Plugging this data into the generalized coset states in Eq.~\eqref{eq:1D-coset-states} yields
    \begin{equation}\label{eq:water-coset}
    \left|\cs=(\alpha\beta\gamma)\right\rangle ={\textstyle \frac{1}{\sqrt{2}}}\big(\left|\gr=(\alpha,\beta,\gamma)\right\rangle - \left|\gr=(\alpha,\beta,\gamma+\pi)\right\rangle \big)~.
    \end{equation}
    Applying \(\lr_{00\pi}\) to this state yields \(-1\), confirming that the state transforms according to the \(\b\) irrep.

    Per Example~\ref{ex:triplet-water}, the rotational states transforming according to the \(\b\) irrep are the asymmetric states \(|^\ell_m,\om\ket\) for which \(\om\) is odd.
    The number of such states for a given \(\ell\) is \(\mult(\ell) = 2\lfloor(\ell+1)/2\rfloor\), and the symmetric rotational states are \(|^\ell_{m\k}\ket = |^\ell_m,\om=2\k-\mult(\ell)-1\ket\) for \(1\leq\k\leq\mult(\ell)\).
    The Fourier transform between position and rotational states for this isomer is then expressible using the Wigner \(D\)-matrix elements,
    \begin{subequations}
        \begin{align}
            |\cs\ket&=\sum_{\ell>0}{\textstyle \sqrt{\frac{2\ell+1}{4\pi^{2}}}}\sum_{|m|\leq\ell}\sum_{\k=1}^{\mult(\ell)}D_{m,2\k-\mult(\ell)-1}^{\ell}(\cs)|_{m\k}^{\ell}\ket\\
            |_{m\k}^{\ell}\ket&={\textstyle \sqrt{\frac{2\ell+1}{4\pi^{2}}}}\int_{\mathsf{L}_{4,1}}\dd\cs\,D_{m,2\k-\mult(\ell)-1}^{\ell\star}(\cs)|\cs\ket~.
        \end{align}
    \end{subequations}
    Note that all momenta participate except \(\ell=0\), which has no odd values of its \(\z\)-axis component.

    A basis for the isomer consists of tensor products of the above states with any triplet-subspace state of the two hydrogen nuclei and an arbitrary state of the oxygen.
\end{example}

\begin{example}[deuterated hydrogen, \(\a_1\) ortho isomer]\label{ex:deuterium-a1-position}
Recalling Example~\ref{ex:deuterium-a1}, \ce{D2} is \(\D\infty\)-symmetric, so its position states are parameterized by points in the projective plane \(\rptwo = \nicefrac{\so 3}{\D \infty}\).
This is the same set of labels as for disulfur in Example~\ref{ex:disulfur-position}.
The rotational-state irrep participating in this isomer is also trivial, \(\l=\a_1\), so the Fourier transform on the rotational part of this isomer is the same as that of \ce{S2} in Eq.~\eqref{eq:homonuclear-position}.

A basis for the isomer consists of tensor products of the above states with any triplet-subspace state of the two deuterium nuclei.

\end{example}

\begin{example}[deuterated hydrogen, \(\a_2\) para isomer]\label{ex:deuterium-a2-position}
Recalling Example~\ref{ex:deuterium-a2}, the other isomer of \ce{D2} is constructed by setting both the rotational and nuclear-spin states to transform according to the sign irrep, \(\l=\t=\a_2\).
While the position-state labels, \(\cs\in\rptwo\), remain the same as those of the \(\a_1\) isomer, the rotational-state irrep is different.

Since the irrep is nontrivial, the coset-state superposition is no longer uniform like it was in Eq.~\eqref{eq:continuous-coset-states}.
Parameterizing \(\z\)-axis rotations by angle \(\gamma\) by the Euler-angle triple \((00\gamma)\) and equatorial rotations by \((0\pi\gamma)\), position states are parameterized by \(\cs=(\alpha\beta)\) (see Example~\ref{ex:disulfur-position}), and position states from Eq.~\eqref{eq:1D-coset-states} are
\begin{equation}\label{eq:deuterium-a2-position}
    \left|(\alpha\beta)\right\rangle \equiv{\textstyle \frac{1}{4\pi}}\int_{0}^{2\pi}\dd\gamma~\left|\gr=(\alpha,\beta,\gamma)\right\rangle -\left|\gr=(\alpha+\pi,\pi-\beta,\gamma)\right\rangle \,,
\end{equation}
where the relative phase is due to the fact \vva{that} all equatorial rotations evaluate to \(-1\) in the irrep.

The Fourier transform between position and momentum states (worked out in Example~\ref{ex:deuterium-a2}) is
    \begin{subequations}\label{eq:deuterium-a1-position}
        \begin{align}
            |\cs\ket&=\sqrt{2}\sum_{\ell\text{ odd}}\sum_{|m|\leq\ell}Y_{m}^{\ell\star}(\cs)|_{m}^{\ell}\ket\\|_{m}^{\ell}\ket&=\sqrt{2}\int_{\rptwo}\dd\cs\,Y_{m}^{\ell}(\cs)|\cs\ket~.
        \end{align}
    \end{subequations}
A basis for the isomer consists of tensor products of the above states with the  singlet state of the two deuterium nuclei.

\end{example}

\begin{example}[ammonia, \(^2\e\) isomer]\label{ex:ammonia-position}
Recalling \vva{Example~\ref{ex:ammonia}}, \ce{NH3} is \(\C3\)-symmetric, so its position states are parameterized by points in the lens space \(\mathsf{L}_{6,1} \cong \nicefrac{\so 3}{\C 3}\).
The symmetry group contains three rotations by \(\frac{2\pi}{3} p\) for \(p\in\{0,1,2\}\), which evaluate to \(\exp(-i\frac{2\pi}{3}p)\) in the \(^2\e\) irrep.

Position states from Eq.~\eqref{eq:1D-coset-states} are
\begin{equation}
    \left|\cs=(\alpha\beta\gamma)\right\ket = \sum_{p=0}^{2}e^{-i\frac{2\pi}{3}p}\left|\gr=(\alpha,\beta,\gamma+{\textstyle \frac{2\pi}{3}}p)\right\rangle ~,
\end{equation}
parameterized by \(\alpha \in [0,2\pi)\), \(\beta \in [0,\pi]\), and \(\gamma \in [0,2\pi/3)\).
States for the \(\C N\)-symmetric \(^2 \e_1\) isomer
are obtained by letting \(3\to N\) and iterating \(p\) from \(0\) to \(N-1\).

Per Example~\ref{ex:ammonia}, the rotational states transforming according to the \(^2\e\) irrep are the asymmetric states \(|^\ell_m,\om\ket\) for which \(\om\equiv2\) modulo 3.
The admissible rotational states are
\begin{equation}
    |^\ell_{m\k}\ket = |^\ell_m,\om = 3\k - 1 - 3 \lfloor (\ell+2)/3 \rfloor \ket
\end{equation}
for \(|m|\leq\ell\) and \(1\leq \k \leq \mult(\ell) = \ell - \lfloor \ell/3 \rfloor\).
The Fourier transform between position and rotational states for this isomer is then expressible using the Wigner \(D\)-matrix elements,
        \begin{align}
            |\cs\ket&=\sum_{\ell>0}{\textstyle \sqrt{\frac{2\ell+1}{4\pi^{2}}}}\sum_{|m|\leq\ell}\sum_{\k=1}^{\mult(\ell)}D_{m,3\k-1-3\lfloor(\ell+2)/3\rfloor}^{\ell}(\cs)|_{m\k}^{\ell}\ket\nonumber\\
            |_{m\k}^{\ell}\ket&={\textstyle \sqrt{\frac{2\ell+1}{4\pi^{2}}}}\int_{\nicefrac{\so 3}{\C 3}}\dd\cs\,D_{m,3\k-1-3\lfloor(\ell+2)/3\rfloor}^{\ell\star}(\cs)|\cs\ket~.
        \end{align}
\end{example}

\subsubsection{$\protect\dm\protect\geq2$ entangled isomers}

The collective rotational and nuclear-spin wavefunction of an entangled
isomer transforms according \vva{to} the irrep $\s$, which is present in
the tensor product of its corresponding $\dm$-dimensional rotational
irrep $\l$ and a nuclear-spin irrep $\t$. We first formulate ``uncoupled''
position states on the rotational factor that transform as $\l$,
and then perform the projection onto the $\s$ irrep according to
the prescription of Sec.~\ref{sec:spin}.

Since the rotational irrep is no longer one-dimensional, $\l(\rg)$
is a $\dm$-dimensional matrix with matrix elements
\begin{equation}
\l^{\m\n}(\rg)=\bra\mu|\l(\rg)|\nu\ket\,,
\end{equation}
where $\m,\n\in\{1,2,\cdots,\dm\}$, and where
$\rg\in\G$. Defining these elements requires a choice of orthonormal basis \(|\mu\ket\) for
the irrep space, but the results below hold for any choice.

Extending the coset-state construction from Eq.~(\ref{eq:1D-coset-states})
to this case yields $\dm^{2}$ states for each representative $\cs\in\nicefrac{\so 3}{\G}$,
\begin{equation}
|\cs;\l^{\m\n}\ket={\textstyle \sqrt{\frac{\dm}{|\G|}}}\sum_{\rg\in\G}\l^{\m\n}(\rg)|\gr=\cs\rg\ket\,,\label{eq:Zak}
\end{equation}
with each state corresponding to a particular matrix element.

Molecule-based rotations in the symmetry group leave the position-state
label $\cs$ invariant, but transform the internal degrees of freedom
at each position. More precisely, rotations act on the $\nu$ index
according to $\l$.
Generalizing
Eq.~(\ref{eq:separable-transform-under-rotations}) yields, for any $\rh\in\G$,
\begin{subequations}
\begin{align}
\lr_{\rh}|\cs;\l^{\m\n}\ket&={\textstyle \sqrt{\frac{\dm}{|\G|}}}\sum_{\rg\in\G}\l^{\m\n}(\rg\rh)|\gr=\cs\rg\ket\\&={\textstyle \sqrt{\frac{\dm}{|\G|}}}\sum_{\rg\in\G}\sum_{\sigma=1}^{\dm}\l^{\m\sigma}(\rg)\l^{\sigma\n}(\rh)|\cs\rg\ket\\&=\sum_{\sigma=1}^{\dm}\l^{\sigma\nu}(\rh)|\cs;\l^{\m\sigma}\ket~,
\end{align}
where we have resolved the identity on the internal
$\l$-irrep space in order to split up the product of $\rg$ and
$\rh$.
\end{subequations}

We now express Eq.~\eqref{eq:Zak} in terms of rotational states $\{|_{m\k}^{\ell},\m\ket\}$
from Eq.~\eqref{eq:asym-adapted-fourier-momentum}. The derivation proceeds analogously to the rotationally
symmetric case in Eqs.~(\ref{eq:1D-interim-position-states}-\ref{eq:sym-position})
whilst taking into account the internal \(\mu,\nu\)-labeled degrees of freedom.

After plugging in the $\G$-adapted expressions for asymmetric position states
$|\gr\ket$ from Eq.~(\ref{eq:asym-adapted-fourier-position}), writing
out the $\G$-adapted matrix elements from Eq.~(\ref{eq:adapted-wigner-ME}),
and splitting up the product between $\cs$ and $\rg$, Eq.~(\ref{eq:Zak})
becomes
\begin{align}
|\cs;\l^{\m\n}\ket & =\sum_{\ell\geq|m|\geq0}{\textstyle \sqrt{\frac{(2\ell+1)/\dm}{8\pi^{2}/|\G|}}}\sum_{\ir\up\ell}\sum_{\sigma=1}^{\dim\ir}\sum_{\k=1}^{\ml_{\ell}\ir}\label{eq:entangled-interim-position-states}\\
 & \,\,\,\Big\langle m\Big|D^{\ell}(\cs)\Big[{\textstyle \frac{\dm}{|\G|}}{\textstyle \sum_{\rg}}\l^{\m\n}(\rg)D^{\ell}(\rg)\Big]\Big|\sigma\k\Big\rangle\,\,|_{m\k}^{\ell},\sigma\ket\,.\nonumber
\end{align}

This equation can be further simplified by noticing that \(\rg\) is evaluated in the \(\ir\) irrep, i.e.,
\begin{equation}\label{eq:evaluating-in-irrep}
    D^{\ell}(\rg)|\sigma\k\ket=\ir^{\star}(\rg)|\sigma\ket\ot|\k\ket~,
\end{equation}
per the convention chosen in the isotypic decomposition from Eq.~\eqref{eq:isotypic-wigner-1}.
Inserting another resolution of identity on the irrep space and simplifying yields
\begin{align}
    |\cs;\l^{\m\n}\ket&=\sum_{\ell\geq|m|\geq0}{\textstyle \sqrt{\frac{(2\ell+1)/\dm}{8\pi^{2}/|\G|}}}\sum_{\ir\up\ell}~\sum_{\sigma,\tau=1}^{\dim\ir}\sum_{\k=1}^{\ml_{\ell}\ir}\\&\,\,D_{m;\tau\k}^{\ell}(\cs)\Big[{\textstyle \frac{\dm}{|\G|}}{\textstyle \sum_{\rg}}\l^{\m\n}(\rg)\ir^{\tau\sigma\star}(\rg)\Big]\,\,|_{m\k}^{\ell},\sigma\ket\,.\nonumber
\end{align}
The sum in square brackets is then evaluated using Schur orthogonality, yielding three \(\delta\)-functions, \(\delta_{\l\ir}\d_{\mu\tau}\d_{\nu\sigma}\).

The first \(\delta\)-function kills the sum over $\ir$ by selecting $\ir=\l$.
Conversely, the sum over $\ell$ is reduced to a sum over only those momenta which contain at least one copy of
the desired irrep; we denote this set by $\ell\down\l$.
The latter two \(\delta\)-functions kill the sums over \(\tau\) and \(\sigma\), respectively, leaving only the multiplicity sum over \(\k\).
Altogether, Eq.~(\ref{eq:entangled-interim-position-states}) is simplified to
\begin{equation}
|\cs;\l^{\m\n}\ket=\sum_{\ell\down\l}\sum_{|m|\leq\ell}{\textstyle \sqrt{\frac{(2\ell+1)/\dm}{8\pi^{2}/|\G|}}}\,\sum_{\k=1}^{\mult(\ell)}D_{m;\m\k}^{\ell}(\cs)|_{m\k}^{\ell},\nu\ket\,.\label{eq:entangled-zak}
\end{equation}
expressing position states precisely in terms of the uncoupled rotational
states $\{|_{m\k}^{\ell},\nu\ket\}$ from Eq.~\eqref{eq:asym-adapted-fourier-momentum}.

We now couple the above position states to the $\t$-irrep
nuclear-spin subspace and restrict the result to the collective $\s$
irrep in order to obtain the correct spin statistics. According to
Sec.~\ref{sec:spin}, we couple internal irrep space (indexed above by $\nu$)
to its partner, the $\t$-irrep space, and project onto the admissible
rotational states from Eq.~\eqref{eq:entangled-state},
\begin{equation}
|_{m\k}^{\ell}\ket=\frac{1}{\sqrt{\dm}}\sum_{\n=1}^\dm s_{\n}|_{m\k}^{\ell},\nu\ket_{\rot}|\nu\ket_{\nat}\,.\label{eq:entangled-state-rotational}
\end{equation}
This coupling takes in the set $\{|\cs;\l^{\m\n}\ket\}_{\nu=1}^{\dm}$
of $\dm$ states and yields a single entangled isomers' position state,
\begin{subequations}
\label{eq:entangled-position-states}
\begin{align}
|\cs,\mu\ket & \equiv\frac{1}{\sqrt{\dm}}\sum_{\n=1}^{\dm}s_{\n}|\cs;\l^{\m\n}\ket_{\rot}|\nu\ket_{\nat}\\
 & =\sum_{\ell\down\l}\sum_{|m|\leq\ell}{\textstyle \sqrt{\frac{(2\ell+1)/\dm}{8\pi^{2}/|\G|}}}\sum_{\k=1}^{\mult(\ell)}D_{m;\m\k}^{\ell}(\cs)|_{m\k}^{\ell}\ket\,,
\end{align}
\end{subequations}
for each $\cs$ and $\mu$. While the $\nu$-labeled degree of freedom
has been eliminated due to the enforcement of spin statistics, the
$\m$-labeled degree of freedom remains!

Together with the nuclear basis states indexing copies of $\t$, the
identity on an entangled isomer $\sp$ decomposes in terms of position
states as
\begin{align}
\id_{\mol}^{\G,\sp} & =\int_{\nicefrac{\so 3}{\G}}\dd\cs\sum_{\mu=1}^{\dm}|\cs,\mu\ket\bra\cs,\mu|\ot\sum_{\chi=1}^{\mst}|\chi\ket\bra\chi|\,,\label{eq:id-pos-rotnuc}
\end{align}
where the tensor product \textit{is not} between the rotational and nuclear-spin factors since the molecular position states contain a nuclear-spin component.
Each molecular
position $\cs\in\nicefrac{\so 3}{\G}$ carries with it, not only the requisite
$\mst$-dimensional nuclear multiplicity space, but also an internal
$\dm$-dimensional pseudo-spin factor, spanned by $\{|\mu\ket\}_{\mu=1}^{\dm}$,
that we call the isomer's \textit{fiber}.

Position states and their fibers form \vva{an} orthonormal ``basis'' for the
isomer, satisfying
\begin{equation}\label{eq:ortho-rotnuc}
\bra\cs,\mu|\cs^{\pr},\mu^{\pr}\ket=\d_{\m\mu^{\pr}}\d^{\nicefrac{\so 3}{\G}}(\cs,\cs^{\pr})\,.
\end{equation}
Orthogonality and completeness can be proven by observing that the uncoupled position states \(\{|\cs;\ir^{\m\n}\ket\}\) from Eq.~\eqref{eq:Zak} --- when collected over all irreps, matrix elements, and coset representatives --- form an orthonormal ``basis'' for the state space of an asymmetric molecule that is a special case of the \textit{Zak basis} \cite{zak_finite_1967,justel_zak_2018,albert_robust_2020,culf_group_2022}.
Orthogonality and completeness relations for the Zak basis \cite[Eq.~(125)]{albert_robust_2020} imply analogous relations for the coupled states from Eqs.~\eqref{eq:id-pos-rotnuc} and \eqref{eq:ortho-rotnuc}.

The reverse Fourier transform expressing rotational states $|_{m\k}^{\ell}\ket$
in terms of position states $|\cs,\mu\ket$ can be derived as follows.
We start with expression (\ref{eq:asym-adapted-fourier-momentum})
of the uncoupled rotational states for $\ir=\l$ in terms of an integral superposition over the asymmetric
position states,
\begin{equation}
    |_{m\k}^{\ell},\nu\ket={\textstyle \sqrt{\frac{2\ell+1}{8\pi^{2}}}}\int_{\nicefrac{\so 3}{\G}}\dd\cs\sum_{\rg\in\G}D_{m;\nu\k}^{\ell\star}(\cs\rg)|\gr=\cs\rg\ket\,,
\end{equation}
where we have split the integral into one over the coset space and a sum over the symmetry group according to Eq.~\eqref{eq:coset-partition}.

The product \(\cs\rg\) inside the matrix element of \(D^{\ell}\) can be split such that Eq.~\eqref{eq:evaluating-in-irrep} can be applied.
This yields
\begin{align}
    |_{m\k}^{\ell},\nu\ket&={\textstyle \sqrt{\frac{2\ell+1}{8\pi^{2}}}}\int_{\nicefrac{\so 3}{\G}}\dd\cs\sum_{\rg\in\G}\sum_{\mu=1}^{\dm}D_{m;\mu\k}^{\ell\star}(\cs)\l^{\mu\nu}(\rg)|\gr\ket\nonumber\\
    &={\textstyle \sqrt{\frac{(2\ell+1)/\dm}{8\pi^{2}/|\G|}}}\int_{\nicefrac{\so 3}{\G}}\dd\cs\sum_{\mu=1}^{\dm}D_{m;\m\k}^{\ell\star}(\cs)|\cs;\l^{\mu\nu}\ket\,.
\end{align}
We then plug in the above into Eq.~(\ref{eq:entangled-state-rotational}),
eliminating the $\nu$ index, and use Eq.~(\ref{eq:entangled-position-states})
to obtain
\begin{equation}\label{eq:entangled-momentum-isomer-in-position}
|_{m\k}^{\ell}\ket={\textstyle \sqrt{\frac{(2\ell+1)/\dm}{8\pi^{2}/|\G|}}}\int_{\nicefrac{\so 3}{\G}}\dd\cs\sum_{\mu=1}^{\dm}D_{m;\m\k}^{\ell\star}(\cs)|\cs,\mu\ket\,,
\end{equation}
completing the Fourier transform for entangled isomers.

Absorbing the constants into a \(\G\)-adapted ``harmonic'',
\begin{equation}\label{eq:harmonic}
    H_{m\kappa}^{\vva*{\ell}}(\cs,\mu)={\textstyle \sqrt{\frac{(2\ell+1)/\dm}{8\pi^{2}/|\G|}}}D_{m;\m\k}^{\ell}(\cs)\,,
\end{equation}
yields Eq.~\eqref{eq:fourier} from Sec.~\ref{sec:summary}.
\nw{
The \(\G\)-adapted orthogonality and completeness relations become
\begin{subequations}\label{eq:explicit-orthocompleteness}
\begin{align}
   {\displaystyle \int_{\nicefrac{\so 3}{\G}}\dd\cs\sum_{\m=1}^{\dm}H_{m\k}^{\ell}(\cs,\m)H_{m^{\pr}\k^{\pr}}^{\ell^{\pr}\star}(\cs,\m)}&=\d_{\ell\ell^{\pr}}\d_{mm^{\pr}}\d_{\k\k^{\pr}}\\{\displaystyle \sum_{\ell\downarrow\l}\sum_{|m|\leq\ell}\sum_{\k=1}^{\mult(\ell)}H_{m\k}^{\ell}(\cs,\m)H_{m\k}^{\ell\star}(\cs^{\pr},\m^{\pr})}&=\d_{\m\m^{\pr}}\d^{\nicefrac{\so 3}{\G}}(\cs,\cs^{\pr})\,.
\end{align}
\end{subequations}
}

\begin{example}[boron \vva{trifluoride}, \(\e^{\ast}\) isomer]
Recalling Example~\ref{ex:BF3}, this \(\D3\)-symmetric molecule is the simplest example of a rotation-spin entangled isomer.
Position states are labeled by points in the prism space \(\nicefrac{\so 3}{\D 3}\).
Because the isomer is entangled, the molecule's position states \(|\cs,\mu\ket\) have a fiber degree of freedom, with internal index \(\mu\in\{1,2\}\).

The isomer's Fourier transform is in Eqs.~\eqref{eq:entangled-position-states} and \eqref{eq:entangled-momentum-isomer-in-position}, with \(\G=\D3\).
We will see in the next section that, since the fiber degree of freedom comes from the rotational state space, it will transform in a nontrivial way under lab-based rotations.
\end{example}

\section{Holonomy of position states}
\label{sec:holonomy-position}

In Sec.~\ref{sec:position}, we developed admissible position states of rotationally and perrotationally symmetric nuclear spin isomers and obtained the rotation-spin states $|\cs,\mu\ket$, tensored with a nuclear-spin factor spanned by appropriately transforming nuclear-spin states $|\chi\ket$.
Here, we derive the holonomy of such states arising from a closed adiabatic path in position state space, showing that it depends only on the ``global'' or ``topological'' details of the path.
\vva{We also determine, in terms of the fundamental group of the position-state space, which induced representations of \(\so 3\) admit a locally flat Berry connection.}

We omit the residual \(|\chi\ket\) factor in the position-state expressions from now on because it is decoupled from and not relevant to the calculation.
Since we deal only with the (dressed) rotational factor, we also drop the ``rot'' subscript, i.e, \(\l\to\Gamma\).

\subsection{Induced representations}

We derive the holonomy only for the case of a general entangled perrotational isomer.
\vva{Its} position states are
\begin{equation}
\big\{~|\cs,\m\ket\,\,\text{s.t.}\,\,\cs\in\nicefrac{\so 3}{\G}\text{ and }\mu\in\{1,\cdots,\dm\}~\big\}\,,\label{eq:position-states-summary}
\end{equation}
where $\m$ labels the internal degrees of freedom of the rotational
$\G$-irrep $\Gamma$, and where \(\dm = \dim\Gamma\).
This general case can then be specialized to all other cases of interest.
States of separable isomer are obtained by setting $\Gamma$ to a one-dimensional irrep and removing
the $\m$ index.
Any rotationally symmetric
case is obtained by setting $\Gamma$ to be the trivial irrep.

Interpreting the above state set from a geometric perspective, each
point $\cs\in\nicefrac{\so 3}{\G}$ can be thought of as housing the fiber's \(\dm\)-dimensional vector space.
Such a space transforms under lab-based rotations \(\rr_\rg\) according to what is known
as the induced representation, $\Gamma\up\so{3}$ \cite{mackey_induced_1952,mackey1968induced,coleman_induced_1968,Inui_group_1990,carter_lectures_1995,chirikjian_engineering_2000,folland_course_2016}.

For all trivial-irrep cases considered here, the vector space has
effectively zero dimension, and each state $|\cs\ket$ can be thought
of as a point. This corresponds to the representation $\irt\up\so{3}$,
encompassing the $\G$-symmetric cases and the $\G=\C 1$ asymmetric
case.

The $\dm=1$ case for a nontrivial one-dimensional irrep $\Gamma$
yields $\Gamma\up\so{3}$, corresponding to separable isomers
of perrotationally symmetric molecules.
In that case, the space is one-dimensional, and lab-based rotations change the global phase of the position states.

The $\dm>1$ case corresponds
to an entangled isomer, and the fiber degrees of freedom transform under lab-based symmetry rotations according to the inducing irrep \cite[Eq.~(131)]{albert_robust_2020}.
More generally, assuming sufficiently slow (i.e., adiabatic) movement along \textit{any} path, the resulting transformations from symmetry rotations depend only on the ``global'' details of the path and not on any small deformations \vva{(see Theorem~\ref{thm:flat})}.

\subsection{Non-Abelian connection and holonomy}

Consider evolving in an adiabatic path in the coset space $\nicefrac{\so 3}{\G}$
of the set of states above parameterized by Eq.~\eqref{eq:position-states-summary}.
We determine the
holonomy for all closed (piece-wise) continuous paths,
\begin{equation}
\{~\cs(t)\,~\text{s.t.}~\,t\in[0,1]~\}\,.
\end{equation}
By closed, we mean that the final point $\cs(1)$ is identified in
the coset space with the initial point $\cs(0)$ because of the symmetry of the molecule.
Following Sec.~\ref{sec:asymmetric-holonomy}, we parameterize the path using the same coset-state parameterization for all points except, potentially, the last point \(\cs(1)\).

Upon traversing a closed path, the molecular basis state $|\cs,\mu\ket$
undergoes a holonomy,
\begin{equation}
|\cs(0),\mu\ket\quad\to\quad U_{\hol}|\cs(0),\mu\ket\,.
\end{equation}
Since there is a vector space for each $\cs$, the holonomy is a $\dm$-dimensional
unitary Berry-Wilczek-Zee matrix \cite{wilczek_appearance_1984}. Generalizing Eq.~\eqref{eq:holonomy-abelian},
\begin{equation}
U_{\hol}=U_{\mon}\mathcal{P}\exp\left(i\int_{\cs(0)}^{\cs(1)}\dd\cs\,A(\cs)\right)\,.\label{eq:holonomy-nonabelian-matrix}
\end{equation}

The second term on the right-hand side is the path-ordered integral
of exponentials of the Berry-Wilczek-Zee connection $A(\cs)$ (we call this the Berry connection from now on). The
connection is now matrix-valued, with elements
\begin{equation}
A_{\mu\nu}(\cs)=\frac{i\bra\cs,\mu|\p\cs,\nu\ket}{\sqrt{\bra\cs,\mu|\cs,\mu\ket\bra\cs,\nu|\cs,\nu\ket}}\,,\label{eq:nonabelian-connection}
\end{equation}
all defined for points \(\cs(t<1)\).
Path-ordering, denoted by \(\cal P\), means that matrix exponentials of $A$ earlier in the
path $\cs(t)$ are written before those later in the path.
\vva{Since position states are not normalizable, we evaluate Eq.~\eqref{eq:nonabelian-connection} on normalizable approximate states with a regularization parameter \(\Delta>0\).
We then define the position-state connection as the \(\Delta\to0\) limit (see Sec.~\ref{subsec:approximate-states}).}

The monodromy $U_{\mon}$ is a compensating matrix that is the result of re-expressing the label $\cs(1)$ of the final state in terms of the label
$\cs(0)$ of the initial state.
The monodromy is only an issue when the path
is non-contractible, but such paths exist in $\nicefrac{\so 3}{\G}$ for any
$\G$ that admits perrotationally symmetric molecules (see Tab.~\ref{tab:isomer-table}).

The set of holonomies for closed paths starting and ending at a particular
\textit{base point} $\cs(0)$ forms a group called the \textit{holonomy group}
at that point. It is sufficient to know the holonomy group at the identity
base point, $\cs(0)=\re$, to obtain the group at all other
basepoints. This is because the space $\so 3$ admits a path connecting any two points,
so if we know the holonomy group at $\re$, we can obtain
other holonomy groups by shifting via lab-based rotations
$\rr_{\rg}$ (for some $\rg$) to some other desired
base point
\cite[Eq.~(16)]{gottesman_fibre_2017}.

\vva{We prove in Theorem~\ref{thm:flat} that the position-state Berry connection
is zero for all isomers considered in this work.}
Given such a locally flat connection, the holonomy group
is equal to the \textit{monodromy group} --- the group formed by
all monodromies.
\vva{We calculate this group in the next subsection and summarize it in Table~\ref{tab:monodromy}.}

\begin{table}
\begin{centering}
\begin{tabular}{cclcc}
\toprule
\multicolumn{3}{c}{molecule~~~~~} & \multicolumn{2}{c}{monodromy}\tabularnewline
symmetry & isomer & $\dm$~~~~ & ~~$\G_{\mon}$~~ & non-Ab.\tabularnewline
\midrule
\multirow{3}{*}{$\C N$} & $\a,\a^{\ast}$ & $1$ & $\C 1$ & \tabularnewline
 & $\b,\b^{\ast}$ & $1$ & $\C 2$ & \tabularnewline
 & $^{j}\e_{i},{}^{j}\e_{i}^{\ast}$ & $1$ & $\C M$ & \tabularnewline
\midrule
\multirow{5}{*}{$\D N$} & $\a,\a_{1},\a_{1}^{\ast}$ & $1$ & $\C 1$ & \tabularnewline
 & $\a_{2},\a_{2}^{\ast}$ & $1$ & $\C 2$ & \tabularnewline
 & $\b_{i},\b_{i}^{\ast}$ & $1$ & $\C 2$ & \tabularnewline
 & $\e_{1},\e_{1}^{\ast}$ & $2$ & $\D N$ & $\largestar$\tabularnewline
 & $\e_{i>1},\e_{i>1}^{\ast}$ & $2$ & $\D M$ & $\filledlargestar$\tabularnewline
\midrule
\multirow{2}{*}{$\D{\infty}$} & $\a_{1},\a_{1}^{\ast}$ & $1$ & $\C 1$ & \tabularnewline
 & $\a_{2},\a_{2}^{\ast}$ & $1$ & $\C 2$ & \tabularnewline
\midrule
\multirow{3}{*}{$\T$} & $\phantom{^{j}}\a\phantom{_{i}}$ & $1$ & $\C 1$ & \tabularnewline
 & $^{j}\e\phantom{_{i}}$ & $1$ & $\C 3$ & \tabularnewline
 & $\phantom{^{j}}\ti\phantom{_{i}}$ & $3$ & $\T$ & $\largestar$\tabularnewline
\midrule
\multirow{5}{*}{$\O$} & $\a_{1},\a_{1}^{\ast}$ & $1$ & $\C 1$ & \tabularnewline
 & $\a_{2},\a_{2}^{\ast}$ & $1$ & $\C 2$ & \tabularnewline
 & $\e,\e^{\ast}$ & $2$ & $\D 3$ & $\filledlargestar$\tabularnewline
 & $\ti_{1},\ti_{1}^{\ast}$ & $3$ & $\O$ & $\largestar$\tabularnewline
 & $\ti_{2},\ti_{2}^{\ast}$ & $3$ & $\O$ & $\filledlargestar$\tabularnewline
\midrule
\multirow{5}{*}{$\I$} & $\a\phantom{_{i}}$ & $1$ & $\C 1$ & \tabularnewline
 & $\ti_{1}$ & $3$ & $\I$ & $\largestar$\tabularnewline
 & $\ti_{2}$ & $3$ & $\I$ & $\filledlargestar$\tabularnewline
 & $\g\phantom{_{i}}$ & $4$ & $\I$ & $\filledlargestar$\tabularnewline
 & $\h\phantom{_{i}}$ & $5$ & $\I$ & $\filledlargestar$\tabularnewline
\bottomrule
\end{tabular}
\par\end{centering}
\caption{\label{tab:monodromy}Table listing \vva{the} monodromy
group $\protect\G_{\protect\mon}$ (\ref{eq:monodromy-group}) for
the perrotationally symmetric nuclear spin isomers from Table \ref{tab:isomer-table}.
\vva{The position-state Berry connection is zero for all of them (see Theorem~\ref{thm:flat}), so each monodromy group is also the isomer's holonomy group.}
The isomer $\protect\sp^{\ast}$ has the same monodromy group as
$\protect\sp$ since the rotational states of both transform under
the same induced representation, $\protect\sp\protect\up\protect\so{3}$.
The parameter $M=\text{GCD}(N,i)$ for the groups $\protect\C M$
and $\protect\D M$.
\nw{Isomers with a star in the last column have non-Abelian monodromy groups, while the fibers of those with a filled-in star also form protected encodings (see Sec.~\ref{sec:fiber-codes}).
}
}
\end{table}

\subsection{Nontrivial monodromy\label{subsec:Monodromy}}

Fixing $\cs(0)=\re$, we begin with the ``identity'' coset subspace from Eq.~\eqref{eq:entangled-position-states},
\begin{equation}
\left|\cs=\re,\mu\right\rangle =\sum_{\ell\down\Gamma}\sum_{|m|\leq\ell}{\textstyle \sqrt{\frac{(2\ell+1)/\dm}{8\pi^{2}/|\G|}}}\sum_{\k=1}^{\mult(\ell)}D_{m;\m\k}^{\ell}(\re)|_{m\k}^{\ell}\ket\,,\label{eq:initial-coset-state}
\end{equation}
where $\m\in\{1,2,\cdots,\dm\}$.
We consider paths obtained by applying lab-based rotations in the
symmetry group, i.e., $\rr_{\rg}$ for any $\rg\in\G$. Such rotations
parameterize a ``direct'' path (w.r.t.~some metric on the group) from $\cs(0)=\re$ to $\cs(1)=\rg$.
Since the underlying molecule is $\G$-symmetric, such paths are in
fact closed in the coset space despite $\rg\neq\re$.

For example, a $\pi$-rotation around an axis perpendicular to the primary
axis of \ce{H2} exchanges the molecule's nuclei and yields a non-contractible
path in its position state space, $\nicefrac{\so 3}{\D \infty}=\rptwo$.
Such a path would be considered open if the nuclei were distinguishable,
but it is instead closed in the state space because the nuclei are
\textit{in}distinguishable.
This rotation maps the spatial coordinates of one nucleus to those of the other, but acts purely on the rotational factor and \textit{does not} permute any nuclear-spin factors.

\vva{Since the connection $A=0$ (see Theorem~\ref{thm:flat}), the holonomy equals the monodromy, which is independent of smooth
path deformations} \cite{read_non-abelian_2009,gottesman_fibre_2017}.
\vva{The monodromy for any other path between $\re$ and $\rg$ is thus the same as that for the ``direct''
path from $\re$ to $\rg$.}

Applying $\rr_{\rg}$ to the state from Eq.~(\ref{eq:initial-coset-state}),
we recall from Eq.~\eqref{eq:active-rot} that lab-based rotations act on the $|_{m}^{\ell}\ket$
factor of each rotational state,
\begin{equation}
\rr_{\rg}|_{m\k}^{\ell}\ket=\sum_{|n|\leq\ell}D_{nm}^{\ell}(\rg)|_{n\k}^{\ell}\ket\,.
\end{equation}
Plugging this in and simplifying yields
\begin{align}
\!\!\!\rr_{\rg}\left|\re,\mu\right\rangle  & =\sum_{\ell\down\Gamma}{\textstyle \sqrt{\frac{(2\ell+1)/\dm}{8\pi^{2}/|\G|}}}\sum_{\k=1}^{\mult(\ell)}\sum_{n,m}D_{nm}^{\ell}(\rg)D_{m;\m\k}^{\ell}(\re)|_{n\k}^{\ell}\ket\nonumber \\
 & =\sum_{\ell\down\Gamma}\sum_{|n|\leq\ell}{\textstyle \sqrt{\frac{(2\ell+1)/\dm}{8\pi^{2}/|\G|}}}\sum_{\k=1}^{\mult(\ell)}D_{n;\mu\k}^{\ell}(\rg)|_{n\k}^{\ell}\ket\nonumber \\
 & =\left|\rg,\mu\right\rangle \,.
\end{align}

Since \(\rg\) is a symmetry rotation, both $\left|\re,\mu\right\rangle $
and $\left|\rg,\mu\right\rangle $ describe a molecule in the same
physical position. Mathematically, both $\rg$ and $\re$ are in the same (identity) coset, $\re\G = \{\rh\in\G\}$.

However, the $\left|\rg,\mu\right\rangle $ state is representing
the coset by $\rg$, while the $\left|\re,\mu\right\rangle $ state
is using the identity $\re$. We have to pick \textit{one} parameterization
to be consistent with our expression of the holonomy, and we pick
the representative of each coset to be the element closest to the
identity (w.r.t.~to some metric on the group). Given such
a parameterization, we have to re-express $\left|\rg,\mu\right\rangle $
in terms of $\left|\re,\mu\right\rangle $.

Applying Eq.~\eqref{eq:evaluating-in-irrep} and using the fact that $\Gamma^{\nu\mu\star}(\rg)=\Gamma^{\mu\nu}(\rg^{-1})$, we obtain
\begin{align}\label{eq:monodromy-identity}
\left|\rg,\mu\right\rangle  & =\sum_{\ell\down\Gamma}\sum_{|n|\leq\ell}{\textstyle \sqrt{\frac{(2\ell+1)/\dm}{8\pi^{2}/|\G|}}}\sum_{\k=1}^{\mult(\ell)}\sum_{\nu=1}^{\dm}D_{n;\nu\k}^{\ell}(\re)\Gamma^{\nu\mu\star}(\rg)|_{n\k}^{\ell}\ket\nonumber \\
 & =\sum_{\nu=1}^{\dm}\Gamma^{\mu\nu}(\rg^{-1})\left|\re,\nu\right\rangle \,,
\end{align}
resulting in an operation on the fiber degrees of freedom $\{|\mu\ket\}$.
This change in the internal state is precisely the monodromy, which
in this case realizes the $\Gamma$ irrep of the symmetry group.

The above is true for any $\rg\in\G$, so there are at most $|\G|$
different types of monodromies. The actual number of distinct monodromies
is equal to the number of distinct irrep elements.
For example, in the case of rotationally symmetric molecules,
all monodromies are $+1$.

\begin{example}[rotationally symmetric molecules]\label{ex:rot-symmetric-holonomy}
In these cases, $\G$ is any group, but the induced representation
is induced by the trivial irrep, $\Gamma=\irt$. The states $\left|\cs,\mu\right\rangle $
reduce to the coset states $|\cs\ket$ from Eq.~\eqref{eq:coset-states}. The fiber is
one-dimensional ($\dm=1$), so there is no $\m$ index. Since $\Gamma(\rg)=\irt(\rg)=1$
for any $\rg$, the states $|\cs=\rg\ket$ and $|\cs=\re\ket$ are
directly identified, and there is no monodromy.

The above applies also to the asymmetric case, for which $\G=\C 1$,
and any perrotationally symmetric isomer for which $\Gamma$ is the trivial
irrep. Hence, the monodromy group in the first line of each symmetry
group in Table \ref{tab:monodromy} is the trivial group, $\G_{\mon}=\C 1$.
\end{example}

On the other hand, perrotationally symmetric nuclear spin isomers exhibit
nontrivial monodromy for all $\Gamma$ but the trivial one. Their monodromy
group is the group formed by the distinct elements $\{\Gamma(\rg)\,,\,\rg\in\G\}$.
This \vva{group is isomorphic to} the quotient
of $\G$ by the kernel of $\Gamma$,
\begin{equation}
\G_{\mon}=\G/\ker\Gamma\,,\label{eq:monodromy-group}
\end{equation}
where $\ker\Gamma$ is the subset of elements of $\G$ which evaluate
to the identity in the $\Gamma$ irrep.

\begin{example}[water, $\b^{\ast}$ ortho isomer]\label{ex:triplet-water-monodromy}
This isomer, with symmetry group $\G=\C 2$ and irrep $\Gamma=\b$, admits
the simplest nontrivial monodromy. The irrep is one-dimensional, so
there is no $\m$ index. The only non-identity symmetry rotation,
$\rg=(00\pi)$ in the Euler-angle prescription,
exchanges the two hydrogen nuclei and evaluates to $-1$ in the $\b$
irrep (see Example~\ref{ex:triplet-water-position}).
This yields a monodromy of $-1$, which is not present in the
para isomer since $\Gamma$ is trivial for that isomer. The monodromy
group is the symmetry group itself, $\G_{\mon}=\C 2$.

The $-1$ monodromy can alternatively be derived using the coset-state
expression from Eq.~\eqref{eq:water-coset}. Beginning with the identity
coset state, $\cs=\re=(000)$, and applying the symmetry rotation yields
\begin{align}
\rr_{(00\pi)}\left|(000)\right\rangle  & =\rr_{(00\pi)}{\textstyle \frac{1}{\sqrt{2}}}(\left|\gr=(000)\right\rangle -\left|\gr=(00\pi)\right\rangle )\nonumber \\
 & ={\textstyle \frac{1}{\sqrt{2}}}(\left|\gr=(00\pi)\right\rangle -\left|\gr=(000)\right\rangle )\nonumber \\
 & =-\left|(000)\right\rangle \,.
\end{align}
Above, we continue to abuse notation and use a rotation's Euler-angle parameterization as a stand-in for the rotation \(\gr\) (or \(\cs\)) itself.
\end{example}

\begin{example}[deuterated hydrogen, $\a_{2}$ para isomer]
\label{ex:deuterium-a2-monodromy}
The coset states of this $\D{\infty}$-symmetric molecule are labeled
by points in the projective plane, $\nicefrac{\so 3}{\D \infty}=\rptwo$
(see Example~\ref{ex:deuterium-a2-position}). The $\a_{2}$ irrep
is one-dimensional, so there is no $\m$ index in the position states
$|\cs\ket$ {[}see Eq.~\eqref{eq:deuterium-a1-position}{]}.

A $\pi$-rotation around any equatorial axis yields a monodromy
of $-1$, while the $\C{\infty}$ subgroup of $\z$-axis rotations
is mapped to $+1$. The monodromy group is $\G_{\mon}=\D{\infty}/\C{\infty}=\C 2$.
This is the same monodromy group as that for the $\a_{2}^{\star}$
(ortho) isomers of ordinary \ce{H2} since the rotational state
subspace --- $\a_{2}\up\so{3}$ --- is the same
for both isomers.

We easily extract the monodromy using momentum states. The identity position state from Eq.~\eqref{eq:deuterium-a1-position} is
\begin{equation}
    \left|\cs=\re\right\rangle =\sum_{\ell\,\text{odd}}{\textstyle \sqrt{\frac{2\ell+1}{2\pi}}}~|_{0}^{\ell}\ket~.
\end{equation}
Applying a \(\pi\)-rotation around the \(\y\) axis and using the fact that \(\rr_{(0\pi0)}|_{0}^{\ell}\ket=\left(-1\right)^{\ell}|_{0}^{\ell}\ket\) yields the \(-1\) phase.
A \(\pi\)-rotation around any equatorial axis yields the same phase since such a rotation can be expressed as a product of \(\rr_{(0\pi0)}\) and a rotation around the \(\z\) axis.

In contrast, position states of ortho deuterium and para hydrogen are of the same form as above, except that only even momenta are present.
The monodromy for those cases is trivial.
\end{example}

\begin{example}[ammonia, $^{2}\e$ isomer]\label{ex:ammonia-monodromy}
This isomer, with symmetry group $\G=\C 3$ and irrep $\Gamma={}^{2}\e$,
is the simplest to realize a monodromy that is not $\pm1$, i.e.,
that is not related to Bose/Fermi spin statistics.

The $^{2}\e$ irrep is one-dimensional, so there is no $\m$ index
in the position states $|\cs\ket$ {[}see Eq.~\eqref{eq:deuterium-a1-position}{]}.
The irrep evaluates to third roots of unity and has a trivial kernel,
yielding the monodromy group $\G_{\mon}=\C 3$. The $\cs=\re=(000)$
position state transforms under a symmetry rotation as
\begin{align}
\rr_{00\frac{2\pi}{3}}\left|(000)\right\rangle  & =\sum_{p=0}^{2}e^{-i\frac{2\pi}{3}p}\left|\gr=(0,0,{\textstyle \frac{2\pi}{3}}p-{\textstyle \frac{2\pi}{3}})\right\rangle \nonumber \\
 & =e^{-i\frac{2\pi}{3}}\left|(000)\right\rangle \,.
\end{align}
\end{example}

\begin{example}[boron trifluoride, \vva{$\e^{\ast}$} isomer]\label{ex:bf3-holonomy}
    Per Example~\ref{ex:BF3}, this perrotationally \(\D3\)-symmetric isomer admits a \(\dm=2\)-dimensional fiber --- the simplest fiber with non-unity dimension.
    Rotations permuting its three fluorine nuclei realize the group's two-dimensional irrep $\e$ from Eq.~\eqref{eq:D3-2D-irrep} via monodromy.
\end{example}

\begin{example}[methane, $\ti$ isomer]\label{ex:methane-holonomy}
    Per Example~\ref{ex:methane-e}, the position state space of methane is parameterized by the octahedral space \(\nicefrac{\so 3}{\T}\), with fiber degrees of freedom \(|\m\ket\) with \(\mu\in\{1,2,3\}\).
    Rotations permuting its four hydrogen nuclei realize the group's three-dimensional irrep $\ti$ via monodromy.
\end{example}

\begin{example}[\ce{$^{13}$C60} fullerene, $\h$ isomer]\label{ex:fullerene-h-holonomy}
    Per Example~\ref{ex:fullerene-h}, the position state space of isotopic fullerene is parameterized by the \vva{Poincar\'e} dodecahedral space \(\nicefrac{\so 3}{\I}\), with fiber degrees of freedom \(|\m\ket\) with \(\mu\in\{1,2,3,4,5\}\).
    Rotations permuting its sixty \vva{\ce{^{13}C}} nuclei realize the group's five-dimensional irrep $\h$ via monodromy.
    This irrep has a trivial kernel, so the monodromy group is \(\I\).

    The remaining isomers, with exception of the trivial-irrep isomer, also realize the entire icosahedral group as their monodromy group.
    This is because this group has no normal subgroups other than itself and the trivial group, which implies that every non-trivial irrep must have trivial kernel.
\end{example}

\subsection{Locally flat connection\label{subsec:Connection}}

\vva{We now determine when the position-state connection~\eqref{eq:nonabelian-connection} is zero.}
\begin{theorem}\label{thm:flat}
\vva{Let \(\G\) be a proper closed subgroup of \(\so 3\), let \(\Gamma\) be a \(\G\)-irrep, and let \(A\) be the position-state Berry connection on \(\Gamma\up\so 3\) in the coset parameterization of Eq.~\eqref{eq:entangled-position-states}.
The following are equivalent: (i) \(A=0\), (ii) the curvature \(F=\dd A-iA\wedge A\) is zero, (iii) \(\Gamma\) is trivial on the identity component of \(\G\), and (iv) there exists a \(\pi_{1}(\nicefrac{\so 3}{\G})\)-irrep \(\rho\) such that
\begin{equation}\label{eq:mathcup}
\frac{\G}{\ker\Gamma}\cong\frac{\pi_{1}(\nicefrac{\so 3}{\G})}{\ker\rho}~.
\end{equation}
If \(\G\) is finite, these conditions hold for every \(\Gamma\), and the connection of the approximate states~\eqref{eq:approx-states} vanishes as \(\Delta^{-1}e^{-\phi_{\G}^{2}/(4\Delta)}\) or faster when \(\Delta\to0\).
Here, \(\phi_{\G}\) is the smallest rotation angle in \(\G\) other than \(0\) and \(\pi\), with \(\phi_{\G}=\infty\) if there is none.}
\end{theorem}
\vva{Every isomer in Table~\ref{tab:monodromy} satisfies condition (iii), since its symmetry group is either finite or \(\D\infty\) with \(\Gamma\in\{\a_{1},\a_{2}\}\).
Thus, each of these isomers has a zero connection.
We compute the connection for finite and continuous \(\G\) in Secs.~\ref{subsec:connection-finite} and \ref{subsec:connection-continuous}, respectively, and complete the proof in Sec.~\ref{subsec:fundamental-group}.}

\subsubsection{Approximate states\label{subsec:approximate-states}}

\vva{Regularization} proceeds by applying a damping operator to the original position states,
\begin{equation}
|\cs,\m\ket~\to~ e^{-\Delta\hat{\mathbf{\ell}}^2/2}|\cs,\m\ket\,,
\label{eq:approx-states}
\end{equation}
where the parameter \vva{\(\Delta>0\)}, and where \vva{\(\hat{\mathbf{\ell}}^2=\sum_{\ell,m,\om}\ell(\ell+1)|_{m}^{\ell},\om\ket\bra_{m}^{\ell},\om|\)} is the total angular momentum operator squared.
The approximate states \vva{are} straightforward generalizations of finite-energy GKP states \cite{gottesman_encoding_2001,menicucci_fault-tolerant_2014} and their molecular-code extensions \cite{albert_robust_2020,culf_group_2022}.
\vva{This regularization converts the delta-function-like states \(|\cs,\mu\ket\) into wavepackets centered at \(\cs\), which are no longer exactly orthogonal for different \(\cs\).
The overlap of states centered at \(\cs\) and \(\cs^{\pr}\) is a weighted sum of the function \(h_{\Delta}\) from Eq.~\eqref{eq:heat-kernel} at the rotation angles of \(\cs^{-1}\cs^{\pr}\rg\) for \(\rg\in\G\).
By Eq.~\eqref{eq:poisson}, the overlap of the normalized states is at most a constant times \(e^{-\phi^{2}/(4\Delta)}\), where \(\phi\) is the smallest such angle (cf.~\cite{albert_robust_2020}).
For continuous \(\G\), the sum becomes an integral, and the bound acquires an extra factor of \(\Delta^{-1/2}\).}
The damping operator commutes with all rotations because it is a function of the total angular momentum, so \vva{approximate states undergo the same monodromies~\eqref{eq:monodromy-identity} as the ideal states}.

\vva{For approximate states, each momentum \(\ell\) in the connection~\eqref{eq:nonabelian-connection} carries the weight \(c_{\ell}=e^{-\Delta\ell(\ell+1)}(2\ell+1)|\G|/(8\pi^{2}\dm)\).
The fiber states at each \(\cs\) remain orthogonal,
\begin{align}
\bra\cs,\m|e^{-\Delta\hat{\mathbf{\ell}}^2}|\cs,\n\ket & =\sum_{\ell\down\Gamma}c_{\ell}\sum_{\k=1}^{\mult(\ell)}\bra\mu\k|D^{\ell\dg}(\cs)D^{\ell}(\cs)|\n\k\ket\nonumber\\
 & =\d_{\m\n}\sum_{\ell\down\Gamma}c_{\ell}\,\mult(\ell)\,,\label{eq:connection-normalization}
\end{align}
and the numerator simplifies to}
\begin{subequations}
\label{eq:connection-me}
\begin{align}
\bra\cs,\mu|\vva*{e^{-\Delta\hat{\mathbf{\ell}}^2}|}\p\cs,\nu\ket & =\sum_{\ell\down\Gamma}\vva*{c_{\ell}}\sum_{\k=1}^{\mult(\ell)}\bra\mu\k|D^{\ell\dg}(\cs)\p D^{\ell}(\cs)|\n\k\ket\\
 & =-i\sum_{\aa}\p w_{\aa}(\cs)\sum_{\ell\down\Gamma}\vva*{c_{\ell}}\sum_{\k=1}^{\mult(\ell)}\bra\mu\k|L_{\aa}^{\ell}|\n\k\ket\,,
\end{align}
\end{subequations}
where
we expand the angular velocity matrix
$D^{\dg}\p D$ in terms of \(\so 3\) angular momentum generators $L_{\aa}^{\ell}$
with $\aa\in\{\x,\y,\z\}$ and coordinate vector $\p w_{\aa}(\cs)$.

Plugging the numerator (\ref{eq:connection-me}) and denominator (\ref{eq:connection-normalization})
into the connection Eq.~(\ref{eq:nonabelian-connection}) yields the
expansion
\begin{equation}
A_{\m\n}^{\vva*{(\Delta)}}(\cs)={\textstyle \frac{\dm}{|\G|}}\sum_{\aa}\p w_{\aa}(\cs)\frac{\sum_{\ell\down\Gamma}\vva*{e^{-\Delta\ell(\ell+1)}}\left(2\ell+1\right)A_{\m\n}^{\aa,\ell}}{\sum_{\ell\down\Gamma}\vva*{e^{-\Delta\ell(\ell+1)}}\left(2\ell+1\right)\mult(\ell)}\,,\label{eq:full-connection}
\end{equation}
with two expressions for the $\ell$-dependent component,
\begin{subequations}
\begin{align}
A_{\m\n}^{\aa,\ell} & =\frac{|\G|}{\dm}\sum_{\k=1}^{\mult(\ell)}\bra\mu\k|L_{\aa}^{\ell}|\n\k\ket\label{eq:component-sum}\\
 & =\sum_{\rg\in\G}\Gamma^{\n\m}(\rg)\tr\left(D^{\ell}(\rg)L_{\aa}^{\ell}\right)\,.\label{eq:component-projection}
\end{align}
\end{subequations}
The second expression is obtained by expressing the sum over $\k$
as a projection onto the $\nu,\m$th matrix element of all copies
of $\Gamma$ present for a given angular momentum \cite[Eq.~(2.63)]{arovas_lecture_nodate}\cite[Eq.~(5.5.20)]{harter_principles_1993}.
\vva{Both sums in Eq.~\eqref{eq:full-connection} converge, and the position-state connection is \(A=\lim_{\Delta\to0}A^{(\Delta)}\).}

\vva{Schur's lemma implies that \(A^{\aa,\ell}=0\) for every \(\ell\) whenever \(\Gamma\ot\Gamma^{\star}\) shares no irreps with the restriction of the \(\ell=1\) irrep of \(\so 3\) to \(\G\).
For real \(\Gamma\), it suffices that the antisymmetric part of \(\Gamma\ot\Gamma\) shares no such irreps.
One of these two conditions holds for all one-dimensional irreps except the complex ones of \(\C N\) and \(\C\infty\), for the \(\e\) irrep of \(\O\) (cf.~\cite{gross_encoding_2021}), and for the \(\ti_{2}\) irrep of \(\I\) (cf.~\cite{Kubischta_family_2023,Kubischta_free_2024}).
Such per-\(\ell\) selection rules may be applicable to quantum computation with fixed-\(\ell\) subspaces \cite{Pereira_anticoherent_2017,Chryssomalakos_toponomic_2022,Omanakuttan_multispin_2023}.}

\vva{Approximate states are elements of the energy-constrained state space \cite{hayashi2016fourier}.
Their mean squared angular momentum is \(\bra\hat{\mathbf{\ell}}^2\ket=-\p_{\Delta}\ln\sum_{\ell}c_{\ell}\mult(\ell)\), which behaves as \(\frac{3}{2\Delta}\) for finite \(\G\) and as \(\frac{1}{\Delta}\) for continuous \(\G\) when \(\Delta\to0\).}

\subsubsection{\vva{Finite symmetry groups}\label{subsec:connection-finite}}

\vva{Let \(\G\) be finite, and let \(\phi_{\rg}\in[0,\pi]\) and \(\vv_{\rg}\) be the rotation angle and axis of \(\rg\in\G\).
The characters \(\chi_{\ell}(\rg)=\tr D^{\ell}(\rg)=\sin[(\ell+\frac{1}{2})\phi_{\rg}]/\sin\frac{\phi_{\rg}}{2}\) depend only on \(\phi_{\rg}\).
Composing \(\rg\) with a rotation by a small angle \(\epsilon\) around the \(\aa\) axis changes \(\phi_{\rg}\) by \(\epsilon\,\vv_{\rg}\cdot\aa\) to first order when \(0<\phi_{\rg}<\pi\), so
\begin{equation}
\tr\left(D^{\ell}(\rg)L_{\aa}^{\ell}\right)=i\,\chi_{\ell}^{\pr}(\phi_{\rg})\,\vv_{\rg}\cdot\aa\,,\label{eq:trace-derivative}
\end{equation}
where both sides vanish for \(\phi_{\rg}\in\{0,\pi\}\).
Since \(A^{\aa,\ell}=0\) whenever \(\mult(\ell)=0\) by Eq.~\eqref{eq:component-sum}, both sums in Eq.~\eqref{eq:full-connection} can run over all \(\ell\).
Plugging Eq.~\eqref{eq:trace-derivative} into Eq.~\eqref{eq:component-projection} and using \(\mult(\ell)=\frac{1}{|\G|}\sum_{\rg\in\G}\chi_{\Gamma}^{\star}(\rg)\chi_{\ell}(\rg)\) expresses both sums in terms of the heat kernel on \(\so 3\) \cite{hall_segal-bargmann_1994},
\begin{equation}
h_{\Delta}(\phi)=\sum_{\ell\geq0}e^{-\Delta\ell(\ell+1)}\left(2\ell+1\right)\chi_{\ell}(\phi)\,,\label{eq:heat-kernel}
\end{equation}
yielding}
\begin{equation}
\vva*{A_{\m\n}^{(\Delta)}(\cs)=i\dm\sum_{\aa}\p w_{\aa}(\cs)\frac{\sum_{\rg\in\G}\Gamma^{\n\m}(\rg)\,h_{\Delta}^{\pr}(\phi_{\rg})\,\vv_{\rg}\cdot\aa}{\sum_{\rg\in\G}\chi_{\Gamma}^{\star}(\rg)\,h_{\Delta}(\phi_{\rg})}\,.}\label{eq:heat-connection}
\end{equation}

\vva{Since \(h_{\Delta}\) is even and \(2\pi\)-periodic in \(\phi\), \(h_{\Delta}^{\pr}(0)=h_{\Delta}^{\pr}(\pi)=0\).
Thus, neither the identity nor any \(\pi\)-rotation contributes to the numerator.
For the identity, this also follows from the tracelessness of each \(L_{\aa}^{\ell}\), as in Eq.~\eqref{eq:current}.
All other terms are exponentially small, as seen by Poisson summation over \(j=\ell+\frac{1}{2}\),
\begin{equation}
h_{\Delta}(\phi)=\frac{\sqrt{\pi}\,e^{\Delta/4}}{2\Delta^{3/2}\sin\frac{\phi}{2}}\sum_{n\in\Z}(-1)^{n}(\phi+2\pi n)\,e^{-(\phi+2\pi n)^{2}/(4\Delta)}\,.\label{eq:poisson}
\end{equation}
The denominator of Eq.~\eqref{eq:heat-connection} is \(|\G|\) times that of Eq.~\eqref{eq:full-connection}, so it is positive.
It is dominated by the identity term, \(\dm\,h_{\Delta}(0)\simeq\dm\sqrt{\pi}\,\Delta^{-3/2}\), and is thus at least a constant times \(\Delta^{-3/2}\) for \(\Delta\leq1\).
For \(0<\phi<\pi\), the \(n=0\) term dominates Eq.~\eqref{eq:poisson}, so that \(h_{\Delta}^{\pr}(\phi)\simeq-\sqrt{\pi}\,\phi^{2}e^{-\phi^{2}/(4\Delta)}/[4\Delta^{5/2}\sin(\phi/2)]\).
To leading order, unless the sum over \(\rg\) below vanishes, the connection is therefore
\begin{equation}
A_{\m\n}^{(\Delta)}(\cs)\simeq\frac{-i\,\phi_{\G}^{2}\,e^{-\phi_{\G}^{2}/(4\Delta)}}{4\Delta\sin(\phi_{\G}/2)}\sum_{\rg:\,\phi_{\rg}=\phi_{\G}}\Gamma^{\n\m}(\rg)\sum_{\aa}\p w_{\aa}(\cs)\,\vv_{\rg}\cdot\aa\,,\label{eq:connection-decay}
\end{equation}
where the angle \(\phi_{\G}\) from Theorem~\ref{thm:flat} is \(2\pi/N\) for \(\C N\) and \(\D N\) with \(N\geq3\), \(2\pi/3\) for \(\T\), \(\pi/2\) for \(\O\), and \(2\pi/5\) for \(\I\).}

\vva{For a bound beyond leading order, note that \(|\phi+2\pi n|\geq\phi\) for \(0<\phi\leq\pi\).
The sum in Eq.~\eqref{eq:poisson} and its derivative are then bounded by constant multiples of \(e^{-\phi^{2}/(4\Delta)}\) and \(\Delta^{-1}e^{-\phi^{2}/(4\Delta)}\), respectively, for \(\Delta\leq1\).
The coefficient of each \(\p w_{\aa}(\cs)\) in Eq.~\eqref{eq:heat-connection} is therefore bounded by a constant multiple of \(\Delta^{-1}e^{-\phi_{\G}^{2}/(4\Delta)}\).
It is zero for every \(\Delta\) if \(\G\) has no rotation angles other than \(0\) and \(\pi\), i.e., for \(\G\in\{\C 1,\C 2,\D 2\}\).
This proves conditions (i) and (ii) of Theorem~\ref{thm:flat}, along with the stated rate, for every irrep of every finite \(\G\).}

\vva{Since \(\bra\hat{\mathbf{\ell}}^2\ket\simeq\frac{3}{2\Delta}\), the holonomy of approximate states approaches the monodromy exponentially in their mean squared angular momentum.
The holonomy is thus robust to finite-energy effects.}

\subsubsection{\vva{Continuous symmetry groups}\label{subsec:connection-continuous}}

\vva{For continuous groups, Eq.~\eqref{eq:component-sum} yields the connection directly for every \(\Delta\).
Here, the group sum in Eq.~\eqref{eq:Zak} becomes an integral, as in Eq.~\eqref{eq:continuous-coset-states}.
The order \(|\G|\) becomes the group volume (\(2\pi\) for \(\C\infty\) and \(4\pi\) for \(\D\infty\)), which cancels in Eq.~\eqref{eq:full-connection}.
Let \(\G=\C\infty\) consist of \(\z\)-axis rotations, and let \(\Gamma=\lambda\in\Z\).
Each momentum \(\ell\geq|\lambda|\) then contains one copy of \(\Gamma\), spanned by \(|\om=\lambda\ket\) (see Example~\ref{ex:cinfty}).
Since \(\bra\om|L_{\aa}^{\ell}|\om\ket=\om\,\d_{\aa\z}\), the momentum sums in the numerator and denominator of Eq.~\eqref{eq:full-connection} cancel, leaving \(A=\lambda\,\p w_{\z}(\cs)\).
In the Euler-angle parameterization \(\cs=(\alpha\beta)\) from Example~\ref{ex:cinfty-position}, this is the one-form \(A=\lambda\cos\beta\,\dd\alpha\).
It is the connection of a monopole of charge \(\lambda\), whose Berry phase depends on the area enclosed by the path \cite{berry_quantal_1984}.
Its curvature \(F=\dd A=\lambda\sin\beta\,\dd\alpha\wedge\dd\beta\) vanishes only for \(\lambda=0\).
In other contexts, \(\lambda\up\so{3}\) is called the spin-weighted representation \cite[Sec.~12.3]{marinucci_random_2011}\cite{fan_Quantum_2023}.}

\vva{For \(\G=\D\infty\), the one-dimensional irreps \(\a_{1}\) and \(\a_{2}\) occur once in each even and odd momentum, respectively, spanned by \(|\om=0\ket\), so \(A=0\).
Each two-dimensional irrep \(\lambda\in\Z_{>0}\) from Table~\ref{tab:groups} occurs once in each momentum \(\ell\geq\lambda\), spanned by \(|\om=\pm\lambda\ket\).
These irreps do not occur for the nuclear spin isomers considered here (see Table~\ref{tab:isomer-table}).
Since \(L_{\x}\) and \(L_{\y}\) change \(\om\) by one and \(2\lambda\geq2\), only \(L_{\z}\) contributes.
This yields \(A=\lambda\,\textrm{diag}(1,-1)\,\p w_{\z}(\cs)\) in the basis \(\{|\om=\lambda\ket,|\om=-\lambda\ket\}\), whose curvature \(F=\lambda\,\textrm{diag}(1,-1)\sin\beta\,\dd\alpha\wedge\dd\beta\) is nonzero.}

\subsection{\vva{Flat connections and the fundamental group}}
\label{subsec:fundamental-group}

Induced representation state spaces, \(\Gamma\up\K\) for Lie group \(\K\) and \(\G\)-irrep \(\Gamma\), are examples of \(\G\)-bundles --- fiber bundles whose base space is \(\K/\G\) and whose fiber houses the irrep.

When \(\G\) is a Lie group, such bundles come with their own canonical or \(\G\)-connection~\vva{\cite{levay_canonical_1996}}.
\vva{In the Hermitian convention of Eq.~\eqref{eq:nonabelian-connection}, it reads \(A^{\G}_{\m\n}=-i[\dd\Gamma(\mathrm{pr}_{\mathfrak{g}}\,\cs^{-1}\p\cs)]_{\n\m}\), where \(\dd\Gamma\) is the derivative of \(\Gamma\).
The map \(\mathrm{pr}_{\mathfrak{g}}\) projects \(\K\)'s Maurer-Cartan form \(\cs^{-1}\p\cs\) onto \(\G\)'s Lie algebra \(\mathfrak{g}\), orthogonally with respect to an \(\mathrm{Ad}\)-invariant inner product.
The position-state connections of Secs.~\ref{subsec:connection-finite} and \ref{subsec:connection-continuous} coincide with \(A^{\G}\), which is zero for finite \(\G\) and equals \(\lambda\,\p w_{\z}\) for the irrep \(\lambda\) of \(\C\infty\).}

\vva{Flat bundles over \(\K/\G\) correspond to homomorphisms from the fundamental group \(\pi_1(\K/\G)\) into the unitary group of the fiber, up to conjugation \cite[Lemma~1]{Milnor_existence_1958}.
Irreducible flat bundles correspond to irreps \(\rho\) of \(\pi_1(\K/\G)\), which generate monodromies after non-contractible paths \cite[Eq.~(1)]{Sudarshan_configuration_1988}, with monodromy group \(\pi_1(\K/\G)/\ker\rho\).
Position-state monodromies instead realize the irrep \(\Gamma\) of \(\G\), with monodromy group \(\G/\ker\Gamma\)~\eqref{eq:monodromy-group}.
Condition (iv) of Theorem~\ref{thm:flat} says that the two monodromy groups are isomorphic for some \(\rho\).
Our conjecture~\cite{Albert_flat_2024} states that the position-state connection on \(\Gamma\up\K\) is flat if and only if there exists a \(\pi_1(\K/\G)\)-irrep \(\rho\) satisfying Eq.~\eqref{eq:mathcup} with \(\so 3\) replaced by \(\K\).
Theorem~\ref{thm:flat} proves this conjecture for \(\K=\so 3\) and every proper closed subgroup \(\G\).}

\begin{proof}[\vva{Proof of Theorem~\ref{thm:flat}}]
\vva{Up to conjugation, \(\G\) is finite, \(\C\infty\), or \(\D\infty\).
For finite \(\G\), conditions (i) and (ii) hold by Sec.~\ref{subsec:connection-finite}, and (iii) holds because the identity component of a finite \(\G\) is \(\{\re\}\).
The universal cover \(\su 2\to\nicefrac{\so 3}{\G}\) has the binary group \(2\G\) as its deck group, so \(\pi_1(\nicefrac{\so 3}{\G})=2\G\)~\cite{mermin_topological_1979}.
The projection \(\varpi\colon 2\G\to\G\) is onto, so \(\rho=\Gamma\circ\varpi\) is an irrep of \(\pi_1(\nicefrac{\so 3}{\G})\).
It satisfies \(\pi_1(\nicefrac{\so 3}{\G})/\ker\rho\cong\Gamma(\G)\cong\G/\ker\Gamma\), so (iv) holds.
For \(\C\infty\) and \(\D\infty\), Sec.~\ref{subsec:connection-continuous} shows that (i) and (ii) hold if and only if \(\Gamma\) is the trivial irrep \(\lambda=0\) of \(\C\infty\) or one of the irreps \(\a_{1}\) and \(\a_{2}\) of \(\D\infty\).
These are exactly the irreps that satisfy (iii).
Since \(\pi_1(\SS^2)=\C 1\) and \(\pi_1(\rptwo)=\C 2\), (iv) holds for these irreps, whose monodromy groups are \(\C 1\), \(\C 1\), and \(\C 2\).
It fails for the irreps \(\lambda\neq0\) of \(\C\infty\) and \(\D\infty\), whose monodromy groups \(\C\infty/\C{|\lambda|}\) and \(\D\infty/\C{\lambda}\) are infinite.}
\end{proof}

\vva{The conjectured equivalence fails for general Lie groups \(\K\).
Consider \(\K=\su 2\times\su 2\), \(\G=\su 2\times\{\re\}\), and \(\Gamma\) the defining irrep of the first factor, with position states defined as in Sec.~\ref{sec:position}.
Position states then factor into a fixed fiber state and a position state on \(\K/\G\cong\su 2\).
The connection is thus zero by the tracelessness argument of Sec.~\ref{sec:asymmetric-holonomy}.
However, \(\pi_1(\K/\G)=\pi_1(\SS^3)\) is trivial, while \(\G/\ker\Gamma\cong\su 2\).
For compact connected \(\K\), condition (iii) remains equivalent to the conjectured condition, since \(\pi_1(\K/\G)\) is countable and maps onto the component group of \(\G\).
If the position-state connection equals \(A^{\G}\), as it does for \(\K=\so 3\), it is flat if and only if \(\dd\Gamma(\mathrm{pr}_{\mathfrak{g}}[\xi,\eta])=0\) for all \(\xi,\eta\perp\mathfrak{g}\).
This criterion is equivalent to condition (iii) whenever such projected brackets span \(\mathfrak{g}\).
They span \(\mathfrak{g}\) for \(\nicefrac{\so 3}{\C\infty}\) but not in the above counterexample.}

\section{Fiber codes}\label{sec:fiber-codes}

\nw{
Following the notation of the previous sections, we define our encoding to be in the fiber degree of freedom of the ``identity'' orientation, \(\cs = \re\), of an entangled isomer \(\l \equiv \Gamma\) (i.e., whose irrep dimension \(\dm =\dim \Gamma > 1\)).
A basis of code states, or codewords, for this qu\(\dm\)it encoding consists of position states~\eqref{eq:initial-coset-state}
\begin{equation}\label{eq:fiber}
    |\overline{\mu}\ket \propto |\cs=\re,\mu\ket\quad\quad\text{for}\quad\quad\m\in \{1,2,\cdots,\dm\}~.
\end{equation}
We omit the residual nuclear \(|\chi\ket\) factor from now on since we do not use it to store information.

The above encoding lies fully in a \textit{single} molecular orientation \(\re\), and we will show that this encoding is comparable to other encodings for asymmetric molecules whose codewords are superpositions of \textit{several} molecular orientations~\cite{albert_robust_2020}.

Errors acting on the molecule can be grouped into two types --- inter-isomer and intra-isomer.
\textit{Inter-isomer} errors cause the information to leak out of the \(\Gamma\) isomer into a different isomer of the same molecule.
Such errors can, in principle, be detected by monitoring the other isomers.
Some errors will also be correctable, but determining correctability requires developing a physical basis of inter-isomer operators --- an open question from both the chemical and quantum-information perspectives that is outside the scope of this work (cf. \cite{Jain_AE_2023}).

We focus on \textit{intra-isomer} errors, which cause transitions within the given isomer.
We generalize the noise operators from Ref.~\cite{albert_robust_2020} for asymmetric and rotationally \(\C\infty\)-symmetric molecules to rotationally and perrotationally \(\G\)-symmetric isomers.

\subsection{Intra-isomer noise operators}

Intra-isomer operators can shift the position of the molecule by applying some lab-frame rotation \(\gr\) and/or kick the molecule's angular momentum by some amount \(\ell\).
General noise operators are products of position shifts \(\rr\) and momentum kicks \(\hat{H}\),
\begin{equation}
\hat{E}_{m\k}^{\ell}(\gr)={\textstyle \sqrt{\frac{\dm}{|\G|}}}~\rr_{\gr}\hat{H}_{m\k}^{\ell}~.
\end{equation}
Here, \(\ell\), \(m\), and \(\k\) are indices associated with the induced representation \(\Gamma\up\so 3\), and only kicks by these specified values keep the system within the given isomer.

Above, the $X$-type operators are lab-frame rotations acting as
\begin{subequations}\label{eq:action}
\begin{equation}
\rr_{\gr}|\cs,\mu\ket=\frac{1}{\sqrt{|\G|}}\sum_{\tau=1}^{\dm}s_{\tau}\sum_{\rg\in\G}\Gamma^{\mu\tau}(\rg)|\gr\cs\rg\ket|\tau\ket~,
\end{equation}
where, for convenience, we have expressed the codeword in its Zak form~(\ref{eq:Zak}) in terms of $\so 3$ position states $|\gr\ket$.
Here and for the remainder of this subsection, every ket that contains only \textsf{sans serif}
text represents an $\so 3$ position state.

The $Z$-type momentum kicks are diagonal in the position-state basis,
\begin{equation}
\hat{H}_{m\k}^{\ell}|\cs,\m\ket=H_{m\k}^{\ell}(\cs,\m)|\cs,\m\ket~,
\end{equation}
whose diagonal matrix elements are $\G$-adapted harmonics~(\ref{eq:harmonic}).
One can show these form an orthonormal basis for functions on the induced representation using the \(\G\)-adapted orthogonality relation~\eqref{eq:explicit-orthocompleteness}.
\end{subequations}

Products of shifts and kicks form a complete basis for the space of all physical operators on the isomer, meaning that any noise operator can be expanded as a sum of such products.
To show this, we need to prove the following,
\begin{widetext}
\begin{equation}
\int_{\so 3}\dd\gr\sum_{\ell\down\Gamma}\sum_{|m|\leq\ell}\sum_{\k=1}^{\mult(\ell)}\bra\cs,\mu|\hat{E}_{m\k}^{\ell}(\gr)|\rb,\nu\ket \bra\rb^{\pr},\nu^{\pr}|\hat{E}_{m\k}^{\ell}(\gr)|\cs^{\pr},\mu^{\pr}\ket^{\star}=\bra\cs,\mu|\cs^{\pr},\mu^{\pr}\ket\bra\rb,\nu|\rb^{\pr},\nu^{\pr}\ket~,\label{eq:proof-of-operator-set}
\end{equation}
\end{widetext}
for any coset representatives $\cs,\cs^{\pr},\rb,\rb^{\pr}\in\so 3/\G$ and $\Gamma$-irrep elements $\m,\m^{\pr},\n,\n^{\pr}$.

Using the action~\eqref{eq:action} of the two types of noise operators, we first simplify the following matrix element to
\begin{equation}
\bra\cs,\mu|\hat{E}_{m\k}^{\ell}(\gr)|\rb,\nu\ket={\textstyle \sqrt{\frac{\dm}{|\G|}}}H_{m\k}^{\ell}(\rb,\nu)\sum_{\rg\in\G}\Gamma^{\nu\mu}(\rg)\bra\cs\rg^{-1}|\gr\rb\ket~.
\end{equation}
Then, we plug this into two places in the left-hand side of Eq.~\eqref{eq:proof-of-operator-set} and apply the $\G$-adapted completeness relation~(\ref{eq:explicit-orthocompleteness}), simplifying the left-hand side to
\begin{equation}
\!\!\!\!\bra\rb,\nu|\rb^{\pr},\nu^{\pr}\ket\!\!\int_{\so 3}\!\!\!\!\dd\gr{\textstyle \frac{\dm}{|\G|}}\sum_{\rg,\rh\in\G}\Gamma^{\nu\mu}(\rg)\Gamma^{\nu\mu^{\pr}\star}(\rh)\bra\cs\rg^{-1}|\gr\rb\ket\bra\gr\rb|\cs^{\pr}\rh^{-1}\ket.
\end{equation}
Finally, we notice that the outer product of position kets $|\gr\rb\ket$
yields the identity on $\so 3$ position-state space when combined with the integral over $\gr$. 
Removing this identity, using Eq.~\eqref{eq:coset-partition}, and applying Schur orthogonality over $\G$ completes the proof.

\subsection{Protected encodings}

A fiber code~\eqref{eq:fiber} detecting intra-isomer operators \(\hat E\) satisfies~\cite{knill_theory_1997}
\begin{equation}
    \bra\overline{\n}|\hat{E}|\overline{\m}\ket=c_{\hat{E}}\d_{\mu\nu}
\end{equation}
for all codeword labels \(\n,\m\).
In words, either the error causes the information to leave the code space (corresponding to \(c_{\hat{E}} = 0\)) or the error acts as the identity on the codespace (up to the constant \(c_{\hat{E}}\)).
The constant can be infinite since we are using the idealized, non-normalizable position states as code states.
Using approximate but normalized position states will yield finite constants\vva{.
Approximate code states at a fixed orientation remain exactly orthogonal, while their overlaps with code states rotated by any \(\gr\notin\G\) are suppressed exponentially in the mean squared angular momentum of the states (see Sec.~\ref{subsec:approximate-states}).}

We can utilize our holonomy calculations from Sec.~\ref{sec:holonomy-position} to show that all but a measure-zero set of position shifts are detectable.
Since all rotations \(\rr_{\gr}\), except for those in \(\G\), map the codeword to a different coset, any rotations \textit{not} in the symmetry group will map the codeword to an error state that is orthogonal to the original codespace.
In other words, intra-isomer noise operators satisfy
\begin{equation}
    \bra\overline{\nu}|\hat{E}_{m\k}^{\ell}(\gr)|\overline{\m}\ket=0\quad\quad\forall\quad\gr\notin\G\,,
\end{equation}
detecting all rotations except those in the symmetry group (which in turn form fault-tolerant monodromy gates).
Moreover, rotations \(\gr\in\nicefrac{\so 3}{\G}\) (i.e., those that are also coset representatives closest to the identity) are correctable since one can undo them by rotating back to the codespace without inducing a monodromy.

Protecting against momentum kicks requires picking an isomer for which transitions between momentum states require more than a single quantum of angular momentum.
The set of momentum kicks \(\hat{H}^{\ell}_{m\k}\) for a given isomer \(\Gamma\) takes values in momenta \(\ell \down \Gamma\), i.e., those momenta which contain the isomer's irrep when restricted to the symmetry group (see Sec.~\ref{sec:spin}).
This set has a minimal element \(\ell_{\text{min}}\), and the corresponding isomers are immune to momentum kicks by \(\ell < \ell_{\text{min}}\)
since kick operators by such amounts do not exist within the isomer.


We catalogue all \vva{multi-dimensional} irreps with \(\ell_{\text{min}}>1\), which consist of all \vva{multi-dimensional} irreps in Table~\ref{tab:monodromy} excluding those that appear at \(\ell = 1\) (namely, \vva{the dihedral \(\e_1\),} the tetrahedral \(\ti\), and the octahedral and icosahedral \(\ti_1\) irreps).
All such irreps yield fiber encodings that protect against both position shifts and momentum kicks and that require only a single molecular orientation.
The corresponding nuclear spin isomers are marked with a filled-in star in the table.
\vva{This immunity is a statement about the kick operators of Eq.~\eqref{eq:action}.
Physical perturbations that depend only on the molecule's orientation, such as stray-field couplings to its multipoles or polarizability, satisfy \(V(\gr\rg)=V(\gr)\) for all \(\rg\in\G\).
By Eq.~\eqref{eq:Zak}, such a perturbation acts on each position state as the scalar \(V(\cs)\), leaving the fiber untouched for any rank of \(V\).
Nuclear-spin couplings that are symmetric under permutations of identical nuclei act only on the nuclear-spin factor \(|\chi\ket\).
The fiber is thus affected only by rotations along non-contractible paths (monodromy gates or tunneling) and by couplings that distinguish identical nuclear spins.
The latter include tensor spin-rotation and spin-spin interactions~\cite{itano1980avoided}.
They also cause inter-isomer leakage, and they can act on the fiber even when their orientation dependence has rank below \(\ell_{\text{min}}\).}

\begin{example}[\(\D N\) symmetry]
Entangled nuclear spin isomers with dihedral perrotational symmetry correspond to two-dimensional rotation irreps \(\Gamma = \e_i\) for some \(i \geq 1\).
Branching rules~\cite{altmann_point-group_1994} reveal that the \(i\)th irrep features first at angular momentum \(i\), corresponding to \(\ell_{\text{min}} = i\).
Fiber codes of such isomers are immune to momentum kicks up to \(i-1\) and  encode a qubit (since \(\dm = 2\)).
Such codes also correct rotations in \(\nicefrac{\so 3}{\D N}\).

The simplest case is \(\D 5\) symmetry, whose group has two 2D irreps, \(\e_1\) and \(\e_2\).
The \(\e_2\) isomer of such molecules is thus immune to single momentum kicks.
Perrotationally \(\D 5\)-symmetric molecules include pentagonal planar molecules like \ce{XeF5^{-}} as well as the cyclopentadienide molecule and the rather large pentaphenyl-cyclopentadienyl molecule.

Dihedral fiber codes are comparable in performance to dihedral molecular codes, which are based on nested groups \(\D{N/2}\subset \D{N}\) (for even \(N\))~\cite{albert_robust_2020}.
Both codes encode a qubit and correct rotations in \(\nicefrac{\so 3}{\D N}\), but the number of detectable momentum kicks is never larger for fiber codes since the irrep index always satisfies \(i < N/2\) (with dihedral molecular codes detecting \(< N/2\) kicks).
However, fiber codes are much less ``quantum'': dihedral molecular codewords consist of a superposition of \(N/2\) position states of any asymmetric molecule, while dihedral fiber codewords consist of a \textit{single} position state of an entangled perrotationally \(\D N\)-symmetric nuclear spin isomer.
Moreover, the dihedral fiber code admits a fault-tolerant dihedral gate set via monodromy (see Table~\ref{tab:monodromy}), while the only fault-tolerant operation admitted by the molecular code is the logical-\(X\) codeword permutation.
\end{example}

\begin{example}[\(\O\) symmetry]
The two-dimensional \(\e\) and three-dimensional \(\ti_2\) irreps of the octahedral group both feature first at angular momentum \(\ell = 2\), meaning that fiber codes of either isomer are immune to single momentum kicks.
The codes encode a qubit and qutrit, respectively, and correct rotations in \(\nicefrac{\so 3}{\O}\).
Examples include the \(\e^{\star}\) and \(\ti_2^{\star}\) isomer of \ce{SF6}, respectively~\cite{harter_bands_1977,harter_orbital_1977,mcdowell1982modern}.

Octahedral fiber codes can be compared to qubit molecular codes based on nested subgroups \(\T \subset \O\).
Both codes protect against rotations in \(\nicefrac{\so 3}{\O}\), but the molecular code detects up to two momentum kicks, while either fiber code is immune to only a single kick.
The molecular code consists of superpositions of 12 molecular positions, while the fiber code encodes using only one.
Moreover, the octahedral fiber code admits either a fault-tolerant dihedral or octahedral gate set via monodromy (see Table~\ref{tab:monodromy}), while the only fault-tolerant operation admitted by the molecular code is the logical-\(X\) codeword permutation.
\end{example}

\begin{example}[\(\I\) symmetry]
The five-dimensional \(\h\) irrep of the icosahedral group features first at momentum \(\ell = 2\), meaning that its corresponding five-dimensional fiber encoding is immune to single momentum kicks.
At \(\ell = 3\), one encounters the three-dimensional \(\ti_2\) and four-dimensional \(\g\) irreps, yielding respective qutrit and quartrit fiber codes immune to single and double momentum kicks.
All such codes can correct rotations in the \vva{Poincar\'e homology} sphere, \(\nicefrac{\so 3}{\I}\).

Icosahedral fiber codes can be compared to molecular codes based on nested subgroups \(\T \subset \I\), which encode a five-dimensional qudit.
Both codes protect against rotations in \(\nicefrac{\so 3}{\I}\) and admit a fault-tolerant implementation of the icosahedral gate group.
The molecular code detects up to two momentum kicks, one more than the \(\h\)-irrep and the same as the \(\ti_2\)-irrep and \(\g\)-irrep fiber codes.
\end{example}
}

\begin{figure}[t]
\includegraphics[width=0.6\columnwidth]{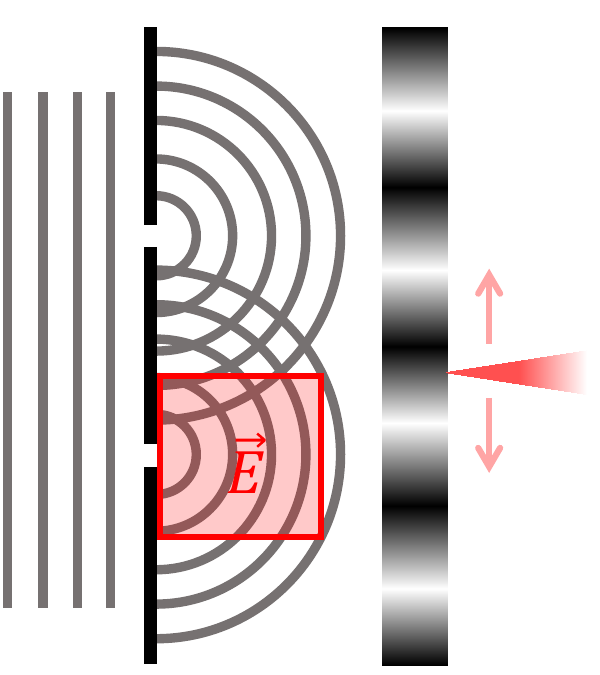}

\caption{\label{fig:interferometry}
Sketch of the matter-wave diffraction approach to induce a monodromy.
A single isomer of a homonuclear diatomic (i.e., either para or ortho) is initialized in the span of states \(|^{\ell}_{m=0}\ket\) and diffracted by a double slit.
Ultrafast laser pulses are applied to the lower arm (in the region outlined by the red square) to re-orient the molecules and induce an isomer-dependent monodromy of \(\pm1\) (see Fig.~\ref{fig:stroboscopic} for details). The phase of the matter-wave interference fringes can be probed by monitoring the photo-ionization probability as a function of distance (red needle). 
%
}
\end{figure}
\section{Monodromy detection via interferometry}
\label{sec:expt}

We turn to the problem of detecting the monodromy phase of homonuclear diatomic isomers. 
One approach is molecular interferometry: conceptually, a cold beam of molecules of a particular symmetry isomer (e.g., ortho or para) is aligned in some fixed position state and then split into two ``arms''. In practice, a fixed position state is difficult to realize, requiring a strong aligning field to be always present. An alternative is transient alignment via excitation by ultrafast laser pulses~\cite{Stapelfeldt_colloquium_2003}. We will comment on this further shortly.

A lab-based rotation, as described earlier, is applied to just one of the two arms. 
Downstream, the two arms are then recombined, whereby they can interfere with each other. 
In practice, the two arms can be separated in real space using matter wave diffraction, or by using the internal degrees of freedom of the molecule such as vibrational or electronic states. 

As mentioned in \vva{Sec.~\ref{subsec:holonomy-summary}}, initializing molecules in states other than position states can be used to simulate a monodromy.
Initializing in the span of \(|^{\ell}_{m=0}\ket\) and applying an equatorial rotation also yields a \(\pm1\) phase (see Example~\ref{ex:deuterium-a2-monodromy}), albeit a phase that can no longer be attributed to a monodromy in position-state space.

\prg{Matter-wave diffraction}
Conceptually, the matter wave diffraction is easier to visualize, but in practice may yield poorer signal and phase contrast due to loss of molecular flux at each ``beamsplitter'' step of the interferometer. 
In this approach, a beam of aligned molecules is incident on a double slit, whose two paths correspond to the arms of the interferometer \vva{(Fig.~\ref{fig:interferometry})}.
\vva{Such an interferometer does not require a quantum-degenerate gas, since its first-order interference involves only each molecule interfering with itself~\cite{chapman_optics_1995,cronin_optics_2009}.
For example, an interferometer for a supersonic beam of homonuclear \ce{Na2} separated its arms by \(38\)~\(\mu\)m, enough to insert a thin barrier between them and a gas cell into one arm~\cite{chapman_optics_1995}.} 

Next, the molecules in the lower arm are rotated according to the stroboscopic sequence of ultrafast pulses described in Fig.~\ref{fig:stroboscopic}.
While hitting the molecule with an ultrafast pulse \vva{shortly after} every rotational revival (\vva{\(T_{\text{rev}}=\pi/B\approx 8\) ps for \ce{N2}}, with \(B\) the rotational constant) can effect a \textit{lab-frame} rotation in a manner compatible with transient alignment, it is also experimentally challenging, as typical pulsed laser repetition rates are too low. 
Furthermore, realistic molecules can support only about 10 good rotational revivals prior to dephasing due to rovibrational coupling \cite{Dooley_direct_2003,Spanner_field-free_2004,Lee_coherent_2006}, so one cannot apply realigning pulses at arbitrarily large integer multiples of \(T_{\text{rev}}\). 

One solution is to split up the aligning laser beam path into $N$ arms, with the \(j\)th arm experiencing a relative path length delay \vva{slightly longer than} \(jT_{\text{rev}}\), before recombining the arms again and directing them onto the molecules. 
Available laser power severely limits the number of times we can split and recombine a beam, meaning that we can afford roughly \(N\sim 3\) arms.

The polarization in the \(j\)th arm would then need to be tilted by \(\pi/3 \sim 60\) degrees relative to the \((j+1)\)-st arm. 
As long as the tilt is less than \(90\) degrees, there should be no ambiguity as to which direction the molecule should rotate.
Within the three revivals, a typical diatomic molecule in a supersonic molecular beam (longitudinal speed \(500\) m/s) only travels \vva{\(\sim 10\) nm}, allowing it to be fully reoriented well within the \(10\) \(\mu\)m waist of a single aligning beam.

Further downstream, those rotated molecules are recombined with the un-rotated molecules in the upper arm due to diffraction. 
The matter-wave interference leads to fringes in the molecular flux. 
The fringes are phase shifted compared to those obtained in a classic double slit experiment (i.e., when no molecular rotation is induced in the lower arm). 
The phase of the molecular interference fringes could be observed by scanning an ionization laser vertically and recording molecular ion counts as a function of distance. 

\bibliography{references}

\end{document}